\documentclass[twocolumn]{aastex62}

\usepackage{gensymb}
\usepackage{natbib}

\usepackage[utf8]{inputenc}
\usepackage[maxfloats=1000]{morefloats}

\shorttitle{The 3.5 keV Line}
\shortauthors{Sicilian et al.}

\turnoffedit

\begin{document}

\title{Probing the Milky Way's Dark Matter Halo for the 3.5 keV Line}

\email{d.sicilian@umiami.edu}

\author{Dominic Sicilian}
\affiliation{University of Miami, Coral Gables, FL}
\affiliation{Center for Astrophysics $|$ Harvard \& Smithsonian, 60 Garden Street, Cambridge, MA}

\author{Nico Cappelluti}
\affiliation{University of Miami, Coral Gables, FL}

\author{Esra Bulbul}
\affiliation{Center for Astrophysics $|$ Harvard \& Smithsonian, 60 Garden Street, Cambridge, MA}
\affiliation{Max Planck Institute for Extraterrestrial Physics, Garching bei München, Bayern, Germany}

\author{Francesca Civano}
\affiliation{Center for Astrophysics $|$ Harvard \& Smithsonian, 60 Garden Street, Cambridge, MA}

\author{Massimo Moscetti}
\affiliation{University of Miami, Coral Gables, FL}
\affiliation{Palmer Trinity School, Miami, FL}

\author{Christopher S. Reynolds}
\affiliation{Institute of Astronomy, University of Cambridge, Madingley Road, Cambridge CB3 OHA, UK}

\begin{abstract}

We present a comprehensive search for the 3.5 keV line, using $\sim$51 Ms of archival \textit{Chandra} observations peering through the Milky Way's Dark Matter Halo from across the entirety of the sky, gathered via the \textit{Chandra} Source Catalog Release 2.0. We consider the data's radial distribution, organizing observations into four data subsets based on angular distance from the Galactic Center. All data is modeled using both background-subtracted and background-modeled approaches to account for the particle instrument background, demonstrating statistical limitations of the currently-available $\sim$1 Ms of particle background data. A non-detection is reported in the total data set, allowing us to set an upper-limit on 3.5 keV line flux and constrain the sterile neutrino dark matter mixing angle. The upper-limit on sin$^2$(2$\theta$) is $2.58 \times 10^{-11}$ \edit2{(though systematic uncertainty may increase this by a factor of $\sim$2)}, corresponding to the upper-limit on 3.5 keV line flux of $2.34 \times 10^{-7}$ ph s$^{-1}$ cm$^{-2}$. \edit2{These limits show consistency with recent constraints and several prior detections}. Non-detections are reported in all radial data subsets, allowing us to constrain the spatial profile of 3.5 keV line intensity, which does not conclusively differ from Navarro-Frenk-White predictions. Thus, while offering heavy constraints, we do not entirely rule out the sterile neutrino dark matter scenario or the more general decaying dark matter hypothesis for the 3.5 keV line. We have also used the non-detection of any unidentified emission lines across our continuum to further constrain the sterile neutrino parameter space.

\end{abstract}


\section{Introduction}

Since its discovery (\citealt{zwicky 1933}; \citealt{zwicky 1937}), dark matter has been shown by numerous observations to be the Universe's dominant source of gravity and composed of undiscovered, non-baryonic material (\citealt{rubin 1970}; \citealt{rubin 1980}; \citealt{clowe 2006}). Its nature and composition, however, remain unknown since the Standard Model does not offer any viable dark matter candidate \citep{boyarsky 2019}.

The solution to the dark matter problem could lie in neutrino cosmology. The three flavors of Standard Model neutrinos are massless and display only left-handed chirality. It is now known that neutrinos oscillate between flavors and therefore are not massless (\citealt{kajita 1999}; \citealt{mcd 2002}), in contrast to the Standard Model, and the probability of oscillation between flavors can be described by the mixing angle, $\theta$ \citep{pal 1982}. 
 
There is currently no explanation for the collective phenomenon of neutrino mass and oscillation, but a possible solution is the existence of hypothetical right-handed neutrinos (\citealt{dodwid 1994}; \citealt{aba 2001a}; \citealt{dolgov 2002}; \citealt{boyarsky 2006b}; \citealt{boyarsky 2009b}; \citealt{boyarsky 2019}). This could give neutrinos more direct correspondence to other known fermions---all of which can exhibit both right- and left-handed chirality---and resolve major problems in modern physics in addition to the neutrino mass problem. Since the weak interaction only couples to left-handed neutrinos (\citealt{drewes 2013}, \citealt{pal 1982}, etc.), right-handed neutrinos are referred to as sterile neutrinos due to their resulting lack of interaction via any forces besides gravity. In accordance with this naming convention, left-handed neutrinos are known as active neutrinos. Incidentally, the sterile neutrino could also solve the matter-antimatter asymmetry problem in the early Universe by giving rise to baryons through the process of leptogenesis (\citealt{asaka 2005}; \citealt{drewes et al 2013}; \citealt{drewes 2016}; \citealt{drewes 2017}; see \citealt{boyarsky 2019} for further discussion).

The most relevant feature of the sterile neutrino to this work is its status as a dark matter candidate. Active neutrino masses are well known for being too small to constitute dark matter \citep{boyarsky 2019}, but sterile neutrinos can have much larger masses. In particular, sterile neutrino dark matter is thought to have mass in the keV range (\citealt{aba 2001a}; \citealt{aba 2001b}; \citealt{boyarsky 2006a}; \citealt{boyarsky 2009a}). Also, in a process separate from neutrino oscillation, any neutrino can decay into another neutrino with lower mass. The probability of this decay depends on the difference in masses of the neutrinos in question, so it rarely happens to active neutrinos \citep{pal 1982}. However, the probability is considerably higher for a sterile neutrino with keV mass, with the decay rate given by:
\begin{equation}\label{decay_rate}
\Gamma_\gamma(m_s,\theta) = 1.38 \times 10^{-29} s^{-1} \bigg(\frac{sin^2(2\theta)}{10^{-7}}\bigg)\bigg({\frac{m_s}{1 keV}}\bigg)^5
\end{equation}
where $m_s$ is the sterile neutrino's mass and $\theta$ is the corresponding mixing angle \citep{pal 1982}. This decay process results in an active neutrino and a photon with E = $m_s$/2, making the decay of a keV sterile neutrino observable by current X-ray telescopes such as \textit{Chandra} (\citealt{aba 2001a}; \citealt{aba 2001b}).

When an unidentified emission line was discovered by \cite{esra 2014} near 3.5 keV at $\sim$4.5$\sigma$ significance in 73 stacked \textit{XMM-Newton} galaxy clusters, in addition to Perseus and other clusters using \textit{Chandra}, it served as the first possible experimental evidence of sterile neutrino dark matter decay. Soon after, the line was detected again with \textit{XMM-Newton} at 3$\sigma$ significance in M31 by \cite{boyarsky 2014}, and in the Milky Way's Galactic Center at 5.7$\sigma$ by \cite{boyarsky 2015}. The line was later detected in the Perseus cluster at $\sim$5$\sigma$ using \textit{Suzaku} by both \cite{urban 2015} and \cite{franse 2016}. Recently, \cite{nico 2018} detected a possible feature resembling the line at $\sim$3$\sigma$ in the \textit{Chandra}-COSMOS Legacy Survey (CCLS) field and \textit{Chandra} Deep Field South (CDFS), and another detection was made using archival \textit{Chandra} observations of the Galactic bulge by \cite{hofmann 2019}.

Due to the use of CCDs in the majority of 3.5 keV line detections, an unknown feature of CCDs has been hypothesized to be the source of the apparent emission. However, the line was also detected by \textit{NuSTAR}'s cadmium zinc telluride (CdZnTe or CZT) detector in its observations of the Bullet cluster \citep{wik 2014}. The line was later detected again using \textit{NuSTAR}, this time at 11$\sigma$ by \cite{neronov 2016a} in the Milky Way's Dark Matter Halo, through the CCLS field and Extended \textit{Chandra} Deep Field South (ECDFS). The \textit{NuSTAR} detections have been questioned due to the proximity of 3.5 keV to the lower bound of \textit{NuSTAR}'s sensitivity and the detection of the line by \cite{perez 2017} in a portion of observations where the FOV contains only Earth, suggesting \edit1{that at least some part of} the line in \textit{NuSTAR} is instrumental.

Despite the many detections by various instruments in dark matter-dominated objects, some studies produced non-detections. These include stacked \textit{Suzaku} clusters \citep{esra 2016}, \textit{XMM-Newton} observations of the Draco dwarf galaxy \citep{ruch 2016}, \textit{Hitomi} observations of the Perseus cluster \citep{hitomi 2017}, \edit2{and \textit{XMM-Newton} observations of various galaxy clusters \citep{bhargava 2020}}. The upper-limits provided by these non-detections are consistent with the original \cite{esra 2014} detection and do not rule out the decaying dark matter interpretation. Recently, \cite{dessert 2020} reported a non-detection in $\sim$31 Ms of archival \textit{XMM-Newton} observations directed through the Milky Way's Dark Matter Halo. The \cite{dessert 2020} results, however, have been questioned in the X-ray astrophysics community due to the unconventional nature of the analysis, which considered an unusually small energy band and failed to properly account for known emission features at 3.3 keV and 3.68 keV, in addition to possible technical errors in data reduction (\citealt{abazajian 2020}; \citealt{boyarsky 2020}). The fiducial constraints reported by \cite{dessert 2020} are in tension with the decaying dark matter interpretation of the 3.5 keV line, while supplemental upper-limits given in that work that account for the 3.3 keV and 3.68 keV emission features are marginally consistent with prior detections and are supported by \cite{boyarsky 2020}. \edit1{Both \cite{boyarsky 2020} and \cite{abazajian 2020} show that the high dark matter density estimate adopted by \cite{dessert 2020} may relax even the supplemental upper-limit by up to a factor of $\sim$3. In particular, \cite{dessert 2020} uses a local dark matter density of 0.4 GeV cm$^{-3}$, much higher than strong empirical values such as the 0.28 GeV cm$^{-3}$ found by \cite{zhang 2013}.} As discussed in \cite{dessert 2020a}, these supplemental results still constrain the sterile neutrino dark matter scenario, but cannot rule out the hypothesis \citep{boyarsky 2020}.

Virtually all non-astrophysical explanations for the 3.5 keV line can be classified as arising from instrumental effects, but due to the variety of X-ray telescopes that have observed the line, this explanation is generally considered unlikely for the majority of detections. The telescopes use different mirror coatings (either gold or iridium) and utilize different detectors, including CCDs and \textit{NuSTAR}'s CZT detector, although as mentioned, the \textit{NuSTAR} detections are likely instrumental. Another non-astrophysical interpretation could be statistical fluctuations, although this is also considered unlikely due to repeated high-significance detections. This prospect will be thoroughly addressed by the size of the data set used in this analysis.

Various non-dark matter, astrophysical explanations for the line have been discussed. Among these interpretations is contamination due to nearby K and Ar dielectric emission, both of which have been evaluated and subsequently ruled out using electron beam ion trap (EBIT) experiments (\citealt{esra 2019}; \citealt{gall 2019}; \citealt{weller 2019}). A current leading interpretation is charge exchange (CX) between bare Sulfur ions and neutral Hydrogen, described by \cite{esra 2014a}, \cite{gu 2015}, \cite{shah 2016}, and others referenced therein. These explanations, and all others that feature baryonic matter, can be tested by considering the flux of the 3.5 keV line as a function of distance from the Galactic Center. This can then be compared to the predictions made for decaying dark matter by the Navarro-Frenk-White (NFW) profile \citep{nfw 1997}. The degree to which the flux profile matches the NFW profile represents the likelihood that the 3.5 keV line arises from decaying dark matter, since baryonic matter has a different distribution function. Recently, \cite{boyarsky 2018} showed rough consistency between the 3.5 keV line flux profile and the NFW profile in the region between 10 arcmin and 35 deg from the Galactic Center.

In this work, we employ a methodology designed to reach the most decisive conclusions on the 3.5 keV line to date. We utilize an extremely comprehensive data set in which we minimize signals from baryonic matter by restricting our search to the Milky Way's Dark Matter Halo. Furthermore, we use data from \textit{Chandra} due to its high angular resolution and stable particle background relative to \textit{XMM-Newton}. These features specific to \textit{Chandra} give us the ability to detect a faint feature such as the putative 3.5 keV line. The results yielded by this ideal data set will then allow us to thoroughly explore the possibility of the 3.5 keV line's decaying dark matter interpretation.

\section{Data Selection}\label{data selection}

\begin{figure}[t!]
\includegraphics[width=9.cm]{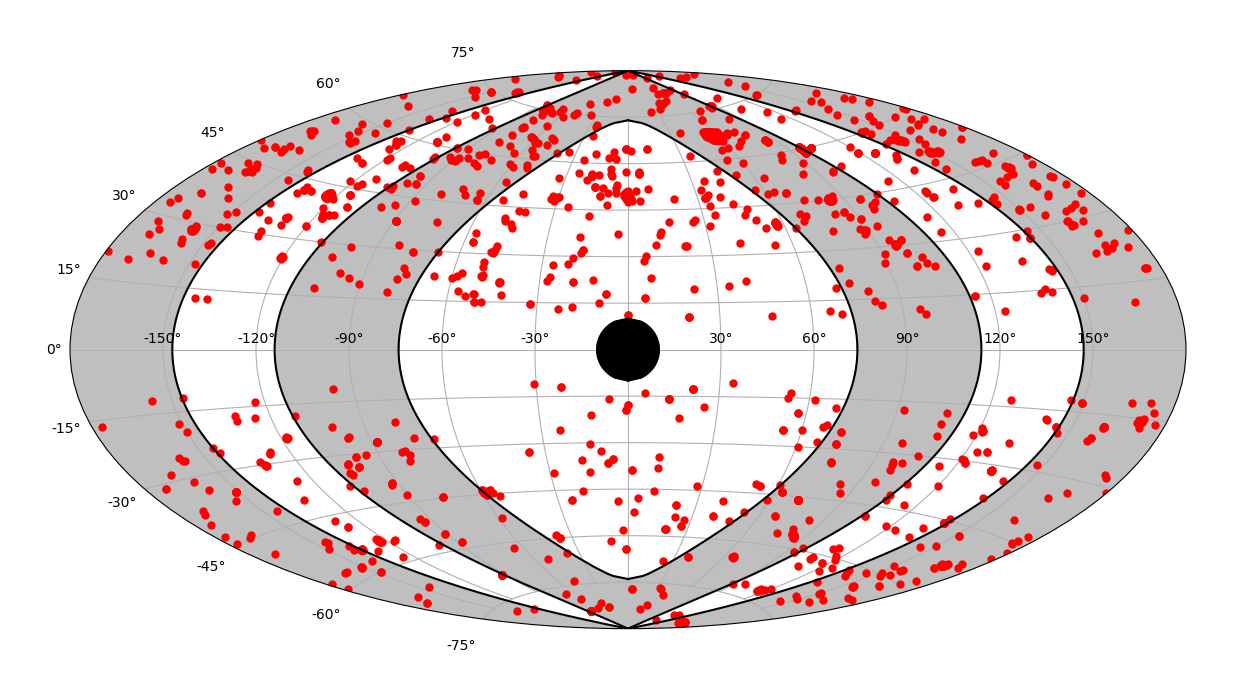}
\caption{The positions of all 1907 observations used in the final data set, plotted in galactic coordinates. The black region represents the excluded Galactic Center (up to an angular distance of 10 deg), while the alternating white and gray regions represent the 4 regions of varying angular distance. Note that no observations appear between $b = -10$ deg and $b = 10$ deg due to the exclusion of the Galactic Disc.}
\label{fig:obsid_map}
\end{figure}

\begin{table*}[t!]
\centering
 \begin{tabular}{c c c c c c c c c}
 \hline
 Bin & $\theta_{GC}$ [deg] & $t_{exp}$ [Ms] & ObsIDs & Counts & Counts & A$_{i}$ [deg$^2$] & A$_{f}$ [deg$^2$] \\ 
 
  & &  &  & (w/sources) & (w/out) &  &  \\ [0.5ex]
 \hline\hline
 
 1 & 10--74 & 8.00 & 306 & 1569344 & 1224559 & 19.18 & 13.44 \\ 

 2 & 74--114 & 14.07 & 715 & 3540528 & 2666281 & 57.96 & 43.98 \\

 3 & 114--147 & 17.58 & 473 & 4374038 & 2785435 & 36.24 & 19.57 \\
 
 4 & $>$147 & 11.00 & 413 & 2466970 & 1855048 & 27.63 & 19.05 \\
 
\hline
 & TOTAL & 50.65 & 1907 & 11950880 & 8531323 & 141.01 & 96.04 \\
 \hline
\end{tabular}
\caption{Breakdown of the angular distance bins showing the distances and amount of data contained in each, with the latter expressed in terms of exposure time ($t_{exp}$), number of observations (ObsIDs), and number of photons (counts) both before and after source removal. A$_{i}$ and A$_{f}$ denote the total detector exposure area before and after source removal, respectively. Note that the counts given here reflect the 2.9--5.6 keV energy band on which we analyze all spectra.
}
\label{tab:bin_info}
\end{table*}

This work considers the entirety of archival \textit{Chandra} observations that peer through the Milky Way's Dark Matter Halo and were documented by the \textit{Chandra} Source Catalog Release 2.0 (CSC 2.0; \citealt{evans 2010}; \citealt{evans 2019}) as of July 2019. This includes observations from 2000--2014, compared to the CSC 1.1, which includes data only up to 2009. The initial data set contains $\sim$94 Ms of exposure time. The final data set (after performing the cleaning process described below) consists of 1907 observations and contains $\sim$51 Ms, making this the most rigorous search for the 3.5 keV line to date and the largest data set in the history of X-ray astronomy. Notably, the data set (originally $\sim$94 Ms, $\sim$51 Ms when cleaned) is larger than that of the similar study of archival \textit{XMM-Newton} data by \cite{dessert 2020} (originally $\sim$86 Ms, $\sim$31 Ms when cleaned). Furthermore, \textit{Chandra} is better-suited for this analysis than \textit{XMM-Newton}. This is due in part to its ability to resolve as much as $\sim$80\% of the Cosmic X-Ray Background (CXB; \citealt{hickox 2007}), thus serving to isolate any possible signal from decaying dark matter. Moreover, \textit{Chandra}'s particle background is substantially more stable than that of \textit{XMM-Newton}, which varies by up to a factor of $\sim$10 on small scales due to solar flares \citep{esra 2020}, thus making \textit{Chandra} more sensitive to weak line searches when considering large data sets.

We consider only observations performed in \texttt{VFAINT} mode by front-illuminated (FI) CCDs of the Advanced CCD Imaging Spectrometer (ACIS; \citealt{garmire 2003}), due to the well-known continuum behavior of the front-illuminated ACIS particle background (see \citealt{hickox 2006} and \citealt{bartalucci 2014}). This includes all of ACIS-I (CCD IDs 0, 1, 2, and 3) and ACIS-S2 (CCD ID 6). To avoid contamination from the Galactic Center or Galactic Disc, we used only observations at latitudes $| b |\geq10$ deg. In addition, we restricted our data to observations with $N_{H} < 10^{22}$ cm$^{-2}$ to minimize the presence of ubiquitous baryonic matter. Upon compiling the list of observations compliant with these criteria, we downloaded those observations directly from the \textit{Chandra} archive. The only Galactic contamination remaining in the final data set is the unavoidable Hot Gas Halo, but its contribution to the X-ray spectrum is notable only in bands much softer than $3.5$ keV (\citealt{kuntz 2000a}; \citealt{kuntz 2000b}), which is therefore inconsequential to our analysis.

Making use of X-ray source detections reported in the CSC, all the point-sources in the selected fields were excised. In particular, the CSC 2.0 detects sources in stacks of observations, meaning that not all sources in a given observation's field of view (FOV) will necessarily contribute to the data collected by that observation. Our treatment, then, is extremely conservative, leaving virtually no contribution from point-sources in the final data set. To prevent contamination from extended emission, we removed all observations belonging to stacks containing extended sources. This was accomplished by first searching the list of extended sources reported in the CSC and removing all observations in stacks with extended emission, then by screening the target names of all remaining observations for low surface brightness extended sources missed by the CSC and removing all observations in the corresponding stacks. We then eliminated all observations containing any instances of instrumental temperature beyond the nominal ACIS operating range, in particular above $159$ K. Lastly, after reducing the data, observations with insufficient statistics were removed. After all data reduction and removal of unsatisfactory observations, we are left with the final $\sim$51 Ms data set. The spatial distribution of our final 1907 observations can be seen in Figure \ref{fig:obsid_map}. 

For the background, ACIS stowed data was utilized instead of the typical blank sky files. This was done for the purpose of considering the detector particle background and the unresolved CXB separately. The blank sky files are not suitable for this analysis due to their intrinsic inclusion of the Dark Matter Halo.

\section{Data Reduction}\label{data reduction}

The data reduction was performed using \textit{Chandra} Interactive Analysis of Observations (CIAO; \citealt{antonella 2006}). First, raw data was calibrated according to the \textit{Chandra} Calibration Database (CALDB) version 4.8.3 using the CIAO tool \texttt{chandra\textunderscore repro}, ensuring the preservation of \texttt{VFAINT} data using \texttt{check\textunderscore vf\textunderscore pha=yes}. Each observation was then matched with a background event file. The background file was produced using the CIAO tool \texttt{dmmerge} to combine the proper ACIS stowed files. These stowed files were identified using two criteria, namely epoch and active CCD numbers. The epoch was determined using the \texttt{DATE-OBS} value in the header of the observation's event file, and the active CCD numbers were determined using the \texttt{DETNAM} value in the header. All stowed files from the same epoch and from any CCD active in the observation were merged to produce the final background file. This ensured proper compatibility of the particle background with all observations, including matching calibration.

Light-curve filtering was applied to the data to remove background flares. This produced good time interval (GTI) files used to create cleaned event files. The stowed file for each observation was reprojected to match the coordinates of the observation.

The CSC was then used to remove all point-sources in each observation. After following the highly conservative protocol described in section \ref{data selection} to identify the point-sources, the point spread function (PSF) width for each source, given in the CSC at 1$\sigma$ confidence, was utilized to establish the effective extension of the point-source in the data. To be conservative, we used the corresponding 5$\sigma$ PSF width. For the numerous sources given in the CSC, but not supplied with PSF widths, we used a width corresponding to a circle of radius 5" to maximize removal of contamination while preserving our data. A region file containing all sources for each observation was made using this data together with the CIAO tool \texttt{dmmakereg}. This region was then inverted, resulting in a region file containing no sources, which was subsequently used as a spatial filter, on both the observation and the background, in the spectral extraction process.

After applying the spatial filter to each stowed file, and hence removing regions corresponding to sources in its associated observation, the exposure time of the stowed file was rescaled to match that of the observation. It should be noted that all observations in the data set had exposure times smaller than 200 ks, while all background exposure times are greater than 240 ks, and therefore there is no case in which the background exposure time was artificially increased. This was done to properly match each observation's particle background spectrum (also referred to as ``stowed spectrum" or ``background spectrum" hereafter) with the data, allowing us to properly weight each contributing background spectrum in the stacked background spectra. Here, ``weight" refers to the influence the spectrum from a particular CCD in a particular ACIS stowed observation has on the final stacked spectrum, which must be based on both its exposure time in observations from that epoch and on the particle background flux at the exact time of the corresponding observations. Our methodology achieves correct weighting by, through the stacking process, inherently scaling the contribution of each particular CCD's stowed spectrum from the ACIS stowed observations to that CCD's same contribution to the data set, resulting in a stacked stowed spectrum displaying the same behavior as the particle background component of the data set's stacked spectrum. 

Note that simply stacking the 3 original stowed files without first matching and rescaling them for each observation (to give 1907 tailored stowed spectra) produces a spectrum incompatible with the particle background contributions to the data set's stacked spectrum, since stacking the original stowed files would inherently and erroneously give equal weights to the stowed spectra of all CCDs used in the analysis and hence produce an incorrectly-shaped particle background spectrum. In contrast, as stated above, our method implicitly assigns the correct weight to each CCD, resulting in a particle background spectrum consistent with the data set. Also note that assigning weights based only on exposure time is similarly insufficient, since it fails to consider the variable particle background flux, while our method inherently accounts for all necessary factors to successfully produce a particle background spectrum correctly-matched to the data set.

The spectrum, redistribution matrix file (RMF), and ancillary response file (ARF) of each observation were obtained using the CIAO tool \texttt{specextract}, and the corresponding stowed spectrum was obtained using \texttt{dmextract}. No ARFs were obtained for the stowed data due to none of its signal passing through \textit{Chandra}'s High Resolution Mirror Assembly (HRMA), but RMFs were generated using methods similar to those employed by \cite{hickox 2006} and \cite{bartalucci 2014}.

Finally, the counts of each stowed spectrum were rescaled to match the corresponding observation using 2 principles of the ACIS particle background spectrum, both of which are detailed in \cite{hickox 2006}. First, the shape of the spectrum is known to stay constant (within $\sim$1--2\%), while the flux varies with the spacecraft's position. And second, \textit{Chandra}'s effective area is very low in the 9.5--12 keV band, and hence the particle background dominates. Exploiting these principles to normalize each stowed spectrum, we thus obtained a final corresponding particle background spectrum for each spectrum in the data set. Upon extracting and normalizing all spectra, \texttt{combine{\textunderscore}spectra} was used to merge them into a total of five stacked spectra to use in the analysis.

First, the spectra from all 1907 observations were stacked to produce one spectrum containing the entirety of the data set. All stowed spectra were stacked to produce the corresponding particle background spectrum.

The other four stacked spectra represent our four bins of angular distance from the Galactic Center, which can be used to study the possible decaying dark matter origins of the 3.5 keV line. The size of each bin is unique and was determined with the goal of producing as many bins as possible that contain sufficient exposure time for making a 3$\sigma$ detection of the 3.5 keV line, calculated based on the 3$\sigma$ detection made using \textit{Chandra} by \cite{nico 2018} and assuming an NFW profile for the line intensity.

To produce the bins, we computed each observation's distance from the Galactic Center based on its pointing coordinates. 
This distance was used to place it in one of the angular distance bins, according to our aforementioned criteria, resulting in 4 bins of angular distances, detailed in Table \ref{tab:bin_info}. Henceforth, the bins will be referred to as bins 1, 2, 3, and 4, numerically ordered by increasing angular distance from the Galactic Center.

Upon obtaining the five final spectra, the corresponding stowed spectra were also stacked to match them. Finally, the exposure time of each stacked stowed spectrum was readjusted from the data set's exposure time (a relic of our weighting method) to its actual value of 1022352.6 s (see Table \ref{tab:stowtime}) to ensure correct statistics, with counts scaled accordingly to preserve proper normalization of the count rate. These particle background spectra can be seen in Figure \ref{fig:stow_plots}, found in the Appendix.\footnote{Note that Figures \ref{fig:stow_plots} and beyond (up to the final Figure \ref{fig:cxb_bins_lineflux_prob}) are located in the Appendix.}

\section{Analysis}

The data analysis was performed using the spectral fitting package XSPEC 12.10.1f \citep{arnaud 1996} via the PyXspec 2.0.3 interface \citep{arnaud 2016}. All spectra were modeled using two approaches. The first involves subtracting the particle background before modeling and the second incorporates a particle background model (``background-subtracted" and ``background-modeled" henceforth, respectively). Gaussian statistics are used throughout due to the sufficient counts contained by all energy bins \citep{protassov 2002}. All background-subtracted spectra are binned such that each bin contains a minimum of 30 counts, while all background-modeled spectra are unbinned (including those analyzed in section \ref{totalcxb}).

\begin{table}[b!]
\centering
 \begin{tabular}{c c}
 \hline
 Year & $t_{exp}$ [Ms] \\ [0.5ex]
 \hline\hline
 
 2000 & 0.415  \\ 

 2005 & 0.367  \\

 2009 & 0.240 \\
\hline
 TOTAL & 1.022 \\
 \hline
\end{tabular}
\caption{Total exposure time of the three \textit{Chandra} ACIS stowed observations.}
\label{tab:stowtime}
\end{table}

\subsection{Treatment of the Background}

As previously mentioned, this analysis requires ACIS stowed spectra rather than blank sky spectra to avoid including a dark matter signal in the background. The ACIS stowed data contains only particle background, so the unresolved CXB was modeled separately.

A pivotal feature of the particle background is its low statistics relative to the data set. In 20 years of \textit{Chandra}, only $\sim$1 Ms of stowed observations has been taken (detailed in Table \ref{tab:stowtime}). However, our data set contains $\sim$51 Ms, putting its high count statistics at risk of inheriting noise from the relatively low particle background exposure. This effect is amplified by the prevalence of \textit{Chandra}'s particle background flux above $\sim$2 keV, which is especially dominant in observations after source-removal. In Table \ref{tab:snr}, we illustrate this using the signal-to-noise ratio in the data set before and after source removal on the band used in our analysis.

The low particle background statistics substantially hinders the background-subtracted results, making the background-modeled method far more statistically robust. However, the simplicity of the background-subtracted models is a possible advantage over the highly more complex background-modeled models, hence providing motivation to include background-subtraction in the analysis. Moreover, it can be useful when placed in the context of similar works, particularly \cite{nico 2018}.

\begin{table}
\centering
 \begin{tabular}{c c c c}
 \hline
  Data & Total & Particle Background & Signal- \\ 
   & Counts & Counts & to-Noise \\
 \hline\hline
 
 Full fields & 151177565 & 139226685 & 971.98 \\ 

 Source-removed & 126560247 & 118028924 & 758.35  \\ [0.5ex]

 \hline
\end{tabular}
\caption{Signal-to-noise ratio before and after excising sources. Counts are calculated on the 2.9--5.6 keV band analyzed in our models. \edit2{Particle background counts are estimated using the rescaled and renormalized $\sim$1 Ms ACIS stowed data.}}
\label{tab:snr}
\end{table}

\begin{figure}[t!]
\includegraphics[width=9.cm]{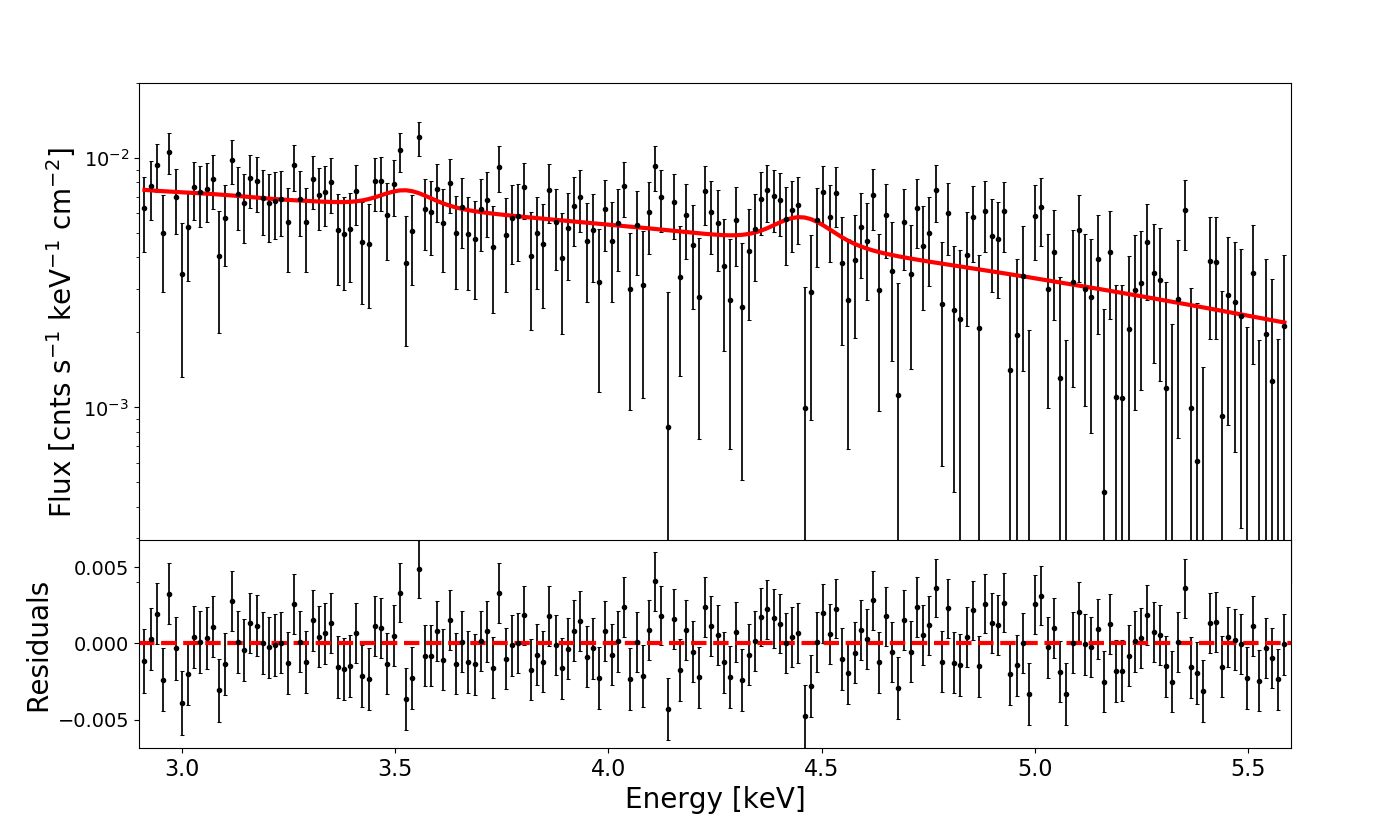}
\caption{Background-subtracted spectrum stacked from all observations in the data set.}
\label{fig:all_sub_line}
\end{figure}

\begin{table*}[t!]
\centering
 \begin{tabular}{ c  c  c  c  c  c }
 
 \hline
  Parameter [Unit] & All & Bin 1 & Bin 2 & Bin 3 & Bin 4 \\
  \hline\hline
  
  $\Gamma_{PL}$ & 1.42$_{-0.18}^{+0.18}$ & 1.38$_{-0.16}^{+0.16}$ & 1.51$_{-0.19}^{+0.19}$ & 1.35$_{-0.2}^{+0.2}$ & 1.45$_{-0.18}^{+0.18}$  \\ 

$I_{PL}$ [10$^{-4}$ ph s$^{-1}$ cm$^{-2}$] & 1.26$_{-0.27}^{+0.33}$ & 1.35$_{-0.26}^{+0.33}$ & 1.45$_{-0.32}^{+0.4}$ & 1.02$_{-0.24}^{+0.31}$ & 1.38$_{-0.3}^{+0.38}$  \\ 

$E_{4.5}$ [keV] & 4.46$_{-0.04}^{+0.08}$ & 4.46$_{-0.04}^{+0.07}$ & 4.47$_{-0.05}^{+0.08}$ & 4.46$_{-0.04}^{+0.08}$ & 4.46$_{-0.04}^{+0.08}$  \\ 

$I_{4.5}$ [10$^{-7}$ ph s$^{-1}$ cm$^{-2}$] & 8.37$_{-4.39}^{+4.82}$ & 9.06$_{-4.75}^{+4.97}$ & 8.24$_{-4.6}^{+5.02}$ & 8.53$_{-4.4}^{+4.62}$ & 9.04$_{-4.72}^{+4.98}$  \\ [0.5ex]
 \hline
\end{tabular}
\caption{Best-fit model parameters for all background-subtracted spectra with 1$\sigma$ errors.}
\label{tab:parameters}
\end{table*}

\begin{table}[b!]
\centering
 \begin{tabular}{c c c}
 \hline
 Bin & $E_{3.5}$ [keV] & $I_{3.5}$ [10$^{-7}$ ph s$^{-1}$ cm$^{-2}$] \\ [0.5ex]
 \hline\hline

 1 & 3.52$_{-0.06}^{+0.04}$ & 5.16$_{-3.35}^{+4.35}$ \\ 

 2 & 3.53$_{-0.06}^{+0.04}$ & 5.58$_{-3.62}^{+4.46}$ \\

 3 & 3.53$_{-0.03}^{+0.03}$ & 8.39$_{-4.28}^{+4.36}$ \\

 4 & 3.52$_{-0.06}^{+0.05}$ & 4.73$_{-3.14}^{+4.12}$ \\ [0.5ex]
 \hline
 ALL & 3.53$_{-0.05}^{+0.04}$ & 6.03$_{-3.69}^{+4.30}$ \\ [0.5ex]
 \hline
\end{tabular}
\caption{Best-fit 3.5 keV line energy and normalization for each background-subtracted spectrum with 1$\sigma$ errors.}
\label{tab:linepars}
\end{table}

\subsection{Background-Subtracted Modeling}\label{bgsub}

Each spectrum was modeled in the 2.9--5.6 keV band, chosen to be wide enough for establishing a reliable power-law component while minimizing total free parameters by avoiding emission features. The unresolved CXB continuum was modeled using an absorbed (\texttt{phabs} in XSPEC) power-law, with $N_H$ fixed at 10$^{20}$ cm$^{-2}$. This value is an approximation of the average column density across all fields used in the analysis, based on \cite{dickey 1990}, and does not contribute to the band of interest. An emission line was fitted at $\sim$4.5 keV, in agreement with \cite{nico 2018}'s detection of a similar feature. As described by \cite{nico 2018}, this feature is consistent with known instrumental lines due to Ti K$_{\alpha_{1,2}}$ at 4.51 keV and 4.50 keV, respectively, within \textit{Chandra}'s energy resolution and the 1$\sigma$ error range of our best-fit line energy values. The line is hard to detect in the $\sim$1 Ms of ACIS stowed data, but appears more clearly in the $\sim$10 Ms of data in \cite{nico 2018} and the $\sim$51 Ms in this work due to the considerably higher statistics of the data sets.

To probe the possibility of a 3.5 keV feature in our data, we added an additional Gaussian emission line component at $\sim$3.5 keV in all spectra, with the energy left free to vary between 3.4--3.6 keV. The best-fit energy and flux of the line in each spectrum is reported in Table \ref{tab:linepars}, with values of 3.53$_{-0.05}^{+0.04}$ keV and 6.03$_{-3.69}^{+4.3}$ 10$^{-7}$ ph s$^{-1}$ cm$^{-2}$, respectively, in the spectrum stacked from the total data set. Each spectrum is plotted with the line (Figures \ref{fig:all_sub_line} and \ref{fig:bins_sub_line}). All parameters were left free to vary, apart from the fixed $N_H$ and widths of emission lines, which were frozen at 0 keV and hence dictated by \textit{Chandra}'s energy resolution after being folded through the response files. This left a total of 6 free parameters, including those describing the 3.5 keV line component. Best-fit parameters for the power-law and $\sim$4.5 keV feature are reported in Table \ref{tab:parameters}. 

All best-fit values were obtained using the Markov-Chain Monte-Carlo (MCMC) in XSPEC. An MCMC was performed for each spectrum, using the Metropolis-Hastings (MH) algorithm and random priors, employing the $\chi^2$ fit statistic on the binned data. Each chain consisted of 1,000,000 runs after discarding the first 40,000 to constitute a burn-in period. For each parameter, the best-fit value is the 0.5 quantile of its MCMC distribution, and 1$\sigma$ lower and upper errors are the 0.16 and 0.84 quantiles, respectively. These values reflect the Gaussian mean for the best-fit, and the Gaussian standard deviation for the 1$\sigma$ errors. The results of the MCMC analysis for each spectrum are given in Figures \ref{fig:all_chain}, and \ref{fig:bins_chain} with confidence contours plotted at 1, 2, and 3$\sigma$ according to the 2-dimensional Gaussian distribution. 

Due to the background-dominated nature of \textit{Chandra} observations, especially in our source-removed data, the MCMC analysis was crucial. High background counts are well-known to cause difficulty in detecting faint emission lines, so the statistically powerful MCMC was employed to combat these statistical issues, since its large volume of runs and Bayesian approach are well-suited for a faint feature such as the putative 3.5 keV line. The $\sim$3.5 keV line energy MCMC probability distribution shows Gaussian convergence to the best-fit values, suggesting the possible presence of a feature.

\subsubsection{Chi-Squared Testing}

All $\chi^2$ testing results are reported in Table \ref{tab:bgsub_chi}. The reduced $\chi^2$ of all models was between $\sim$0.80 and $\sim$0.90, both with and without the $\sim$3.5 keV feature, suggesting the models fit the data very well. The significance of the $\sim$3.5 keV feature is low, at only 1.08$\sigma$ in the total data set, estimated via the $\Delta\chi^2$ method. Bin 3, which includes the region where \cite{nico 2018} made a possible detection, shows the highest significance of all spectra but is consistent only with a 1.72$\sigma$ statistical fluctuation. Hence, the $\chi^2$ analysis suggests a non-detection in the total data set and in all bins.  

\subsubsection{Bayesian Information Criterion}

The Bayesian Information Criterion (BIC; \citealt{schwarz 1978}; \citealt{wit 2012}) is a powerful tool for evaluating and comparing models. A lower BIC is favorable \citep{kass 1995}, and the difference ($\Delta$BIC) between two models can be used to estimate the significance of a feature in a nested model such as ours. Thus, to ensure a robust and composite statistical treatment, we employed the BIC in addition to our $\chi^2$ analysis, the results of which are reported in Table \ref{tab:bgsub_bic}. The significance of the line yielded by the BIC analysis is substantially lower than that of $\Delta\chi^2$ in all spectra, at only 0.02$\sigma$ in the total data set and only 0.06$\sigma$ in bin 3. This firmly supports the claim that the apparent feature is a statistical fluctuation and allows us to conclude a non-detection for the background-subtracted results.

\subsubsection{Peculiarities in ACIS Stowed Spectra}\label{pbk dip section}

An important step of performing background subtraction was examining the particle background spectra for artifacts that could result in a spurious feature in our analysis and others that utilize ACIS stowed data. There appears to be an artifact around 3.5 keV in some of the stowed spectra (shown in Figures \ref{fig:stow2000}, \ref{fig:stow2005}, and \ref{fig:stow2009}). In particular, there appear to be anomalously scattered points below the continuum in various such spectra.

\begin{figure}[h!]
\includegraphics[width=9.cm]{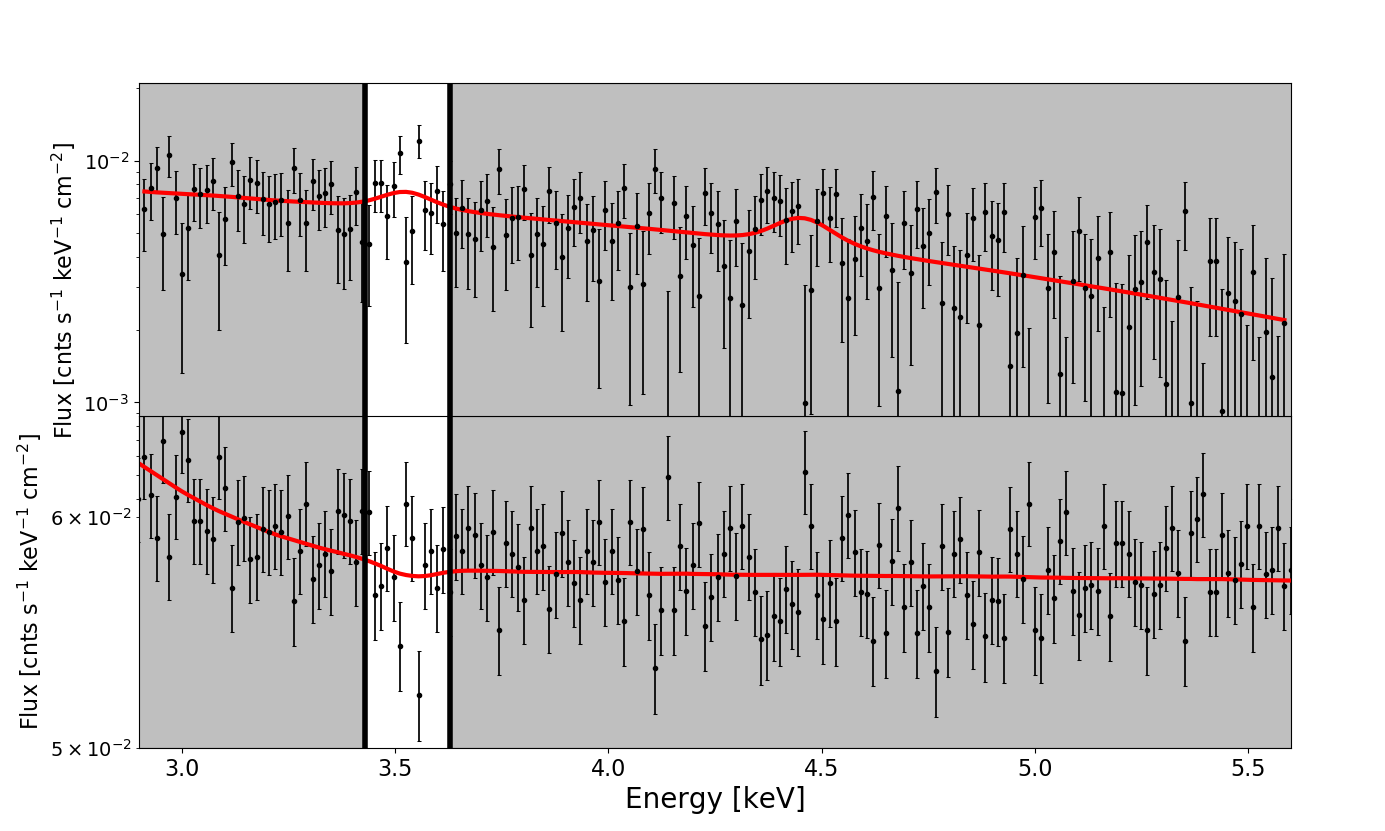}
\caption{Background-subtracted spectrum of the entire data set modeled with a 3.5 keV line (top), compared to the corresponding background spectrum containing the 3.5 keV dip (bottom).}
\label{fig:dip_alldata}
\end{figure}

Thorough checks were performed on the background data, and it was found that the spurious points appear in both the PHA and PI spectra, even when filtering for only good event grades (0, 2, 3, 4, and 6). Moreover, the feature appears prominently when all stowed spectra are stacked to form the data set's particle background spectra, and hence was modeled as a Gaussian absorption line in each stacked background spectrum (Figures \ref{fig:dip_alldata} and \ref{fig:dip_bins}).

\begin{table*}[t!]
\centering
 \begin{tabular}{c c c c c c}
 \hline
  Bin & $\chi^2$ w/line & $\chi^2$ w/out & $\Delta\chi^2$ & Detection & Line  \\
  
   & (DOF$=$178) & (DOF$=$180) & ($\Delta$DOF$=$2) & Probability & Significance \\
   
 \hline
 \hline

 1 & 156.56  & 158.10 & 1.54 & 0.538 & 0.74$\sigma$\\
 
 2 & 147.94 & 149.49 & 1.55 & 0.539 & 0.74$\sigma$ \\
 
 3 & 158.62 & 163.54 & 4.91  & 0.914 & 1.72$\sigma$ \\
 
 4 & 159.03 & 160.01 & 0.98 & 0.388 & 0.51$\sigma$ \\
 
 \hline
 ALL & 143.12  & 145.66 & 2.53 & 0.718 & 1.08$\sigma$ \\
 \hline
\end{tabular}
\caption{The results of $\chi^2$ testing on background-subtracted spectra. Note that DOF denotes degrees of freedom.}
\label{tab:bgsub_chi}
\end{table*}

\begin{table*}[t!]
\centering
 \begin{tabular}{c c c c c c}
 \hline
 Bin & $\Delta$BIC & Evidence \textbf{against}  & Bayes & Detection & Line \\
  & & Model w/Line & Factor & Probability & Significance \\ [0.5ex]
 \hline
 \hline

 1 & 9.07 & Strong & 93.23 & 0.011 & 0.01$\sigma$ \\
 
 2 & 9.00 & Strong & 89.85 & 0.011 & 0.01$\sigma$ \\
 
 3 & 6.10 & Strong & 21.09 & 0.045 & 0.06$\sigma$ \\
 
  4 & 9.52 & Strong  & 116.79 & 0.008 & 0.01$\sigma$ \\
  
  \hline
  
  ALL & 8.15 & Strong & 58.86 & 0.017 & 0.02$\sigma$ \\

 \hline
\end{tabular}
\caption{The results of BIC testing on background-subtracted spectra. The strength of evidence against the model with a line at 3.51 keV is determined via the standard scale used to qualitatively interpret the $\Delta$BIC and corresponding Bayes Factor, originally established by \cite{kass 1995}.}
\label{tab:bgsub_bic}
\end{table*}


The data model's emission feature and the background model's dip occur within the same error range of energy, suggesting that subtracting the background's dip could cause the false appearance of emission feature. In fact, the two highest data points above the continuum in the modeled emission line occur at the exact energies as the most spuriously low data points in the particle background dip, leading to the conclusion that even our statistically insignificant modeled feature here is impacted by this background artifact.

The dip does not follow a Gaussian absorption profile, as it is defined only by the two anomalous data points above an otherwise smooth continuum. Furthermore, the energies of the two points are in close proximity, rivaling the spectral resolution of \textit{Chandra}. Hence, after being folded through the RMF and thus accounting for the \textit{Chandra} energy resolution, a Gaussian dip cannot properly fit the data. Indeed, as seen in Figure \ref{fig:dip_alldata}, the modeled feature does not properly account for the peculiarity, considerably underestimating the depth of the dip. This results in the dip model component having a negligible statistical significance. Due to \textit{Chandra}'s energy resolution, the Gaussian absorption line will always be too wide to fit a sharp spike consisting of only two nearby data points. Regardless, the spurious points will inevitably exaggerate any possible feature at $\sim$3.5 keV in a background-subtracted spectrum, casting doubt on any possible detection.

In this case, the background artifact has clearly amplified the 3.5 keV feature, though with the low significance estimates yielded by both the $\chi^2$ and BIC analyses, it is nonetheless clear that any such feature is consistent only with a statistical fluctuation. It is unclear, however, what role these spurious particle background points may have played in previous works, including the possible \cite{nico 2018} detection.

Due to the unknown extent of the spurious artifact's influence on the data, in addition to the lowered statistics after subtracting the background, we do not consider the results of the background-subtracted analysis definitive. We conclude that the background-modeled analysis is needed to properly utilize our $\sim$51 Ms data set, and hence opt to treat the background-subtracted results as a preliminary test of the data. \edit2{Modeling the background overcomes the low exposure of ACIS stowed data, yielding a high-statistics continuum free from the influence of statistical fluctuations, including the spurious dip data points, as such fluctuations are mitigated by our $\sim$51 Ms of exposure time. Moreover, the particle background normalization method employed here introduces a $\sim$2\% systematic uncertainty in the background-subtracted continuum \citep{hickox 2006}. This, too, is avoided in the background-modeled procedure, which yields a model accounting for all spectral components observed in the $\sim$51 Ms of data, including the particle background. The resulting model's continuum, then, is a model of the true continuum, and hence the systematic uncertainty becomes negligible, contributing only to the particle background component's priors.} The conclusions reached by this work, and the computation of upper-limits in the case of a non-detection, will thus be predicated upon the statistically superior background-modeled results.

\begin{figure}[b!]
\includegraphics[width=9.cm]{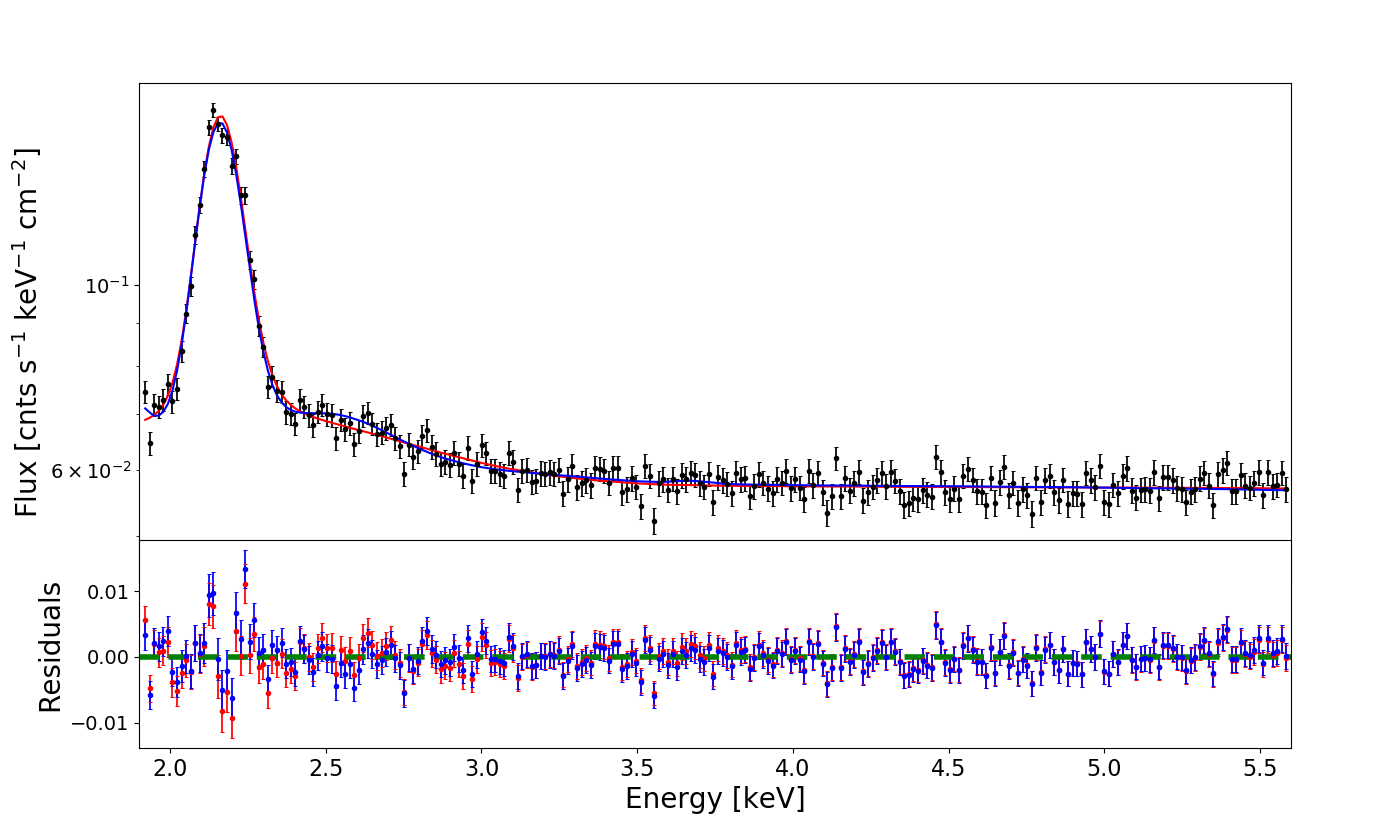}
\caption{Two models of the particle background plotted against the $\sim$1 Ms of ACIS stowed data. \textbf{Red}: The simpler model with 2 Gaussian components. \textbf{Blue}: The more complex model necessary for our background-modeled analysis containing 4 Gaussian components, in addition to the 3.3 keV and 3.68 keV emission features.}
\label{fig:stow_model_comparison}
\end{figure}

\subsection{Background-Modeled Modeling}

Each spectrum was analyzed on the 2.9--5.6 keV band, where models produced highly reliable fits, to maintain consistency with the band analyzed in the background-subtracted procedure. It was necessary to initially include the 1.9--2.9 keV band in the models to achieve good fits between 2.9 keV to $\sim$3.3 keV due to the contribution of a known mother-daughter emission line system starting at $\sim$2 keV \citep{bartalucci 2014}.

\begin{table*}[t!]
\large
\centering
 \begin{tabular}{  c  c  c  c  c  c  }
 
 \hline
  Parameter [Unit] & All & Bin 1 & Bin 2 & Bin 3 & Bin 4 \\
 \hline\hline
 
 $\Gamma_{PL}$ & 1.41$_{-0.08}^{+0.13}$ & 1.46$_{-0.01}^{+0.03}$ & 1.5$_{-0.08}^{+0.08}$ & 1.42$_{-0.1}^{+0.26}$ & 1.33$_{-0.07}^{+0.07}$  \\ 

$I_{PL}$ [10$^{-4}$ ph s$^{-1}$ cm$^{-2}$] & 1.25$_{-0.11}^{+0.18}$ & 1.46$_{-0.03}^{+0.05}$ & 1.42$_{-0.12}^{+0.13}$ & 1.13$_{-0.12}^{+0.36}$ & 1.19$_{-0.1}^{+0.11}$  \\ [0.5ex]

\hline\hline

$\Gamma_{PBK}$ & 11.34$_{-0.53}^{+0.48}$ & 12.96$_{-0.04}^{+0.04}$ & 10.1$_{-0.71}^{+0.75}$ & 12.13$_{-0.57}^{+0.58}$ & 11.66$_{-0.96}^{+0.81}$  \\ 

$I_{PBK}$ [ph s$^{-1}$ cm$^{-2}$] & 12.0$_{-2.06}^{+1.76}$ & 11.27$_{-0.07}^{+0.09}$ & 6.95$_{-2.09}^{+2.71}$ & 14.09$_{-4.75}^{+3.06}$ & 12.23$_{-3.92}^{+3.58}$  \\

$E_{4.5}$ [keV] & 4.51$_{-0.01}^{+0.01}$ & 4.5$_{-0.02}^{+0.02}$ & 4.53$_{-0.02}^{+0.02}$ & 4.51$_{-0.02}^{+0.02}$ & 4.5$_{-0.03}^{+0.03}$  \\ 

$I_{4.5}$ [10$^{-7}$ ph s$^{-1}$ cm$^{-2}$] & 6.22$_{-0.79}^{+0.79}$ & 6.28$_{-0.96}^{+1.91}$ & 6.77$_{-1.74}^{+1.7}$ & 7.34$_{-1.24}^{+1.05}$ & 5.24$_{-1.61}^{+1.83}$  \\ 

$E_{5.4}$ [keV] & 5.41$_{-0.01}^{+0.01}$ & 5.41$_{-0.01}^{+0.02}$ & 5.42$_{-0.01}^{+0.01}$ & 5.43$_{-0.02}^{+0.01}$ & 5.4$_{-0.01}^{+0.01}$  \\ 

$I_{5.4}$ [10$^{-7}$ ph s$^{-1}$ cm$^{-2}$] & 9.92$_{-1.15}^{+1.36}$ & 9.57$_{-1.16}^{+1.16}$ & 12.86$_{-2.55}^{+2.67}$ & 9.46$_{-1.06}^{+1.49}$ & 9.45$_{-1.04}^{+1.83}$  \\ 

$E_1$ [keV] & 2.78$_{-0.09}^{+0.09}$ & 2.74$_{-0.13}^{+0.13}$ & 2.63$_{-0.12}^{+0.13}$ & 2.67$_{-0.15}^{+0.16}$ & 2.63$_{-0.19}^{+0.19}$  \\ 

$\sigma_{1}$ [keV] & 0.42$_{-0.05}^{+0.05}$ & 0.44$_{-0.05}^{+0.1}$ & 0.41$_{-0.07}^{+0.09}$ & 0.49$_{-0.05}^{+0.01}$ & 0.5$_{-0.08}^{+0.09}$  \\

$I_{1}$ [10$^{-2}$ ph s$^{-1}$ cm$^{-2}$] & 0.33$_{-0.07}^{+0.09}$ & 0.38$_{-0.12}^{+0.11}$ & 0.47$_{-0.14}^{+0.16}$ & 0.42$_{-0.11}^{+0.14}$ & 0.56$_{-0.19}^{+0.24}$ \\ 

$E_2$ [keV] & 2.16$_{-0.0}^{+0.0}$ & 2.16$_{-0.0}^{+0.0}$ & 2.16$_{-0.0}^{+0.0}$ & 2.16$_{-0.0}^{+0.0}$ & 2.16$_{-0.0}^{+0.0}$  \\ 

$\sigma_{2}$ [10$^{-2}$ keV] & 5.02$_{-0.04}^{+0.04}$ & 4.94$_{-0.08}^{+0.06}$ & 4.96$_{-0.06}^{+0.06}$ & 5.06$_{-0.07}^{+0.07}$ & 5.03$_{-0.08}^{+0.08}$  \\ 

$I_{2}$ [10$^{-2}$ ph s$^{-1}$ cm$^{-2}$] & 1.87$_{-0.01}^{+0.01}$ & 1.78$_{-0.02}^{+0.01}$ & 2.07$_{-0.02}^{+0.02}$ & 1.73$_{-0.02}^{+0.02}$ & 1.89$_{-0.03}^{+0.03}$  \\ 

$E_3$ [keV] & 2.5$_{-0.0}^{+0.0}$ & 2.49$_{-0.01}^{+0.01}$ & 2.51$_{-0.01}^{+0.01}$ & 2.51$_{-0.01}^{+0.01}$ & 2.5$_{-0.01}^{+0.01}$  \\ 

$\sigma_{3}$ [keV] & 0.19$_{-0.01}^{+0.01}$ & 0.2$_{-0.02}^{+0.02}$ & 0.17$_{-0.01}^{+0.01}$ & 0.18$_{-0.02}^{+0.02}$ & 0.19$_{-0.02}^{+0.02}$  \\ 

$I_{3}$ [10$^{-2}$ ph s$^{-1}$ cm$^{-2}$] & 0.5$_{-0.05}^{+0.05}$ & 0.52$_{-0.12}^{+0.09}$ & 0.42$_{-0.07}^{+0.08}$ & 0.42$_{-0.08}^{+0.08}$ & 0.48$_{-0.09}^{+0.1}$  \\ 

$E_4$ [keV] & 2.87$_{-0.26}^{+0.29}$ & 3.04$_{-0.36}^{+0.25}$ & 2.85$_{-0.25}^{+0.29}$ & 3.0$_{-0.34}^{+0.25}$ & 2.98$_{-0.31}^{+0.23}$  \\ 

$\sigma_{4}$ [keV] & 13.39$_{-1.43}^{+1.61}$ & 14.9$_{-3.94}^{+3.33}$ & 12.09$_{-1.26}^{+1.68}$ & 15.37$_{-3.57}^{+2.97}$ & 15.61$_{-2.32}^{+2.59}$  \\ 

$I_{4}$ [ph s$^{-1}$ cm$^{-2}$] & 1.95$_{-0.2}^{+0.23}$ & 2.06$_{-0.54}^{+0.45}$ & 1.95$_{-0.2}^{+0.26}$ & 2.06$_{-0.47}^{+0.39}$ & 2.32$_{-0.34}^{+0.38}$  \\ 

$I_{3.7}$ [10$^{-5}$ ph s$^{-1}$ cm$^{-2}$] & 7.5$_{-3.12}^{+2.85}$ & 7.11$_{-4.09}^{+2.85}$ & 7.31$_{-3.54}^{+4.07}$ & 8.34$_{-2.86}^{+3.9}$ & 10.65$_{-4.43}^{+4.15}$  \\ 

$I_{3.3}$ [10$^{-5}$ ph s$^{-1}$ cm$^{-2}$] & 0.97$_{-0.72}^{+1.33}$ & 0.06$_{-0.01}^{+1.11}$ & 2.0$_{-1.46}^{+2.92}$ & 0.0$_{-0.0}^{+0.5}$ & 6.45$_{-2.18}^{+2.87}$  \\  [0.5ex]
 
 \hline
\end{tabular}
\caption{Best-fit model parameters for the background-modeled spectra with 1$\sigma$ errors. The astrophysical parameters are reported in the upper panel, while particle background parameters are reported in the lower panel.}
\label{tab:mod_parameters}
\end{table*}

\begin{figure}[t!]
\includegraphics[width=9.cm]{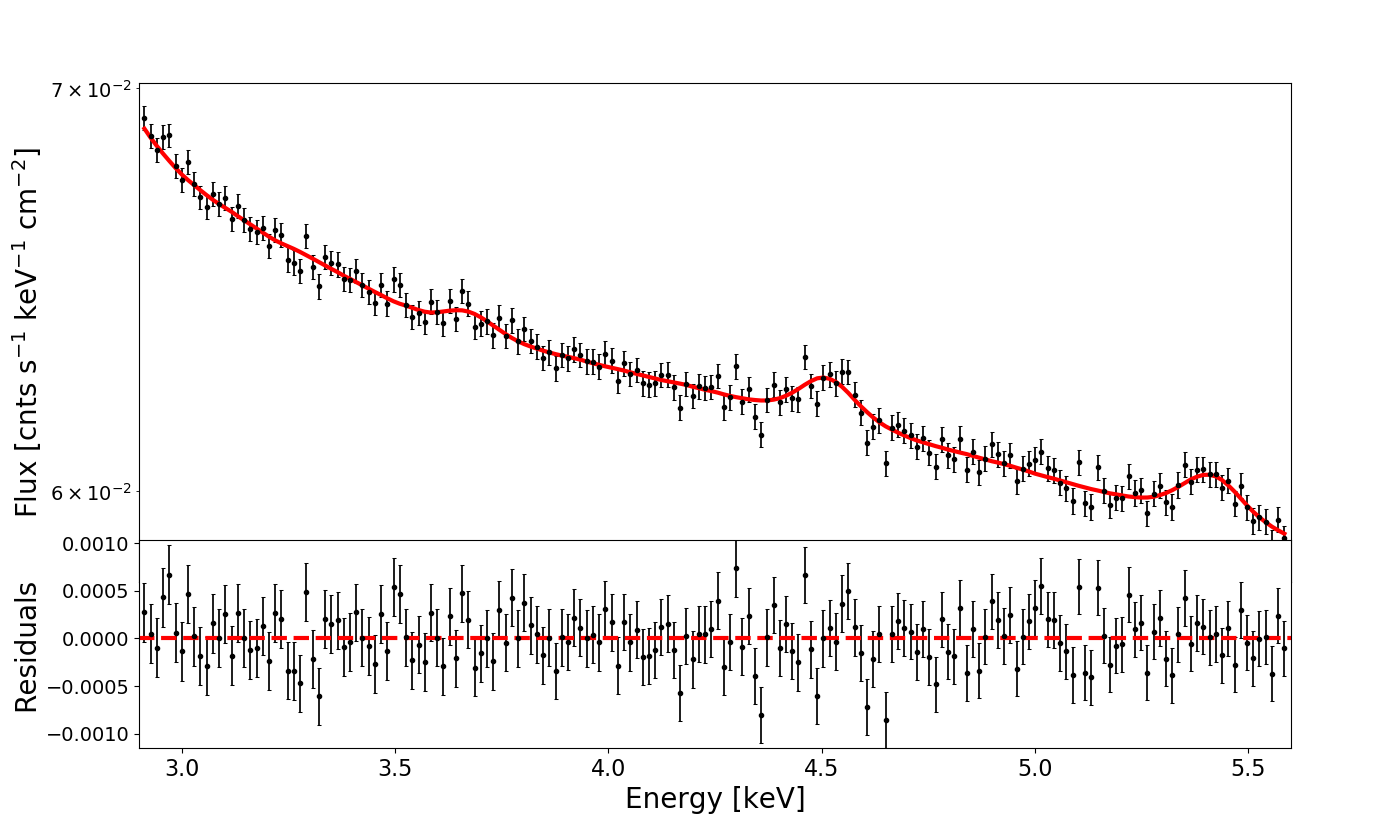}
\caption{Background-modeled spectrum stacked from all observations in the data set.}
\label{fig:all_mod_line}
\end{figure}

\begin{figure}[t!]
\centering
\includegraphics[width=9.cm]{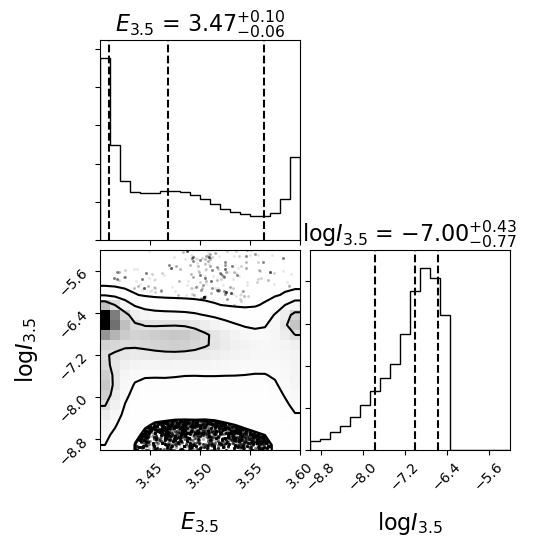}
\caption{MCMC contour plot for 3.5 keV line parameters when energy is left free to vary between 3.4--3.6 keV.}
\label{fig:all_chain_bgmodel}
\end{figure}

The total model is the sum of a particle background model and an astrophysical model. The particle background model was folded through its RMF, while the astrophysical model was folded through the data's RMF and ARF, with the latter accounting for the photons' passage through \textit{Chandra}'s HRMA.

Best-fit parameters for each model were obtained, again, via MCMC in XSPEC. However, while the background-subtracted procedure utilized the MH algorithm due to its ability to set random priors, the background-modeled method employs the Goodman-Weare (GW) algorithm. Unlike MH, GW's priors are obtained from the covariance matrix of a preliminary C-statistic (CSTAT; \citealt{cash 1979}) fit. It also operates via any number of random walkers that can be run in parallel, greatly decreasing computing costs. These features are highly advantageous to the large and degenerate nature of the background-modeled parameter space, allowing the chain to use reliable priors and perform a large volume of runs on substantially shorter time scales than MH. The chain for each model consisted of 2,700,000 runs, with 30 walkers and a burn-in period of 216,000, offering highly robust statistics to the modeling process. The algorithm employed the CSTAT fit statistic in favor of $\chi^2$ to properly accommodate the unbinned data. Best-fit values and 1$\sigma$ errors were again computed to correspond to the Gaussian means and standard deviations of the resulting parameter value distributions, respectively, using the same procedure described in section \ref{bgsub}.


\subsubsection{Particle Background Model}

The particle background model consists of a power-law, in addition to Gaussian components to fit various emission features. This includes all observed effects of the mother-daughter system described by \cite{bartalucci 2014} as an artifact resulting from corrections for Charge Transfer Inefficiency (CTI, described extensively by \citealt{grant 2005}). Emission lines belonging to this system are the only lines in our models with non-zero and free-to-vary widths. All others have widths fixed at 0 keV before being folded through the relevant response files, thus dictating them according to instrumental energy resolution.

However, whereas \cite{bartalucci 2014} found a mother-daughter system containing two Gaussians, our analysis finds two additional Gaussians in the system. In particular, \cite{bartalucci 2014} models Gaussians at $\sim$2.16 keV and $\sim$2.26 keV, while our analysis includes Gaussians at 2.16 keV, 2.50 keV, 2.78 keV, and 2.87 keV, though these energies vary (apart from 2.16 keV) between models. Moreover, the widths of both \cite{bartalucci 2014} Gaussians are $\sim$10$^{-1}$ keV while ours vary depending on the best-fit values obtained via our MCMC methodology. Moreover, \cite{bartalucci 2014} only observed the artifact from CTI correction to exist between 2--3 keV in the same ACIS stowed files used for this analysis, whereas it extends to $\sim$3.3 keV in our spectra. The mother-daughter feature is consistent with \cite{bartalucci 2014} in our lower-statistics spectra, namely our background spectra (Figure \ref{fig:stow_plots}) and background-subtracted models (Figures \ref{fig:all_sub_line} and 
\ref{fig:bins_sub_line}).

The new extension of the CTI correction artifact only appears in our high-statistics background-modeled spectra (Figures \ref{fig:all_mod_line} and \ref{fig:bins_mod_line}). This suggests that the feature is simply difficult to distinguish from the continuum when dealing with spectra that do not contain the extremely high $\sim$51 Ms of data seen in our data set. Indeed, this is exemplified by Figure \ref{fig:stow_model_comparison}, where two models are able to fit the $\sim$1 Ms ACIS stowed spectrum. One model's CTI correction artifact is composed of the two \cite{bartalucci 2014} mother-daughter Gaussians, while the other is the model used in our background-modeled analysis, with four Gaussians. Both are plotted against the $\sim$1 Ms of stowed data in the figure, and the feature modeled in our $\sim$51 Ms spectrum can be seen blending with the continuum.

Two faint emission lines were modeled at 3.3 keV and 3.68 keV, with energies fixed at the known values also used in \cite{abazajian 2020} and \cite{boyarsky 2020}. These features are believed to arise from both astrophysical and instrumental emission lines, and hence to minimize our model's large parameter space we choose to treat the features as single emission lines (as in both \citealt{abazajian 2020} and \citealt{boyarsky 2020}) in the instrumental background model. The 3.3 keV line is thought to be a blend of Ar XVIII, S XVI, and K K$\alpha$, while the 3.68 keV line is thought to be due to Ar XVII or Ca K$\alpha$ (\citealt{boyarsky 2018}; \citealt{abazajian 2020}; \citealt{boyarsky 2020}).

The instrumental line seen in the background-subtracted spectra at $\sim$4.5 keV was again modeled here. An additional emission line was detected and fitted at $\sim$5.4 keV. This is another known instrumental line due to Cr K$\alpha$ and has been seen in \textit{Suzaku} \citep{sekiya 2016} and \textit{XMM-Newton} \citep{esra 2020}. Like the $\sim$4.5 keV line, it is not detectable in the ACIS stowed data due to the limited statistics, and unlike the $\sim$4.5 keV line it is difficult to resolve in the background-subtracted data. However, with the full statistics of our $\sim$51 Ms data set, it appears much more prominently.

In this analysis, the particle background model includes all particle background components with the exception of the $\sim$4.5 keV and $\sim$5.4 keV lines. These two components, though instrumental, are not detected in the $\sim$1 Ms of ACIS stowed data. In particular, the CSTAT fitting of a $\sim$4.5 keV emission line yields zero statistical significance, and the CSTAT fitting of a $\sim$5.4 keV emission line yields a small significance ($\sim$1.4$\sigma$, as estimated by $\Delta\chi^2$) consistent only with a fluctuation. The lack of detection in the particle background spectrum of features that appear prominently in our stacked spectra thus prevents us from obtaining useful priors from the CSTAT fitting process and ultimately results in unreliable models with poor fits, even after running the MCMC procedure. To combat this, we included the $\sim$4.5 keV and $\sim$5.4 keV line components in the astrophysical model, which allowed us to properly account for their contributions.

\subsubsection{Astrophysical Model}

In addition to the $\sim$4.5 keV and $\sim$5.4 keV emission components, the astrophysical model contains the unresolved CXB continuum, again modeled using an absorbed power-law with $N_H$ fixed at 10$^{20}$ cm$^{-2}$, in addition to the possible 3.5 keV feature. The astrophysical and instrumental models thus combine for a total of 22 free parameters, before accounting for the addition of a 3.5 keV component.

\subsubsection{Modeling Without the 3.5 keV Line}\label{433}

The spectra were initially modeled without the 3.5 keV feature, plotted in Figures \ref{fig:all_mod_line} and \ref{fig:bins_mod_line} with best-fit parameters reported in Table \ref{tab:mod_parameters}. The models produced considerably good fits, with that of the total data set achieving a reduced $\chi^2$ of exactly the ideal 1.00, with all models exhibiting reduced $\chi^2$ values between 1.00 and 1.12.

\edit1{A major result from these models is the flux and significance of the line at 3.68 keV, detected at 4.06$\sigma$ in the total data set. This firmly refutes the \cite{dessert 2020} fiducial model, which omits the 3.68 keV feature. Moreover, due to the thorough removal of baryonic material from our observations, particularly from the Galactic Disc, the prominence of the 3.68 keV line in our data suggests that it may be largely instrumental and hence likely due to Ca K$\alpha$ emission.}

\edit1{The 3.68 keV line flux in our total data set is higher than in the \cite{boyarsky 2018} \textit{XMM-Newton} analysis by a factor of $\sim$2. The \cite{boyarsky 2018} data set has smaller particle background statistics and hence yields a correspondingly smaller 3.68 keV line flux due to the large size of our data set. As observed via various other features in our particle background model, including the extension of the mother-daughter emission system and the 5.4 keV line, our robust particle background statistics reveal particle background features with previously unseen clarity. The prominence of the 3.68 keV line in our data, and its similar prominence in \cite{boyarsky 2018}, further suggests it is largely instrumental. The consistency of this interpretation across works using both \textit{Chandra} and \textit{XMM-Newton}, namely this work and \cite{boyarsky 2018}, suggests the instrumental emission at 3.68 keV is due to an effect common to both observatories, hence strengthening the refutation of the \cite{dessert 2020} analysis.}

\edit1{We also find a low flux and low significance for the 3.3 keV line in all models. This differs from \cite{boyarsky 2018}, where the 3.3 keV line is found to have a higher flux than the 3.68 keV feature. The \cite{boyarsky 2018} data set also excises point-sources, but includes the Galactic Center and the Galactic Disc, yielding a higher density of baryonic matter. The contrasting low flux of the 3.3 keV feature in our data, then, suggests that it may be largely astrophysical.}

\subsubsection{Free-to-vary 3.5 keV Line Energy}\label{free energy}

To evaluate the possible presence of a feature at $\sim$3.5 keV, a Gaussian emission component was added to the model and the MCMC fitting procedure was repeated. Both the energy and normalization of the line were initially left free to vary, with energy again restricted to the interval 3.4--3.6 keV. Despite the statistical power offered by GW's CSTAT priors, it was critical to utilize random priors for the 3.5 keV component to avoid any bias from the CSTAT fitting process. Hence, for the free 3.5 keV line parameters, we employed a new method of generating Goodman-Weare priors in XSPEC. 

We first performed our preliminary CSTAT fit to produce a set of initial priors for the method. Subsequently, we ran a Goodman-Weare MCMC chain with length 30 (one for each walker). The chain file was then edited, replacing the 3.5 keV energies and fluxes with appropriately randomized values, hence keeping all other initial priors while replacing the 3.5 keV component's priors with random values. The CSTAT values recorded in the chain of the 30 initial runs were adjusted to match the updated sets of parameters. Finally, a new chain was appended to this file with the properties mentioned above. This effectively offered random priors for the $\sim$3.5 keV line parameters, while still otherwise utilizing the advantages of GW's priors from CSTAT fitting. Also note that we evaluated a preliminary chain for correlations within the fit between the $\sim$3.5 keV line parameters and all other parameters, with the intention of randomizing any other parameters that showed high correlations (i.e., with Pearson Correlation Coefficient magnitudes greater than 0.8). None were found, and hence only the $\sim$3.5 keV line priors were randomized.

The parameter value distributions and contour plots for 3.5 keV line energy and normalization are shown in Figures \ref{fig:all_chain_bgmodel} and \ref{fig:bins_chain_bgmodel}, with confidence contours at 1, 2, and 3$\sigma$ again in accordance with a 2-dimensional Gaussian distribution. The shape of the line energy distribution for the total data set indicates a non-detection, with the probability density reaching its minimum around 3.5 keV. Bins 1 and 4 exhibit this same behavior, while bins 2 and 3 have minor peaks near 3.5 keV. The probability density is highest around 3.5 keV, but in both cases, there appear to be other peaks near the edges of the parameter space, suggesting that any feature at 3.5 keV may not be significant.

We can thus reach preliminary conclusions that the 3.5 keV line was not detected in the total data set, nor was it detected in bins 1 and 4, and that there may be a feature in bins 2 and 3. We rigorously assess these hypotheses by fixing the line energy while leaving the line flux free to vary, thereby isolating the putative emission component. This can thus confirm non-detections and set upper-limits in the total data set, bin 1, and bin 4, while further evaluating the status of the possible feature in bins 2 and 3.

\begin{figure}[t!]
\centering
\includegraphics[width=9.cm]{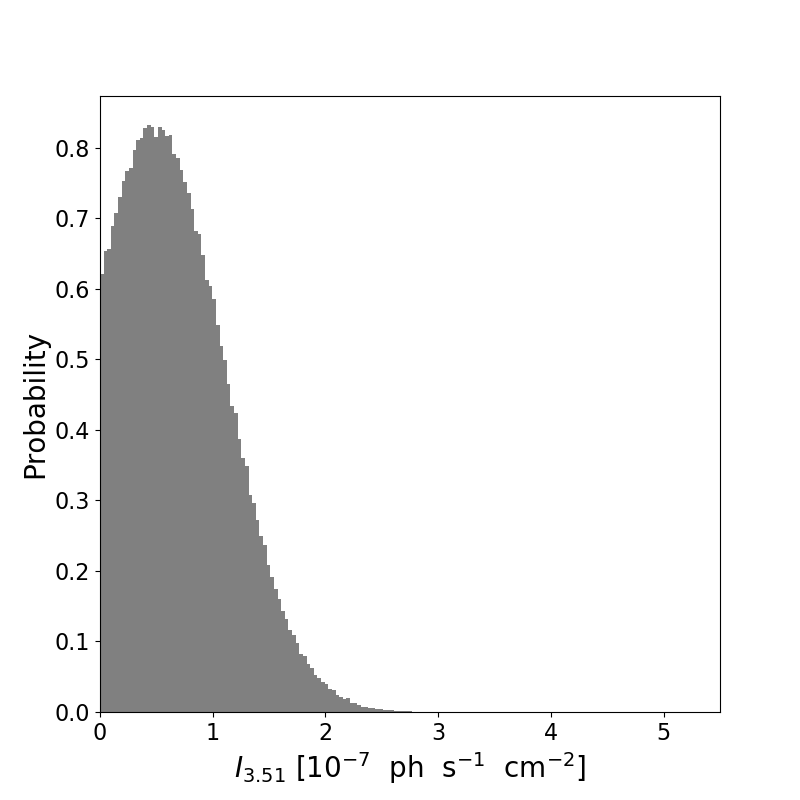}
\caption{Probability density distribution for 3.51 keV line flux in the spectrum stacked from the total data set.}
\label{fig:alldata_lineflux_prob}
\end{figure}

\subsubsection{Fixed 3.5 keV Line Energy}
 
The energy of the 3.5 keV emission component was fixed at 3.51 keV. This value was chosen to be consistent within the 1$\sigma$ error ranges of various previous detections, including \cite{boyarsky 2014}, \cite{urban 2015}, and \cite{nico 2018}, and of the line energy probability distribution peaks in bins 2 and 3. Moreover, in a later section of this work (\ref{totalcxb}), the same procedure used in section \ref{free energy} was applied to our data set without point-source removal, and the line energy probability distribution in bin 3 was found to exhibit the sharpest central peak seen in the entirety of our background-modeled analyses. This peak is found at 3.51 keV, hence solidifying our choice of line energy.

Running another MCMC chain with random priors for the line flux yields a probability density for each bin, shown in Figures \ref{fig:alldata_lineflux_prob} and \ref{fig:bins_lineflux_prob}, with the best-fit values and 1$\sigma$ errors reported in Table \ref{tab:mod_linepars}. Using the best-fit value for the line flux in each model, we can evaluate the statistical significance of the possible feature in all spectra.

\begin{table}[b!]
\centering
 \begin{tabular}{c c}
 \hline
 Bin & $I_{3.51}$ [10$^{-7}$ ph s$^{-1}$ cm$^{-2}$] \\ [0.5ex]
 \hline\hline

 1 & 0.83$_{-0.59}^{+1.02}$ \\ 

 2 & 1.30$_{-0.84}^{+1.06}$  \\
 
 3 & 1.66$_{-0.88}^{+0.95}$ \\

 4 & 0.57$_{-0.41}^{+0.75}$ \\ [0.5ex]
 \hline
 ALL & 0.65$_{-0.42}^{+0.53}$ \\ [0.5ex]
 \hline
\end{tabular}
\caption{Best-fit 3.51 keV line flux for each background-modeled spectrum with 1$\sigma$ errors.}
\label{tab:mod_linepars}
\end{table}

\begin{table*}[t!]
\centering
 \begin{tabular}{c c c c c c}
 \hline
  Bin & $\chi^2$ w/line & $\chi^2$ w/out & $\Delta\chi^2$ & Detection & Line  \\
  
   & (DOF$=$161) & (DOF$=$162) & ($\Delta$DOF$=$1) & Probability & Significance \\
   
 \hline
 \hline

 1 & 182.17 & 181.35 & -0.82 & 0.000 & 0.00$\sigma$\\
 
 2 & 170.09 & 170.63 & 0.54 & 0.538 & 0.74$\sigma$ \\
 
 3 & 167.88 & 170.38 & 2.50 & 0.886 & 1.58$\sigma$ \\
 
 4 & 163.07 & 162.08 & -1.00 & 0.000 & 0.00$\sigma$ \\
 
 \hline
 ALL & 161.38 & 161.90 & 0.52 & 0.529  & 0.72$\sigma$ \\
 \hline
\end{tabular}
\caption{The results of $\chi^2$ testing on background-modeled spectra. Note that DOF denotes degrees of freedom.}
\label{tab:chi_results}
\end{table*}

\begin{table*}[t!]
\centering
 \begin{tabular}{c c c c c c}
 \hline
 Bin & $\Delta$BIC & Evidence \textbf{against}  & Bayes & Detection & Line \\
  & & Model w/Line & Factor & Probability & Significance \\ [0.5ex]
 \hline
 \hline

 1 & 11.18 & Very Strong & 267.95 & 0.004 & 0.00$\sigma$ \\
 
 2 & 9.82 & Strong & 135.92 & 0.007 & 0.01$\sigma$ \\
 
 3 & 7.76 & Strong & 48.40 & 0.020 & 0.02$\sigma$ \\
 
  4 & 11.39 & Very Strong & 297.13 & 0.003 & 0.00$\sigma$ \\
  
  \hline
  
  ALL & 9.88 & Strong & 139.56 & 0.007 & 0.01$\sigma$ \\

 \hline
\end{tabular}
\caption{The results of BIC testing on background-modeled spectra. The strength of evidence against the model with a line at 3.51 keV is determined via the standard scale used to qualitatively interpret the $\Delta$BIC and corresponding Bayes Factor, originally established by \cite{kass 1995}.}
\label{tab:bic_results}
\end{table*}

\subsubsection{Chi-Squared Testing}

The results of all $\chi^2$ testing are reported in Table \ref{tab:chi_results}. The significance of the feature at 3.51 keV is low, estimated using the $\Delta\chi^2$ test at 0.72$\sigma$ in the total data set. The feature exhibits zero statistical significance in both bin 1 and bin 4. In the bins with line energy probability density peaks at $\sim$3.5 keV, namely bins 2 and 3, the significance is estimated at only 0.74$\sigma$ and 1.58$\sigma$, respectively, indicating that the possible feature is consistent only with a statistical fluctuation. Thus, the $\chi^2$ analysis suggests a non-detection in the total data set and in all bins.

\subsubsection{BIC Testing}

The BIC testing results are reported in Table \ref{tab:bic_results}. The significance of the line is estimated by the $\Delta$BIC test at 0.01$\sigma$ in the total data set, a considerably smaller value than that yielded by $\Delta\chi^2$ testing. The feature, again, exhibits zero statistical significance in bins 1 and 4, and in this case has an estimated significance of only 0.01$\sigma$ and 0.02$\sigma$ in bins 2 and 3, respectively. In all cases of non-zero $\Delta\chi^2$ significance, $\Delta$BIC yields substantially smaller estimates, strongly supporting the conclusion that any $\sim$3.5 keV feature in our data is consistent only with statistical fluctuations.

\subsubsection{Upper-Limits}

From these non-detections and the corresponding line flux probability distributions, we can set upper-limits on 3.5 keV line flux. To ensure the strength of the upper-limit for each spectrum, we compute the value using the 0.9985 quantile of the corresponding line flux probability distribution, representing the upper 3$\sigma$ confidence interval\footnote{Note that the correct upper 3$\sigma$ quantile is 0.9985 and not 0.997. The lower 3$\sigma$ quantile is 0.0015, which means a fraction 0.997 of the data correctly lies between the lower and upper 3$\sigma$ quantiles.}. The upper-limits are reported in Table \ref{tab:upperlimits} and will be used to constrain the sterile neutrino mixing angle and line flux radial profile.

\begin{table}[b!]
\centering
 \begin{tabular}{c c}
 \hline
  & $I_{3.51}$ \textbf{Upper-Limit}   \\
 Bin & [10$^{-7}$ ph s$^{-1}$ cm$^{-2}$]
   \\[0.5ex]
 
 \hline\hline

 1 & 4.45  \\

 2 & 4.67  \\

 3 & 4.54  \\

 4 & 3.42 \\ [0.5ex]
 
 \hline

ALL & 2.34 \\[0.5ex]

 \hline
\end{tabular}
\caption{Upper-limits on 3.51 keV line flux computed from the non-detections.}
\label{tab:upperlimits}
\end{table}

\subsection{Data Without Point-Source Removal}\label{totalcxb}


\begin{table*}[t!]
\centering
 \begin{tabular}{c c c c c c}
 \hline
  Bin & $\chi^2$ w/line & $\chi^2$ w/out & $\Delta\chi^2$ & Detection & Line  \\
  
   & (DOF$=$161) & (DOF$=$162) & ($\Delta$DOF$=$1) & Probability & Significance \\
   
 \hline
 \hline

 1 & 169.07 & 168.13 & -0.94 & 0.000 & 0.00$\sigma$\\
 
 2 & 160.63 & 160.04 & -0.59 & 0.000 & 0.00$\sigma$ \\
 
 3 & 164.56 & 165.06 & 0.50 & 0.520 & 0.71$\sigma$ \\
 
 4 & 157.73 & 156.89 & -0.84 & 0.000 & 0.00$\sigma$ \\
 
 \hline
 ALL & 165.78 & 165.21 & -0.57 & 0.000 & 0.00$\sigma$ \\
 \hline
\end{tabular}
\caption{\textbf{Without removing point-sources}: The results of $\chi^2$ testing on background-modeled spectra. Note that DOF denotes degrees of freedom.}
\label{tab:chi_totalcxb}
\end{table*}

\begin{table*}[t!]
\centering
 \begin{tabular}{c c c c c c}
 \hline
 Bin & $\Delta$BIC & Evidence \textbf{against}  & Bayes & Detection & Line \\
  & & Model w/Line & Factor & Probability & Significance \\ [0.5ex]
 \hline
 \hline

 1 & 11.32 & Very Strong & 287.25 & 0.003 & 0.00$\sigma$ \\
 
 2 & 10.99 & Very Strong & 244.05 & 0.004 & 0.00$\sigma$ \\
 
 3 & 9.84 & Strong & 137.31 & 0.007 & 0.01$\sigma$ \\
 
  4 & 11.23 & Very Strong & 274.45 & 0.004 & 0.00$\sigma$ \\
  
  \hline
  
  ALL & 10.98 & Very Strong & 242.71 &  0.004 & 0.00$\sigma$ \\

 \hline
\end{tabular}
\caption{\textbf{Without removing point-sources}: The results of BIC testing on background-modeled spectra. The strength of evidence against the model with a line at 3.51 keV is determined via the standard scale used to qualitatively interpret the $\Delta$BIC and corresponding Bayes Factor, originally established by \cite{kass 1995}.}
\label{tab:bic_totalcxb}
\end{table*}


As shown earlier in Table \ref{tab:bin_info}, the point-source removal process greatly reduced the exposure area of our observations, which resulted in a less favorable signal-to-noise ratio (Table \ref{tab:snr}). To account for this and ensure point-source removal did not mask a $\sim$3.5 keV signal in our data, we thus applied our background-modeled procedure to the data set without removing point-sources, repeating the same methodology and utilizing the same models. Due to the higher exposure area, counts, and signal-to-noise ratio, these spectra offer a higher likelihood of detecting faint $\sim$3.5 keV emission from decaying dark matter in the Dark Matter Halo. If such a signal was detected in the models without source-removal, it would, of course, be difficult to disentangle from baryonic source emission within the scope of this work, and would require further study. However, non-detections would greatly reinforce the results of our analysis, and hence these source-intact models are highly valuable.

\begin{figure}[t!]
\includegraphics[width=9.cm]{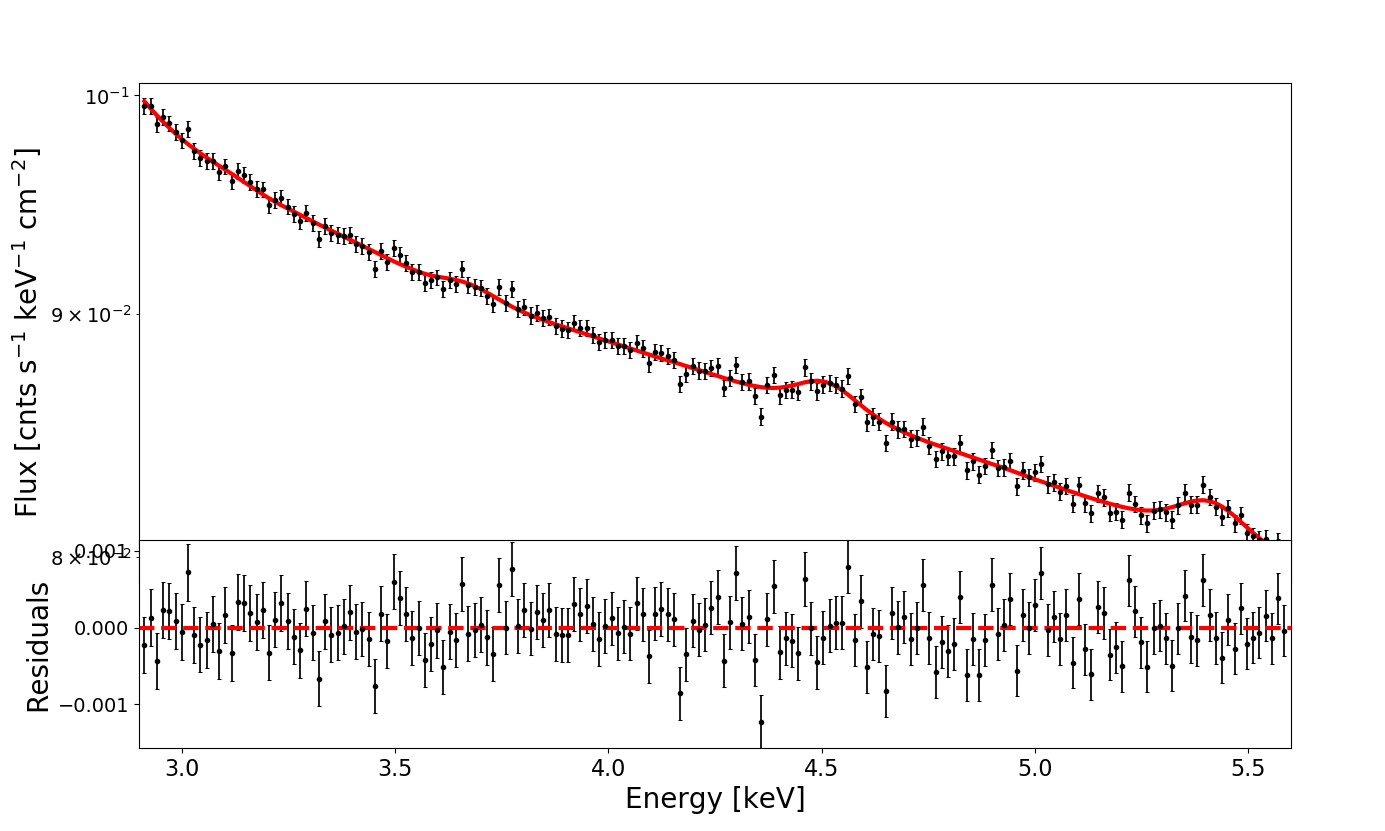}
\caption{\textbf{Without removing point-sources}: Background-modeled spectrum stacked from all observations in the data set.}
\label{fig:all_cxb}
\end{figure}

\begin{figure}[b!]
\centering
\includegraphics[width=8.cm]{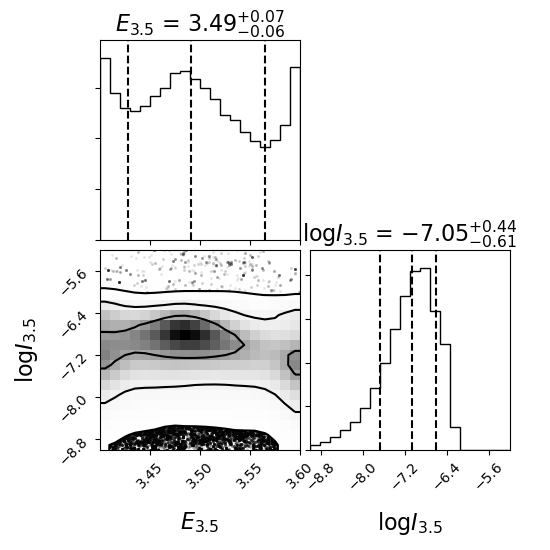}
\caption{\textbf{Without removing point-sources}: MCMC contour plot for the background-modeled spectrum stacked from all observations with 3.5 keV line energy free to vary between 3.4--3.6 keV.}
\label{fig:all_chain_cxb}
\end{figure}

\begin{table*}[t!]
\centering
 \begin{tabular}{  c  c  c  c  c  c  }
 
 \hline
  Parameter [Unit] & All & Bin 1 & Bin 2 & Bin 3 & Bin 4 \\
 \hline\hline
 
 $\Gamma_{PL}$ & 1.63$_{-0.04}^{+0.01}$ & 1.47$_{-0.02}^{+0.03}$ & 1.47$_{-0.01}^{+0.02}$ & 1.47$_{-0.02}^{+0.03}$ & 1.48$_{-0.02}^{+0.03}$  \\ 

$I_{PL}$ [10$^{-4}$ ph s$^{-1}$ cm$^{-2}$] & 3.33$_{-0.13}^{+0.03}$ & 3.21$_{-0.06}^{+0.1}$ & 3.46$_{-0.06}^{+0.09}$ & 3.55$_{-0.07}^{+0.1}$ & 3.45$_{-0.09}^{+0.11}$  \\ [0.5ex]

\hline
\hline

$\Gamma_{PBK}$ & 11.28$_{-0.05}^{+0.02}$ & 12.4$_{-0.63}^{+0.42}$ & 10.78$_{-0.2}^{+0.37}$ & 11.67$_{-0.55}^{+0.59}$ & 11.12$_{-0.44}^{+0.69}$  \\ 

$I_{PBK}$ [ph s$^{-1}$ cm$^{-2}$] & 13.68$_{-0.01}^{+0.44}$ & 11.01$_{-0.93}^{+0.85}$ & 8.56$_{-0.89}^{+1.04}$ & 12.9$_{-1.74}^{+1.42}$ & 9.97$_{-1.52}^{+1.97}$  \\

$E_{4.5}$ [keV] & 4.52$_{-0.01}^{+0.01}$ & 4.5$_{-0.03}^{+0.03}$ & 4.53$_{-0.02}^{+0.01}$ & 4.52$_{-0.01}^{+0.01}$ & 4.51$_{-0.02}^{+0.02}$  \\ 

$I_{4.5}$ [10$^{-7}$ ph s$^{-1}$ cm$^{-2}$] & 8.1$_{-0.54}^{+0.55}$ & 6.37$_{-2.37}^{+1.62}$ & 9.02$_{-1.92}^{+2.08}$ & 9.69$_{-1.86}^{+1.81}$ & 7.35$_{-1.78}^{+1.78}$  \\ 

$E_{5.4}$ [keV] & 5.42$_{-0.01}^{+0.01}$ & 5.42$_{-0.02}^{+0.02}$ & 5.41$_{-0.01}^{+0.01}$ & 5.43$_{-0.01}^{+0.01}$ & 5.4$_{-0.01}^{+0.01}$  \\ 

$I_{5.4}$ [10$^{-7}$ ph s$^{-1}$ cm$^{-2}$] & 14.39$_{-1.21}^{+0.77}$ & 13.33$_{-3.31}^{+2.84}$ & 18.47$_{-2.94}^{+3.09}$ & 13.57$_{-2.7}^{+3.1}$ & 13.49$_{-2.81}^{+3.18}$  \\

$E_1$ [keV] & 2.69$_{-0.04}^{+0.01}$ & 2.63$_{-0.07}^{+0.05}$ & 2.61$_{-0.08}^{+0.07}$ & 2.63$_{-0.07}^{+0.05}$ & 2.58$_{-0.06}^{+0.08}$  \\ 

$\sigma_{1}$ [keV] & 0.47$_{-0.0}^{+0.05}$ & 0.44$_{-0.05}^{+0.08}$ & 0.45$_{-0.05}^{+0.05}$ & 0.53$_{-0.06}^{+0.06}$ & 0.49$_{-0.05}^{+0.05}$  \\ 

$I_{1}$ [10$^{-2}$ ph s$^{-1}$ cm$^{-2}$] & 0.58$_{-0.01}^{+0.1}$ & 0.57$_{-0.08}^{+0.1}$ & 0.68$_{-0.12}^{+0.14}$ & 0.72$_{-0.09}^{+0.11}$ & 0.77$_{-0.12}^{+0.11}$  \\ 

$E_2$ [keV] & 2.16$_{-0.0}^{+0.0}$ & 2.16$_{-0.0}^{+0.0}$ & 2.16$_{-0.0}^{+0.0}$ & 2.16$_{-0.0}^{+0.0}$ & 2.16$_{-0.0}^{+0.0}$  \\ 

$\sigma_{2}$ [10$^{-2}$ keV] & 4.67$_{-0.02}^{+0.01}$ & 4.57$_{-0.09}^{+0.09}$ & 4.57$_{-0.06}^{+0.06}$ & 4.68$_{-0.05}^{+0.05}$ & 4.72$_{-0.07}^{+0.07}$  \\ 

$I_{2}$ [10$^{-2}$ ph s$^{-1}$ cm$^{-2}$] & 2.47$_{-0.01}^{+0.0}$ & 2.13$_{-0.03}^{+0.03}$ & 2.76$_{-0.02}^{+0.02}$ & 2.61$_{-0.02}^{+0.02}$ & 2.41$_{-0.02}^{+0.02}$  \\ 

$E_3$ [keV] & 2.52$_{-0.0}^{+0.0}$ & 2.48$_{-0.01}^{+0.01}$ & 2.52$_{-0.01}^{+0.01}$ & 2.52$_{-0.01}^{+0.01}$ & 2.52$_{-0.01}^{+0.01}$  \\ 

$\sigma_{3}$ [keV] & 0.17$_{-0.0}^{+0.0}$ & 0.2$_{-0.02}^{+0.02}$ & 0.16$_{-0.01}^{+0.01}$ & 0.17$_{-0.01}^{+0.01}$ & 0.16$_{-0.01}^{+0.01}$  \\ 

$I_{3}$ [10$^{-2}$ ph s$^{-1}$ cm$^{-2}$] & 0.59$_{-0.02}^{+0.01}$ & 0.52$_{-0.07}^{+0.09}$ & 0.54$_{-0.06}^{+0.06}$ & 0.61$_{-0.05}^{+0.04}$ & 0.5$_{-0.06}^{+0.06}$  \\ 

$E_4$ [keV] & 2.67$_{-0.02}^{+0.33}$ & 3.03$_{-0.31}^{+0.21}$ & 2.95$_{-0.29}^{+0.26}$ & 2.96$_{-0.3}^{+0.24}$ & 3.05$_{-0.35}^{+0.2}$  \\ 

$\sigma_{4}$ [keV] & 10.44$_{-0.61}^{+0.14}$ & 14.34$_{-2.78}^{+3.93}$ & 17.32$_{-2.68}^{+1.83}$ & 16.52$_{-3.36}^{+2.39}$ & 16.78$_{-2.71}^{+2.24}$  \\ 

$I_{4}$ [ph s$^{-1}$ cm$^{-2}$] & 2.07$_{-0.13}^{+0.02}$ & 2.39$_{-0.46}^{+0.64}$ & 3.58$_{-0.55}^{+0.37}$ & 3.31$_{-0.67}^{+0.47}$ & 3.12$_{-0.5}^{+0.41}$  \\ 

$I_{3.7}$ [10$^{-5}$ ph s$^{-1}$ cm$^{-2}$] & 7.88$_{-1.87}^{+0.48}$ & 10.32$_{-5.21}^{+6.99}$ & 8.57$_{-5.4}^{+5.78}$ & 7.05$_{-3.98}^{+4.9}$ & 7.32$_{-4.14}^{+6.46}$ \\ 

$I_{3.3}$ [10$^{-5}$ ph s$^{-1}$ cm$^{-2}$] & 0.00$_{-0.00}^{+0.00}$ & 3.06$_{-1.43}^{+5.48}$ & 0.00$_{-0.00}^{+0.00}$ & 0.82$_{-0.32}^{+2.8}$ & 6.24$_{-3.2}^{+4.58}$  \\ [0.5ex]
 
 \hline
\end{tabular}
\caption{\textbf{Without removing point-sources}: Best-fit model parameters for the background-modeled spectra with 1$\sigma$ errors. The astrophysical parameters are reported in the upper panel, while particle background parameters are reported in the lower panel.}
\label{tab:cxb_pars}
\end{table*}

\begin{table}[h!]
\centering
 \begin{tabular}{c c}
 \hline
 Bin & $I_{3.51}$ [10$^{-7}$ ph s$^{-1}$ cm$^{-2}$] \\ [0.5ex]
 \hline\hline

 1 & 0.86$_{-0.62}^{+1.10}$ \\ 

 2 & 0.94$_{-0.66}^{+1.05}$ \\

 3 & 1.40$_{-0.90}^{+1.14}$ \\

 4 & 0.83$_{-0.60}^{+1.02}$ \\ [0.5ex]
 
 \hline
 ALL & 0.47$_{-0.33}^{+0.52}$ \\
 \hline
\end{tabular}
\caption{\textbf{Without removing point-sources}: Best-fit 3.51 keV line flux for each background-modeled spectrum with 1$\sigma$ errors.}
\label{tab:cxb_linepars}
\end{table}

\subsubsection{Modeling without the 3.5 keV Line}\label{441}

The models without a 3.5 keV feature are plotted in Figures \ref{fig:all_cxb} and \ref{fig:bins_cxb} and again produced highly effective fits, achieving a reduced $\chi^2$ of 1.02 for the total data set, with all models exhibiting highly ideal reduced $\chi^2$ values between 0.97 and 1.04. The best-fit parameter values are reported in Table \ref{tab:cxb_pars}.

\edit1{Here, regarding the features at 3.3 keV and 3.68 keV, we find results consistent with those of the source-removed data described in section \ref{433}. The 3.68 keV feature exhibits a slightly higher flux and comparable significance ($\sim$4$\sigma$ in the total data set), again thoroughly refuting the fiducial \cite{dessert 2020} analysis that omitted the feature. This is also consistent with the hypothesis that the feature is at least largely instrumental. We again find a low flux for the 3.3 keV line in all models, except in bin 1 where its flux is greater by a factor of $\sim$5. While the 3.3 keV feature in bin 1 still has a low flux that is considerably lower than the 3.68 keV line flux in the same bin, its higher value in that region may be related to the bin's proximity to the Galactic Center, possibly containing traces of baryonic material and hence remaining consistent with the hypothesis that it is largely astrophysical.}

\subsubsection{Free-to-vary 3.5 keV Line Energy}

The parameter value distributions and contour plots for 3.5 keV line energy and normalization are shown in Figures \ref{fig:all_chain_cxb} and \ref{fig:bins_chain_cxb}. In this case, the total data set exhibits a small peak at $\sim$3.5 keV in the line energy distribution. Bins 1 and 4 again have low line energy probability densities around 3.5 keV, with bin 2 showing the same behavior, unlike in the source-removed analysis. Bin 3 displays the sharpest $\sim$3.5 keV peak of line energy probability density seen in this work, with the peak occurring at 3.51 keV (which, as mentioned, is among the justifications for our choice of fixed line energy when assessing significance and setting upper-limits). These results lead to the preliminary conclusions that there is a non-detection in bins 1, 2, and 4, while there may be features in bin 3 and in the total data set, and hence we fix line energy at 3.51 keV to assess these hypotheses.

\begin{figure}[t!]
\centering
\includegraphics[width=9.cm]{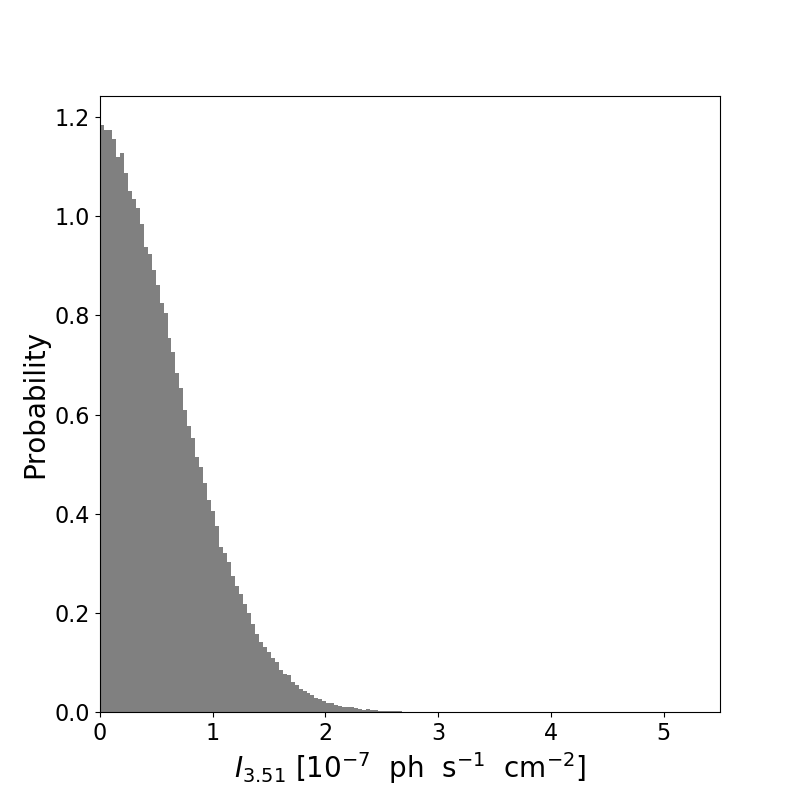}
\caption{\textbf{Without removing point-sources}: Probability density distribution for 3.51 keV line flux in the spectrum stacked from the total data set.}
\label{fig:cxb_alldata_lineflux_prob}
\end{figure}

\subsubsection{Fixed 3.5 keV Line Energy}

The line flux probability densities yielded by MCMC are given in Figures \ref{fig:cxb_alldata_lineflux_prob} and \ref{fig:cxb_bins_lineflux_prob}, with the best-fit values and 1$\sigma$ errors reported in Table \ref{tab:cxb_linepars}. Using these best-fit values, we can evaluate the statistical significance of the possible feature in all spectra.

\subsubsection{Chi-Squared Testing}

The results of the $\chi^2$ testing are reported in Table \ref{tab:chi_totalcxb}. The feature exhibits zero statistical significance in the total data set and in all bins, except bin 3. There, the significance is estimated using $\Delta\chi^2$ to be 0.71$\sigma$, suggesting the possible feature is consistent only with a statistical fluctuation.

\subsubsection{BIC Testing}

The BIC testing results are reported in Table \ref{tab:bic_totalcxb}. The feature again displays zero statistical significance in the total data set and in all bins, except bin 3. In this case, $\Delta$BIC estimates the significance to be 0.01$\sigma$, again consistent only with a statistical fluctuation. 

\subsubsection{Upper-Limits}

Considering the $\chi^2$ and BIC testing results together allows us to conclude non-detections in all source-intact spectra. We hence follow our procedure of obtaining upper-limits on line flux via the upper 3$\sigma$ confidence bound on the line flux probability distribution. The upper-limits, in units of 10$^{-7}$ ph s$^{-1}$ cm$^{-2}$, are 4.96, 4.59, 5.01, and 4.53 in bins 1 through 4 respectively, and 2.27 in the total data set. These are consistent with the upper-limits set by the source-removed data within up to $\sim$11\% in bins 1 through 3, and within $\sim$32\% in bin 4. The upper-limit for the total data set matches the source-removed upper-limit within $\sim$3\%, further reinforcing the results.

\section{Discussion}\label{discussion}

Here we offer an interpretation of the analysis results and a comparison between the surface brightness profile of the line flux's upper-limits and the predicted NFW flux. Additionally, we constrain the $\sim$7 keV sterile neutrino mixing angle and decay rate to compare with previous works.

\subsection{Interpretation of Results}

As mentioned earlier, the background-subtracted results are substantially less statistically viable than the background-modeled results, due to the heavily reduced statistics and the observed background artifact at $\sim$3.5 keV. Therefore, to ensure a thorough interpretation using the full statistics of our $\sim$51 Ms of data, we will draw conclusions only from the background-modeled results.

When 3.5 keV line energy was left free to vary, some spectra yielded very low line flux probability densities around 3.5 keV suggesting non-detections, while others exhibited marginal peaks at $\sim$3.5 keV suggesting a possible feature. Significance testing via both the $\chi^2$ and BIC methods confirmed hypothesized non-detections and showed that all potential 3.5 keV features are consistent only with statistical fluctuations. These results were decisively solidified by applying the same procedure to the source-intact data, allowing us to reach the confident conclusion that no 3.5 keV emission feature was detected in this work. Furthermore, as we considered increasingly high-statistics data, we saw a general decrease in significance estimates via both methods of testing, reinforcing the evidence that any 3.5 keV feature in our data is merely a statistical fluctuation. We can thus use our non-detections to constrain the line flux radial profile, so we choose the upper-limits set by the source-removed analysis. As discussed, those upper-limits are closely consistent with the upper-limits set by the source-intact data. Our choice to utilize the source-removed upper-limits is due to fact that the source-removed data is free of virtually all sources and hence contains minimal foreground signals, allowing for a more thorough and reliable application to the Dark Matter Halo.

\subsection{Comparison to Dessert et al. 2020}

While the fiducial upper-limits set by \cite{dessert 2020} have been questioned in part due to the omission of the 3.3 keV and 3.68 keV lines in the analysis, the publication provided supplemental materials detailing an additional modeling process. This supplemental model accounts for the 3.3 keV and 3.68 keV lines, finding a new upper-limit on the sin$^2$(2$\theta$) a factor of $\sim$8 times higher than that of the fiducial analysis. \cite{boyarsky 2020} finds results consistent with this supplemental upper-limit. \cite{abazajian 2020} argues the true upper-limit may exceed the fiducial upper-limit by a factor of $\sim$20 or more, \edit1{and \cite{boyarsky 2020} supports this claim, pointing out that the reliance of the constraint on local dark matter density introduces a systematic uncertainty of up to a factor of 3 on even the supplemental constraint}. Furthermore, \edit1{as discussed in sections \ref{433} and \ref{441},} the 3.68 keV line was detected at high significance in our total data set, at 4.06$\sigma$ in the source-removed data and at 3.58$\sigma$ in the source-intact data. This suggests the feature is highly significant and hence our results thoroughly refute the fiducial \cite{dessert 2020} constraints found using models that fail to account for the 3.68 keV feature. Thus, when comparing our results to \cite{dessert 2020}, we will henceforth exclusively consider that work's higher supplemental upper-limit in favor of its reported fiducial constraint. \edit1{The supplemental upper-limit will be referred to as the \textit{XMM-Newton} Milky Way Halo upper-limit to avoid confusion with results of the \cite{dessert 2020} fiducial analysis.}

\begin{figure}[b!]
\includegraphics[width=9.cm]{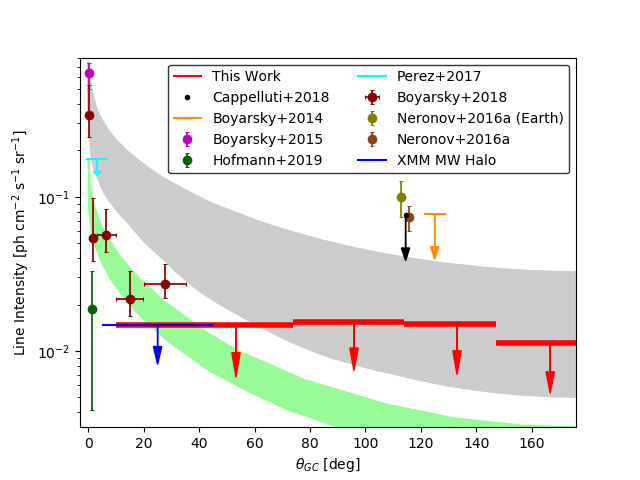}
\caption{The surface brightness profile of the 3.5 keV line upper-limits. The gray extended curve indicates our NFW profile's 2$\sigma$ bounds. The green extended curve represents the 2$\sigma$ bounds of the \cite{boyarsky nfw} NFW profile used in \cite{boyarsky 2018}. Red arrows indicate upper-limits from this work, with each arrow plotted at the average position of observations in the corresponding bin. The blue arrow represents the \edit1{\textit{XMM-Newton} Milky Way Halo upper-limit}. Horizontal error bars indicate the angular distances contained in a given bin.}
\label{fig:nfw}
\end{figure}

\subsection{Radial Profile of 3.5 keV Line Flux}

To constrain the decaying dark matter scenario of the 3.5 keV line, we compare the upper-limit flux distribution to the predicted line flux profile for the Milky Way's Dark Matter Halo by the Navarro-Frenk-White (NFW) distribution of dark matter \citep{nfw 1997}. Here, we follow the NFW profile method of \cite{nico 2018}, similar to that which was later used by \cite{boyarsky 2018}. According to the NFW surface brightness profile, the line intensity should be given by
\begin{equation}\label{nfw_intensity}
    I_{\nu}(\theta_{GC}) = I_{DM,GC}\frac{\int \rho_{NFW}[r(l,0^{\circ})] dl d\Omega}{\int \rho_{NFW}[r(l,\theta_{GC})] dl d\Omega}
\end{equation}
where $I_{DM,GC}$ is the expected dark matter decay signal from the Galactic Center and is empirically determined \citep{nico 2018}. Here, the radial distance from the galactic center ($r$) is formulated as a function of the distance along the line of sight ($l$) and the angular distance from the Galactic Center ($\theta_{GC}$). This formulation and integration shown above must be performed to account for all contributions from decaying dark matter along the line of sight. The function $r$, its variables $l$ and $\theta_{GC}$, and the distance from the Galactic Center to Earth ($d$) are related by the Law of Cosines, given by:
\begin{equation}\label{r_define}
    r(l,\theta_{GC}) = \sqrt{l^2 + d^2 - 2ld cos(\theta_{GC})}
\end{equation}
The NFW profile, describing the density of dark matter halos as a function of distance from their respective galactic centers ($\rho_{NFW}(r)$ in Equation \ref{nfw_intensity}) is established by \cite{nfw 1997} as
\begin{equation}\label{nfw}
    \rho_{NFW}(r) = \frac{\rho_0}{ \frac{r}{R_s}\left(1 + \frac{r}{R_s}\right)}
\end{equation}
and is formulated in (\ref{nfw_intensity}) as $\rho_{NFW}[r(l,\theta_{GC})]$, a functional of $r(l,\theta_{GC})$. The $\rho_0$ and scale radius ($R_s$) parameters are galaxy-specific, and their exact values in the Milky Way are still contested \citep{bland 2016}. Like \cite{nico 2018}, \edit2{we have opted to primarily use the parameters measured by \cite{nesti 2013}, i.e., $R_s \sim 16$ kpc with a local dark matter density $\rho_\odot \sim 0.67$ GeV cm$^{-3}$ at solar position $r_\odot \sim 8.0$ kpc (Table \ref{tab:nfw_pars})}. The resulting NFW profile is plotted with data from various prior works in Figure \ref{fig:nfw}, where the \cite{boyarsky nfw} NFW profile adopted in \cite{boyarsky 2018} \edit2{to optimize its correspondence to that work's data (using the conventionally-approximated $R_s=20$ kpc in contrast to our empirical $R_s$, and a more conservative dark matter density $\rho_\odot \sim 0.4$ GeV cm$^{-3}$ at solar position $r_\odot \sim 8.1$ kpc)} is also shown.

\begin{table}[t!]
\centering
 \begin{tabular}{c c}
 \hline
 Parameter [Units] & Value \\
 \hline\hline
 $d$ [kpc] & 8.02$_{-0.2}^{+0.2}$ \\
 
 $\rho_0$ [10$^6$ M$_{\odot}$ kpc$^{-3}$] & 13.8$_{-6.6}^{+20.7}$ \\
 
 $R_s$ [kpc] & 16.10$_{-5.6}^{+12.2}$ \\
 
 $I_{DM,GC}$ [ph s$^{-1}$ cm$^{-2}$ sr$^{-1}$] & 0.63$_{-0.11}^{+0.11}$ \\ [0.5ex]
 \hline
\end{tabular}
\caption{Milky Way Dark Matter Halo parameters used in Figure \ref{fig:nfw}, originally measured by \cite{nesti 2013}.}
\label{tab:nfw_pars}
\end{table}

The upper-limits appear marginally consistent with the \edit2{empirical \cite{nesti 2013}} NFW profile, though the profile appears to more closely resemble a zero-slope line. This is consistent with a wholly non-existent feature whose upper-limits have no dependence on radial position in the Milky Way, but rather on the procedure responsible for generating the line flux probability distribution. Moreover, the upper-limit set in bin 1 is in exact agreement with the \edit1{\textit{XMM-Newton} Milky Way Halo upper-limit}. The possible detection in \cite{nico 2018} finds a flux considerably higher than our upper-limits in the region, and hence we plot the reported upper-limit from that work, to which our results offer heavy further constraints. Our upper-limits also firmly exclude the \textit{NuSTAR} fluxes, strongly evidencing the hypothesis that the \textit{NuSTAR} detections are instrumental effects.

However, due to the wide range of angular distances contained in each bin, our flux profile lacks the spatial resolution to evaluate whether it matches the NFW profile throughout the galaxy, particularly the highly marginal consistency seen in bin 1. \edit2{Furthermore, the systematic uncertainty introduced by the choice of model parameters prevents us from excluding correspondence to all NFW profiles, such as the \cite{boyarsky nfw} profile shown in Figure \ref{fig:nfw}. Dark matter may follow a profile other than NFW (see \citealt{burkert 1995}, \citealt{einasto 1965}, \citealt{read 2016}, \citealt{abazajian 2020a}, and others referenced therein for extensive discussions)}, rendering an NFW profile comparison model-dependent and not entirely conclusive. Therefore, despite our strong constraints, we cannot definitively rule out the decaying dark matter interpretation of the 3.5 keV line based on the radial profile of its upper-limits.

\edit1{\subsubsection{Profile Comparison to Boyarsky et al. 2018}}

\edit1{Figure \ref{fig:nfw} appears to show tension between our results and those of \cite{boyarsky 2018}, in which the positions of two data points fall within this work's bin 1. In particular, it appears that the two outer regions analyzed by \cite{boyarsky 2018} exhibit a 3.5 keV line flux that is excluded by our bin 1 upper-limit. This, however, is not the case, and in fact our results are consistent with \cite{boyarsky 2018}.}

\edit1{While the two overlapping data points from \cite{boyarsky 2018} are clearly above our bin 1 upper-limit in the figure, they are comparable values, with the lower 1$\sigma$ error bound of the inner point differing from our constraint only by $\sim$15\%. When making such close comparisons towards a spatial boundary of our bin, the coarseness of the bin must be taken into consideration. As described in section \ref{data reduction}, the bin sizes were chosen to include enough exposure time for a significant detection of the 3.5 keV line. As a result, the bins span large distances, across which the predicted NFW dark matter density can vary considerably. While providing the best possible flux upper-limit profile given the limitations of the available data, and thus allowing the most robust possible comparison to NFW predictions, the width of the bins do not yield accurate comparisons to nearby data points at their boundaries, such as the \cite{boyarsky 2018} points in question. Bin 1, for example, contains data from between 10 and 74 degrees of angular distance from the Galactic Center, with its constituent observations averaging a distance of 53.3 degrees, while data from the two \cite{boyarsky 2018} points comes exclusively from the region between 10 and 35 degrees from the Galactic Center. Therefore, a more thorough comparison is necessary.}

\begin{figure}[h!]
\includegraphics[width=9.cm]{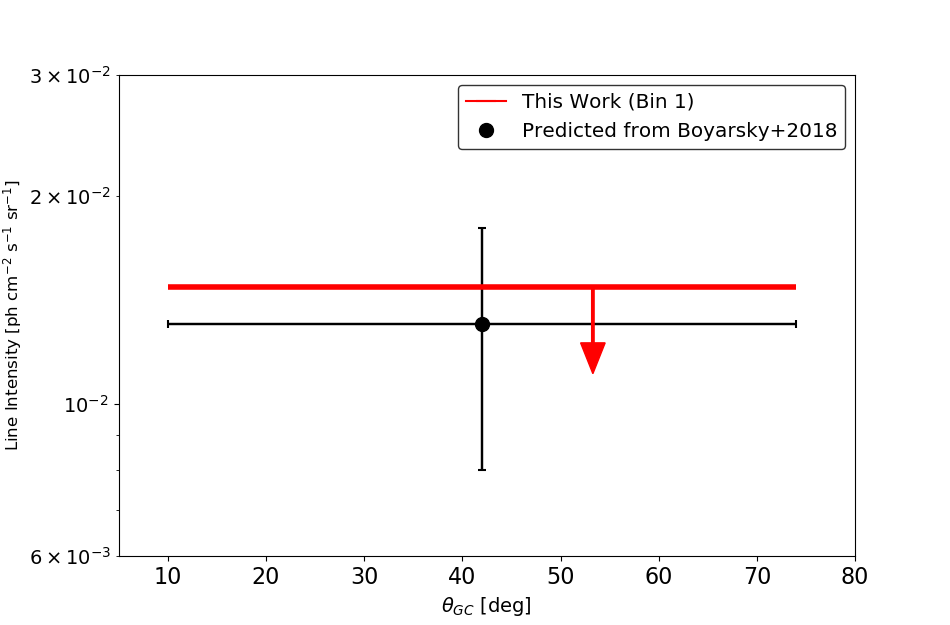}
\caption{\edit1{Comparing this work's upper-limit in bin 1 (see Figure \ref{fig:nfw}) to a 3.5 keV line flux prediction made using data from \cite{boyarsky 2018}. The predicted flux uses the best-fit dark matter decay rate computed in \cite{boyarsky 2018} from that work's total data set, located at angular distances within 35 degrees from the Galactic Center, together with the NFW profile used in that work (represented by the light green region in Figure \ref{fig:nfw}) to yield the expected 3.5 keV line flux in this work's bin 1.}}
\label{fig:nfw_boyarsky}
\end{figure}

\edit1{To perform this comparison, we use the best-fit decay rate from \cite{boyarsky 2018}, obtained in that work via the entirety of its data set. Assuming the NFW profile employed in that work (represented by the light green region in Figure \ref{fig:nfw}) and using the relation between flux and decay rate given by Equation \ref{DM_flux} (found below in section \ref{constrain mixing angle}), we can compute the expected 3.5 keV line flux in bin 1. The results of this computation are shown and compared to our bin 1 upper-limit in Figure \ref{fig:nfw_boyarsky}, which shows consistency between our results.}

\edit1{This consistency is reinforced when considering the differences between our data set and that of \cite{boyarsky 2018}. Here, we exclude the Galactic Disc, whereas a considerable portion of the \cite{boyarsky 2018} data set is dominated by the Galactic Disc. While dark matter profiles such as NFW describe a spherically-symmetric dark matter halo, it has been speculated that disc galaxies such as the Milky Way each have a ``dark disc" such that the galaxy's galactic disc contains a higher dark matter density than otherwise predicted by spherical dark matter profiles (\citealt{read 2008}; \citealt{bruch 2009}; \citealt{read 2010}; \citealt{read 2014}). In particular, \cite{bruch 2009} attempted to measure the density of the Milky Way's dark disc, finding that, in Earth’s local region of the disc, the dark matter density could be 20--100\% more dense than the portion of the Dark Matter Halo outside the Galactic Disc at the same distance from the Galactic Center. In the case of the 3.5 keV line's decaying dark matter interpretation, this would produce a correspondingly higher line flux in a data set that includes the Galactic Disc, such as that of \cite{boyarsky 2018}, than in a data set that excludes the Galactic Disc such as ours.}

\edit1{Accounting for cross-calibration between \textit{XMM-Newton} (used in \citealt{boyarsky 2018}) and \textit{Chandra} further strengthens the agreement between our results. All putative 3.5 keV emission occurs in the hard band (2--7 keV), in which \textit{XMM-Newton} fluxes tend to be greater than those of \textit{Chandra} by $\sim$5--10\%, as shown by \cite{n10} and thoroughly supported by \cite{gerrit 2015}. Therefore, considering this and all other aspects of the comparison between this work and \cite{boyarsky 2018}, we thus reach the robust conclusion that our results are consistent.}

\subsection{Constraints on Sterile Neutrino Mixing Angle}\label{constrain mixing angle}

\edit2{In addition to further constraining the sterile neutrino dark matter parameter space, determining the mixing angle associated with our flux upper-limits allows us to directly compare our results from the Milky Way Halo with the results of works from extra-galactic sources such as galaxies and clusters. Total dark matter mass in the FOV must be considered to calculate the mixing angle from line flux, and therefore the value for a Milky Way Halo analysis such as ours is dependent on the choice of dark matter profile. Thus, we have employed a careful procedure to account for this model-dependent uncertainty in our calculation of the mixing angle upper-limit.}

For our computation, we use the sterile neutrino decay rate from \cite{pal 1982} ($\Gamma_\gamma$; see Equation \ref{decay_rate}), the \edit2{empirical \cite{nesti 2013}} NFW profile, and the predicted flux of decaying dark matter contained in the FOV, given by \cite{neronov 2016b} as:

\begin{equation}\label{DM_flux}
F_{DM} = \frac{\Gamma_\gamma}{4\pi m_s}\frac{M_{DM,FOV}}{D^2}
\end{equation}
where $\Gamma_\gamma$ is the decay rate given in Equation \ref{decay_rate}, $m_s$ is sterile neutrino mass, $M_{DM,FOV}$ is the total mass of dark matter contained in the FOV, and $D$ is the distance from Earth to the mass of dark matter. From this, one can solve for the factor sin$^2$(2$\theta$), where $\theta$ is the mixing angle, which becomes the only variable quantity in $\Gamma_\gamma$ when $m_s$ is known (or, in this case, set to $m_s=2E_{3.51}=7.02$ keV from the model's fixed line energy), obtaining:
\begin{equation}\label{mixing_angle}
sin^2(2\theta) = \frac{F_{3.5}}{C_\Gamma}\frac{1}{m_s^4}\left(\frac{M_{DM,FOV}}{D^2}\right)^{-1}
\end{equation}
where $F_{3.5}$ is the 3.5 keV line flux and the constant $C_\Gamma$ is given by:
\begin{equation}\label{c_gamma}
C_\Gamma = \frac{1.38\times10^{-22}}{4\pi} s^{-1} keV^{-5}
\end{equation}

For a given angular distance from the Galactic Center, the surface mass density factor $\frac{M_{DM,FOV}}{D^2}$ can be obtained by integrating the NFW profile (using Equation \ref{nfw} and \edit2{applying the desired angular distance to} Equation \ref{r_define}) over the FOV's solid angle and line of sight. Integrating accordingly, out to the virial radius $r_{200}=$200 kpc \citep{dehnen 2006}\footnote{Note that the choice of $r_{200}$ does not substantially impact the results due to the asymptotic nature of the NFW profile at large radii.} to obtain the surface mass density factor $\frac{M_{DM,FOV}}{D^2}$, then evaluating Equation \ref{mixing_angle} for the flux in each bin, we can put constraints on the $\sim$7 keV sterile neutrino mixing angle. The results of this calculation are reported in Table \ref{tab:mixing_angles}. Using the results for sin$^2$(2$\theta$) and the corresponding decay rate (Equation \ref{decay_rate}), we obtain the average lifetime ($\tau_{DM}$) of the sterile neutrino, also reported in Table \ref{tab:mixing_angles}.

\begin{table}[h!]
\centering
 \begin{tabular}{c c c c}
 \hline
 Bin & sin$^2$(2$\theta$) [10$^{-11}]$ & $\Gamma_\gamma$ [10$^{-28}$ s$^{-1}$] & $\tau_{DM}$ [10$^{27}$ s]\\ [0.5ex]
 \hline\hline
 1 & 2.31$_{-0.04}^{+0.07}$ & 0.53$_{-0.01}^{+0.02}$ & 18.91$_{-0.49}^{+0.81}$ \\
 
 2 & 4.39$_{-0.08}^{+0.13}$ & 1.00$_{-0.03}^{+0.04}$ & 9.97$_{-0.26}^{+0.43}$ \\
 
 3 & 5.69$_{-0.1}^{+0.16}$ & 1.30$_{-0.03}^{+0.06}$ & 7.68$_{-0.2}^{+0.33}$ \\

 4 & 4.81$_{-0.08}^{+0.14}$ & 1.10$_{-0.03}^{+0.05}$ & 9.10$_{-0.23}^{+0.39}$ \\[0.5ex]
 \hline
 
 ALL & 2.58$_{-0.04}^{+0.07}$ & 0.59$_{-0.02}^{+0.03}$ & 16.95$_{-0.44}^{+0.73}$ \\[0.5ex]
 
 \hline
 
\end{tabular}
\caption{Upper-limit constraints on the $\sim$7 keV sterile neutrino mixing angle $\theta$. Values were computed for the total data set using the average distance of all observations from the GC, while values in each bin used the average distance from the GC of observations in that bin. 1$\sigma$ errors represent propagated uncertainties in parameters used for the calculations. Here, we also report the corresponding sterile neutrino decay rates ($\Gamma_\gamma$) and average lifetimes ($\tau_{DM}$).}
\label{tab:mixing_angles}
\end{table}

\begin{figure}[t!] 
\includegraphics[width=9.cm]{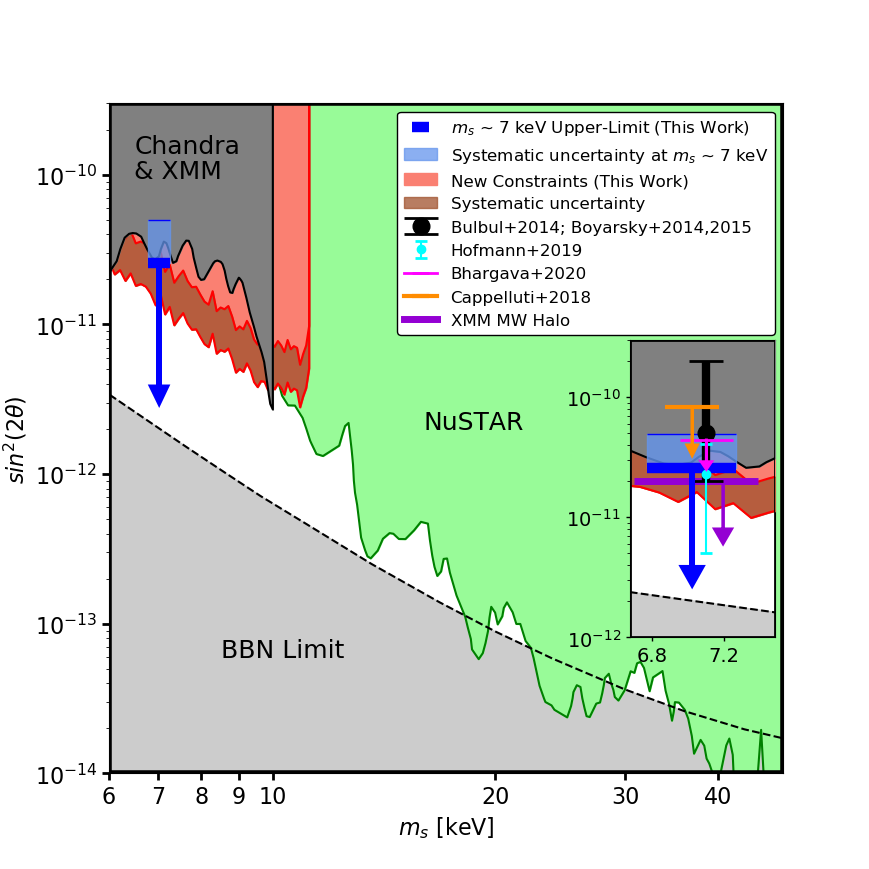}
\caption{Constraints on the sin$^2$(2$\theta$), $m_s$ parameter space adapted from \cite{roach 2020} and updated with the results of this work. Here, we plot our upper-limit, showing its marginal consistency with previous detections, including \cite{esra 2014}, \cite{boyarsky 2014}, \cite{boyarsky 2015}, and moderate consistency with \cite{hofmann 2019}. \edit2{For clarity, all previous $m_s \sim 7$ keV data points and upper-limits are shown in a second panel, enlarged for optimal data visualization. Also shown are our additional continuum constraints. Systematic uncertainties on both our $m_s \sim 7$ keV upper-limit and continuum constraints are shown as corresponding shaded regions.} Note that a similar figure originally appeared in \cite{ng 2019}.}
\label{fig:roach}
\end{figure}


We choose the mixing angle obtained via our most widely spatially-distributed and most statistically robust spectrum in the source-removed data---the $\sim$51 Ms stack containing the total data set---as our final value to constrain the sin$^2$(2$\theta$), $m_s$ parameter space with an upper-limit of sin$^2(2\theta) = 2.58 \times 10^{-11}$ at $m_s = 7.01$ keV. 

\edit2{To account for the systematic uncertainty resulting from the choices of $R_s$ and local dark matter density, we performed a second computation, replacing our NFW parameters with the most conservative up-to-date values (detailed in \citealt{abazajian 2020a}), namely $R_s = 26$ kpc and a local dark matter density $\rho_\odot \sim 0.28$ GeV cm$^{-3}$. This yielded an upper-limit of $sin^2(2\theta) = 4.96 \times 10^{-11}$, higher than our initial value by a factor of $\sim$2. The model-dependent uncertainty on the upper-limit is represented in the figure by a shaded region of uncertainty between the two models' upper-limits. Since the \cite{nesti 2013} parameters yield a fairly strong profile, the true value likely falls between its resulting upper-limit and that produced by the ultra-conservative \cite{abazajian 2020a} parameters. The \cite{boyarsky nfw} profile, for example, yields an upper-limit of $sin^2(2\theta) = 3.85 \times 10^{-11}$, falling almost exactly in the middle of our uncertainty region.}

\subsubsection{Continuum Constraints on $\theta$ and m$_{s}$}

We have also added further constraints to the parameter space, derived from the fact that we did not observe any new, unidentified emission lines in our analysis of the 2.9--5.6 keV energy band. To compute these constraints, we simulated spectra in XSPEC from the model of our total data set and with exposure time equal to the data set's total exposure time. For each simulation, we added a faint emission line ($I=10^{-8}$ ph s$^{-1}$ cm$^{-2}$) at a given energy on the continuum. The significance of the line was then computed using the $\Delta\chi^2$ test. This process was repeated for the same energy, gradually increasing the flux by $5\times10^{-10}$ ph s$^{-1}$ cm$^{-2}$, until a significance of at least 3$\sigma$ was obtained. The procedure was performed for all energies on the 2.9--5.6 keV band, spanned by intervals of 50 eV, a value chosen to be smaller than \textit{Chandra}'s energy resolution and hence to avoid gaps in our constraints across the continuum. \edit2{When computing mixing angles from fluxes, to account for the same systematic effects encountered in section \ref{constrain mixing angle}, we again perform computations for both the \cite{nesti 2013} and the \cite{abazajian 2020a} NFW parameters, represented in Figure \ref{fig:roach} by a similar region of uncertainty.}

\subsubsection{Comparison to Other Works}

We have plotted our $\sim$7 keV sterile neutrino upper-limit, along with the \edit1{\textit{XMM-Newton} Milky Way Halo upper-limit} (and various recent data points), onto the data presented by \cite{roach 2020} detailing other recent constraints on the sin$^2$(2$\theta$), $m_s$ parameter space (Figure \ref{fig:roach}). \edit1{The dark gray upper-limit region represents prior results from \textit{Chandra} and \textit{XMM-Newton} (see \citealt{ng 2019}, \citealt{roach 2020}, and others referenced therein), while the green region represents results from \textit{NuSTAR} (all of which are detailed or reported by \citealt{ng 2019}, \citealt{roach 2020}, and others referenced therein). The lower-limit region of the parameter space arises from theoretical values derived by \cite{roach 2020} from big bang nucleosynthesis (BBN) and the resulting limit placed on lepton asymmetry per unit entropy density (\citealt{dolgov 2002a}; \citealt{serpico 2005}; \citealt{laine 2008}; \citealt{boyarsky 2009b}; \citealt{ven 2016}; \citealt{roach 2020}).} The upper-limit from our analysis is lower than the value obtained from the possible detection in \cite{nico 2018}, suggesting that the 3.51 keV feature found in that work is not associated with sterile neutrino dark matter, and hence we again plot that work's upper-limit. Our upper-limit does, however, remain marginally consistent within 1$\sigma$ errors with \cite{esra 2014}, \cite{boyarsky 2014}, and \cite{boyarsky 2015}, and is moderately consistent with \cite{hofmann 2019}, potentially leaving room for the $\sim$7 keV sterile neutrino scenario. Additionally, the upper-limit \edit2{shows consistency with both} the \edit1{\textit{XMM-Newton} Milky Way Halo upper-limit} \edit2{and the \cite{bhargava 2020} constraint, appearing marginally above the \textit{XMM-Newton} Milky Way Halo value and containing the \cite{bhargava 2020} upper-limit within its systematic uncertainty range}.

Also plotted in Figure \ref{fig:roach} are the new continuum constraints from our simulations, \edit1{offering tighter restrictions on the parameter space for masses between $\sim$6 keV and $\sim$12 keV. except $\sim$7 keV. While the continuum constraints at $\sim$7 keV appear lower than the upper-limit set by our 3.5 keV line analysis, the continuum constraints are computed under the assumption that no feature exists at $\sim$3.5 keV, and hence do not apply to the region of the parameter space at $\sim$7 keV. Hence, we have plotted all data points, error bars, and upper-limit arrows for $m_s \sim 7$ keV strictly in front of the red continuum constraint region to appropriately break its continuity on the plot.}

\edit1{An important difference between our figure and the original version in \cite{roach 2020} is the omission of the ``Milky Way satellite counts" constraints. The values are simulation-dependent and, as detailed by \cite{boyarsky 2019}, cannot be used to constrain the sterile neutrino dark matter parameter space (see also \citealt{lovell 2017} for further discussion). As seen in the figure, a substantial portion of the parameter space remains allowed, particularly at $m_s \sim 7$ keV.} Thus, while the cases for both $\sim$7 keV sterile neutrino dark matter and the general decaying dark matter scenario for the 3.5 keV line have been considerably narrowed in this work, they cannot be ruled out completely.

\subsection{Closing Remarks}

This work provides the most comprehensive search to date for decaying dark matter in the Milky Way Halo using one of the various existing large archives of current X-ray telescopes. A follow-up analysis with new technology and data is required to improve the upper-limits reported here. All-sky coverage and comparable grasp in the keV energy range of the \textit{eROSITA} X-ray telescope will improve current constraints from the Milky Way Halo (\citealt{merloni 2012}; \citealt{hofmann 2019}). The high spectral resolution of observations using \textit{XRISM} and \textit{Athena} are required to measure the velocity dispersion and resolve the 3.5 keV line from nearby astrophysical lines to disentangle the origin of this feature (\citealt{zhong 2020}; \citealt{xrism 2020}).

\section{Acknowledgements}
The authors kindly acknowledge \textit{Chandra} Grant AR-19023B; Paul P. Plucinsky for his assistance in studying the low statistics of the \textit{Chandra} particle background; \edit1{and the anonymous reviewer who helped improve the original manuscript}. Dominic Sicilian acknowledges the University of Miami for supplying funding; Terrance J. Gaetz for helpful discussions on calibration of the \textit{Chandra} particle background; Keith Arnaud for introducing the method of setting random priors for Goodman-Weare MCMC chains in XSPEC; Iacopo Bartalucci for offering further technical insights into modeling the ACIS particle background; Marco Drewes for enlightening discussions on sterile neutrino dark matter; and Kevork Abazajian for beneficial explanations of sterile neutrino oscillation \edit2{and dark matter profiles in the Milky Way}.

\edit1{\software{Astropy (\citealt{astropy 2013}; \citealt{astropy 2018}), CIAO \citep{antonella 2006}, corner.py \citep{corner}, Matplotlib \citep{matplotlib}, NumPy \citep{numpy}, PyXspec 2.0.3 \citep{arnaud 2016}, XSPEC 12.10.1f \citep{arnaud 1996}}}



\appendix

\begin{figure*}[ht] 
\centering
\includegraphics[width=16.cm]{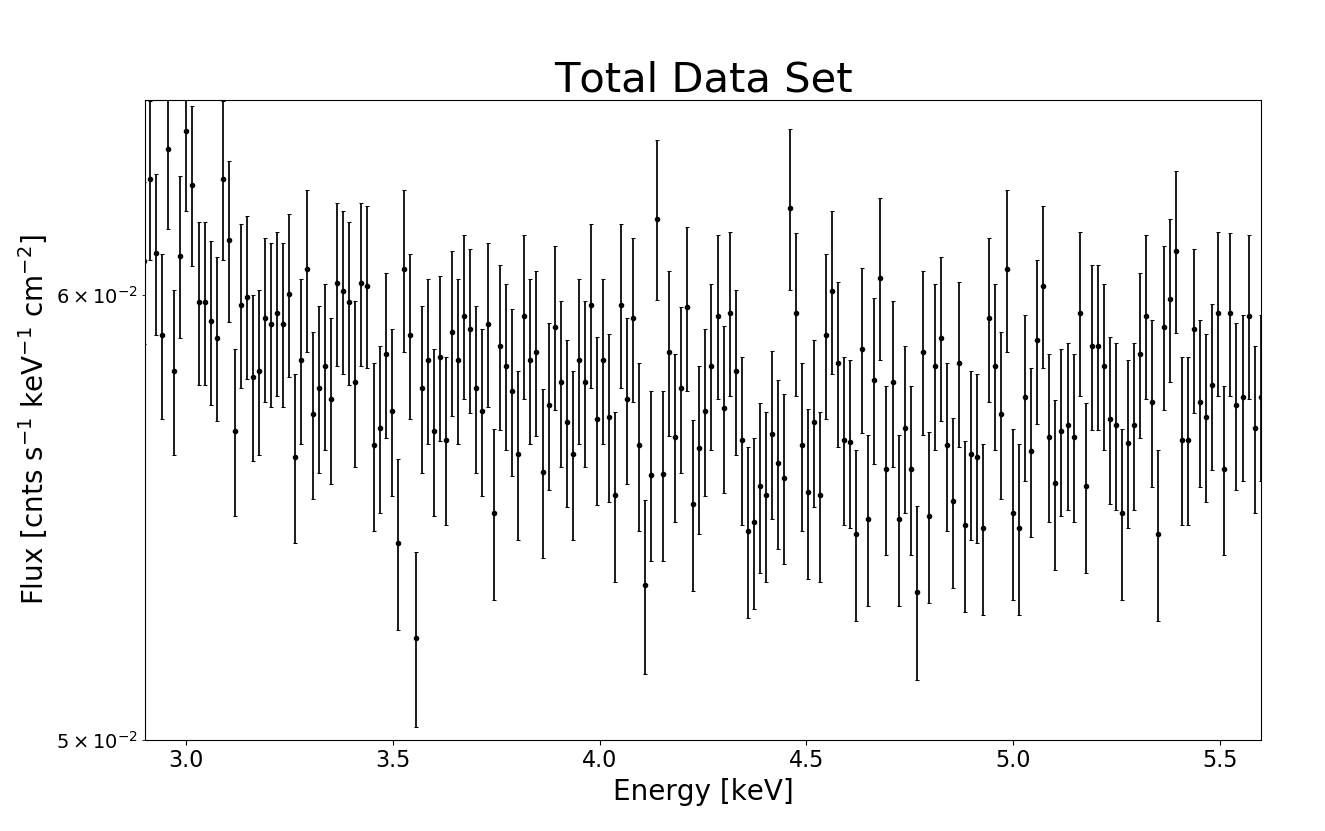}
\includegraphics[width=8.cm]{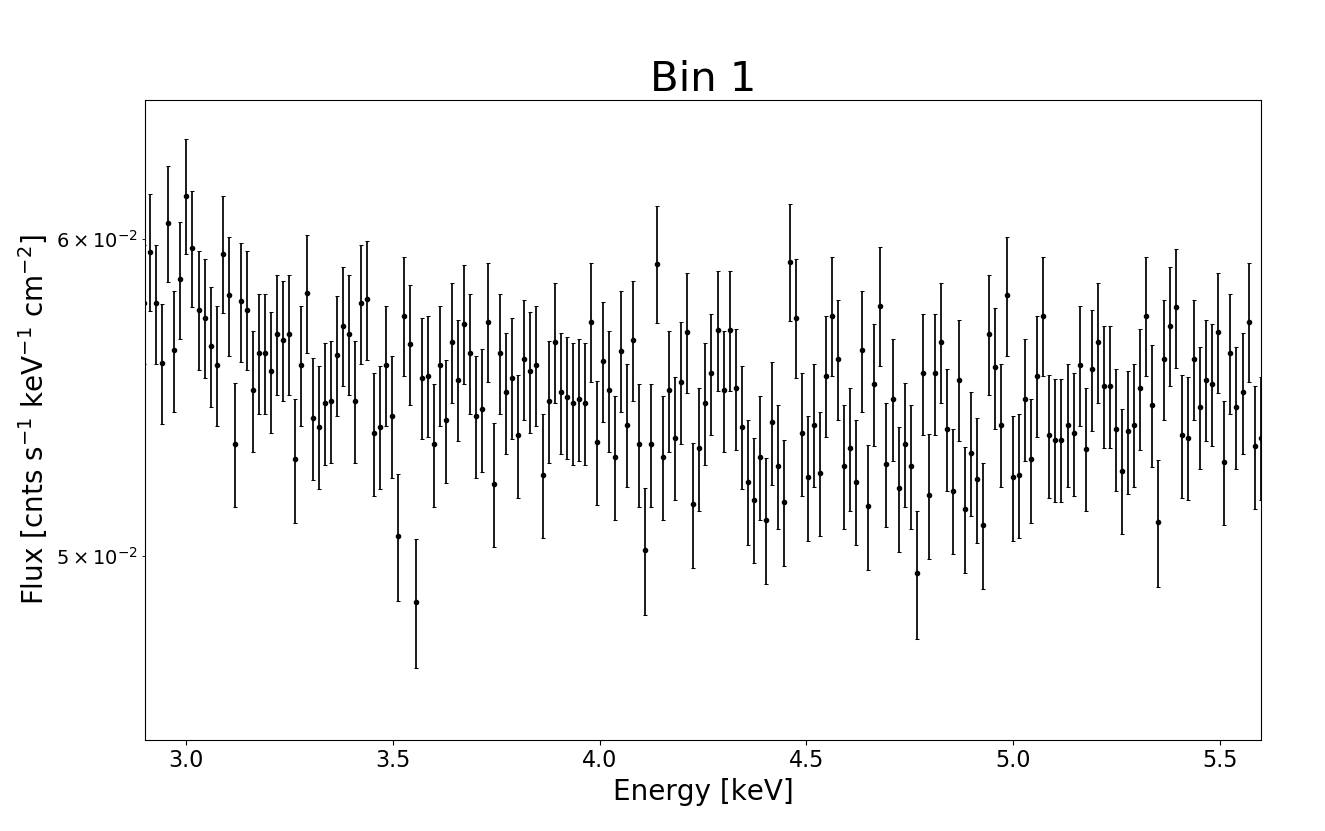}
\includegraphics[width=8.cm]{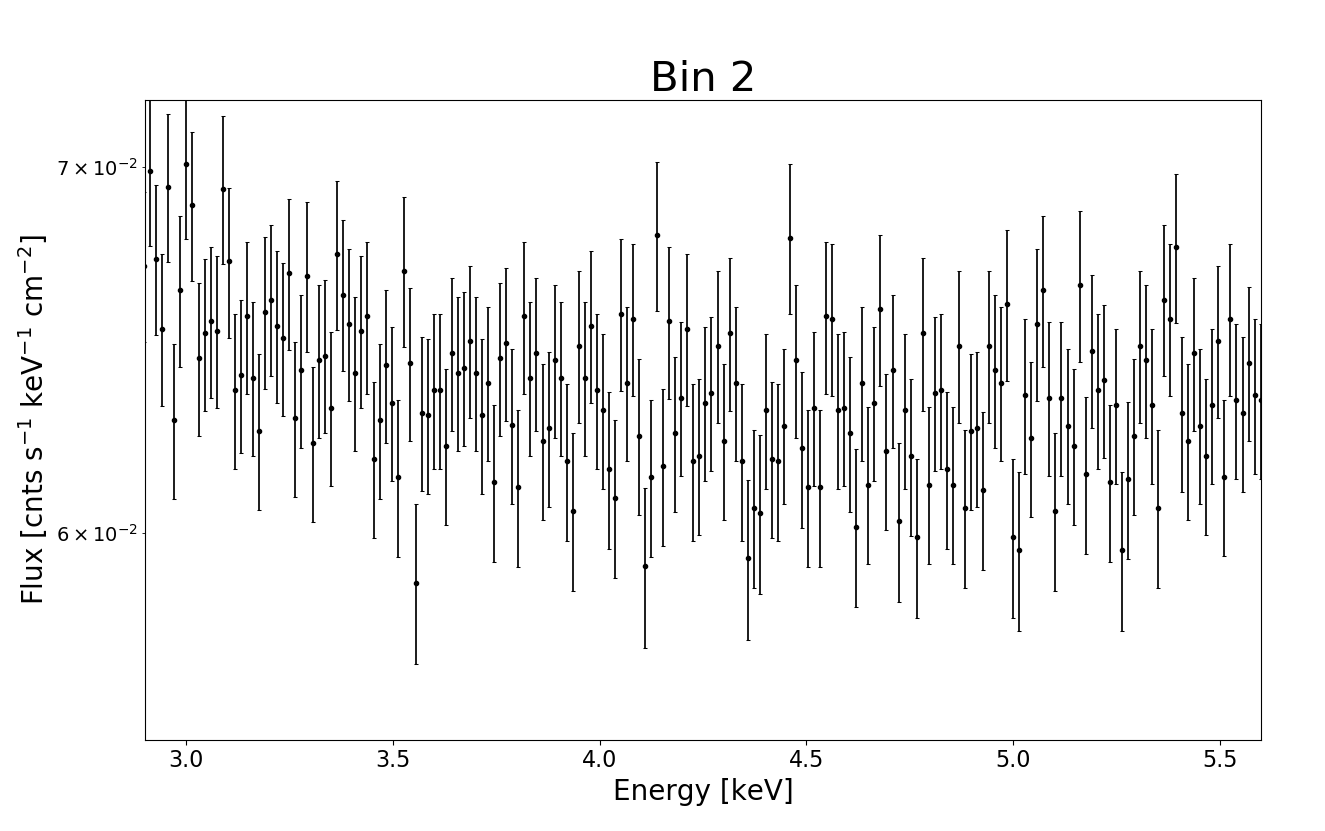}
\includegraphics[width=8.cm]{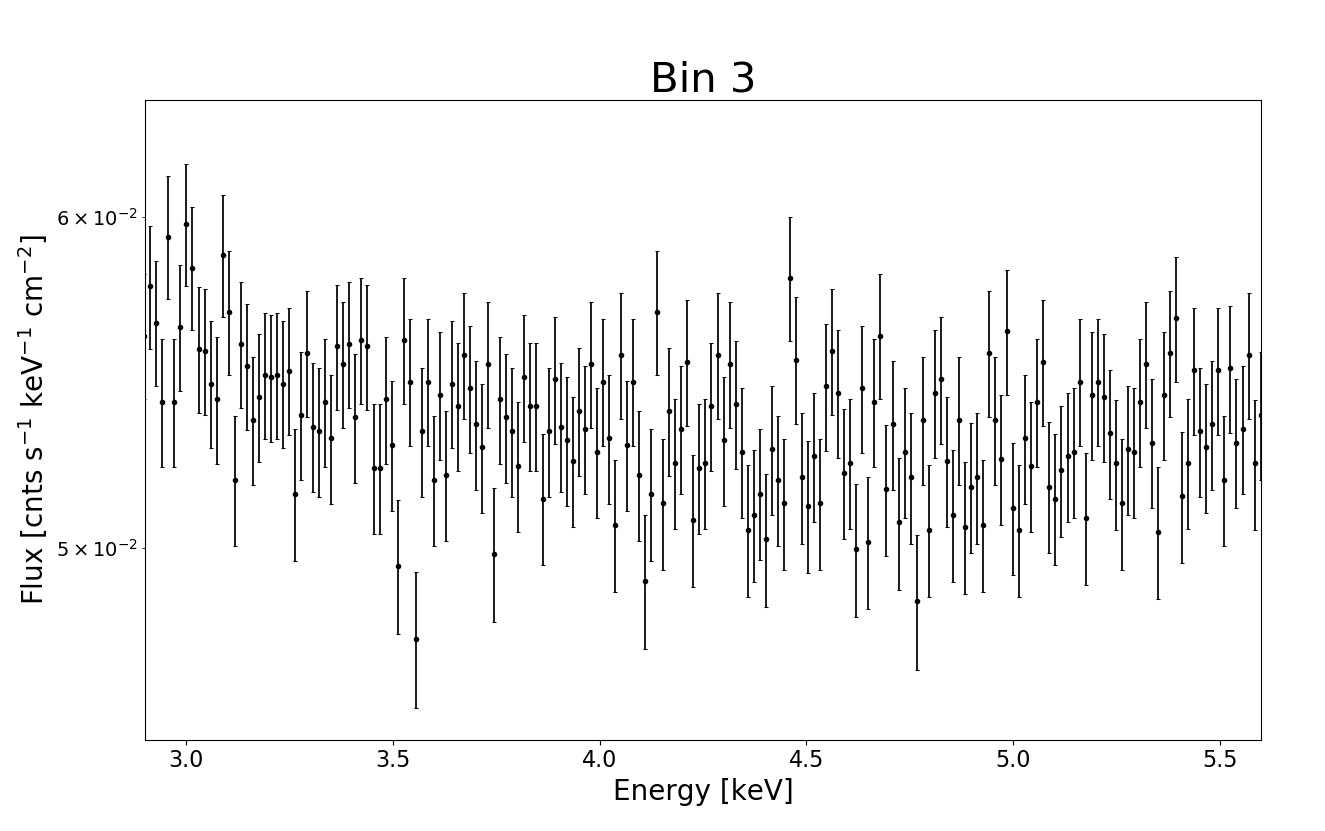}
\includegraphics[width=8.cm]{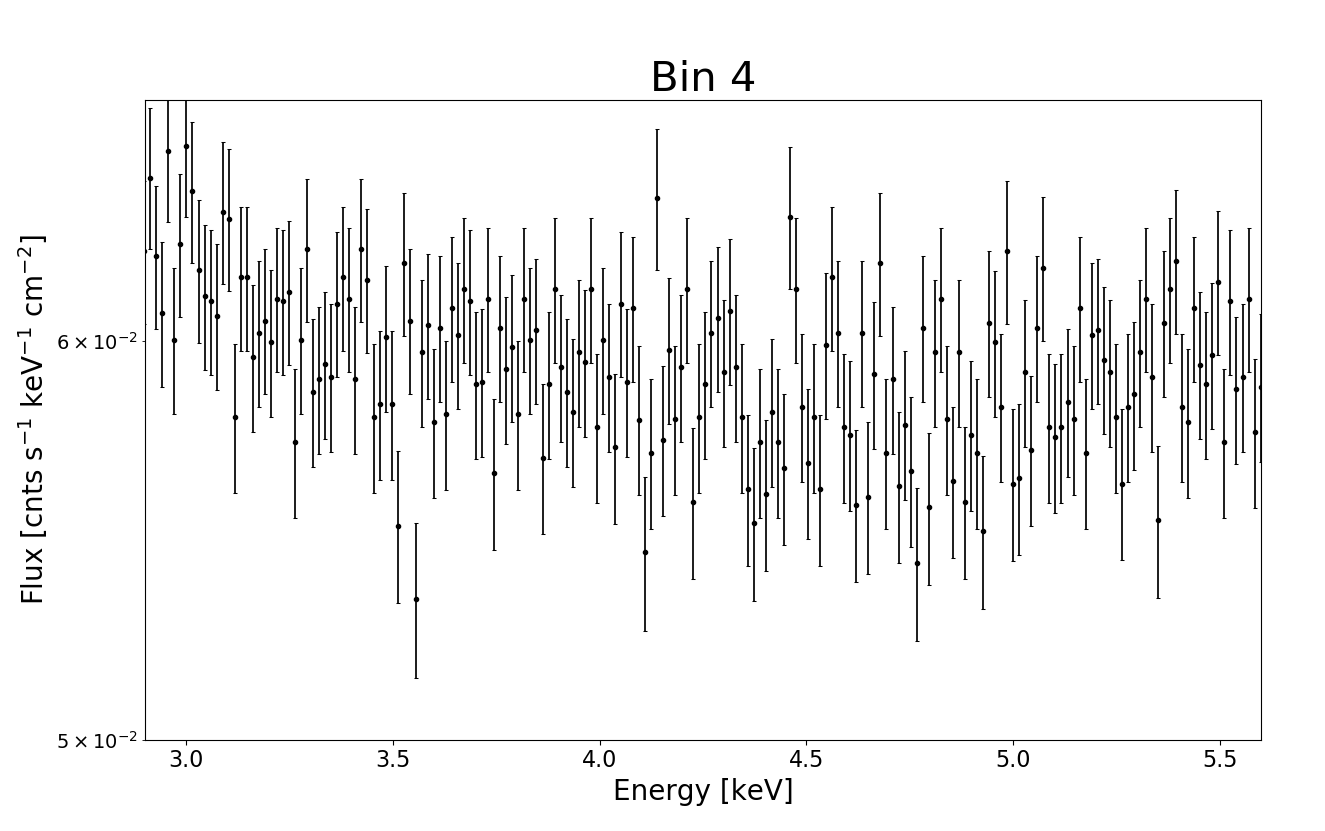}
\caption{Particle background spectra. Each title indicates to which data spectrum the particle background spectrum corresponds.}
\label{fig:stow_plots}
\end{figure*}


\begin{figure*}[ht] 
\includegraphics[width=9.cm]{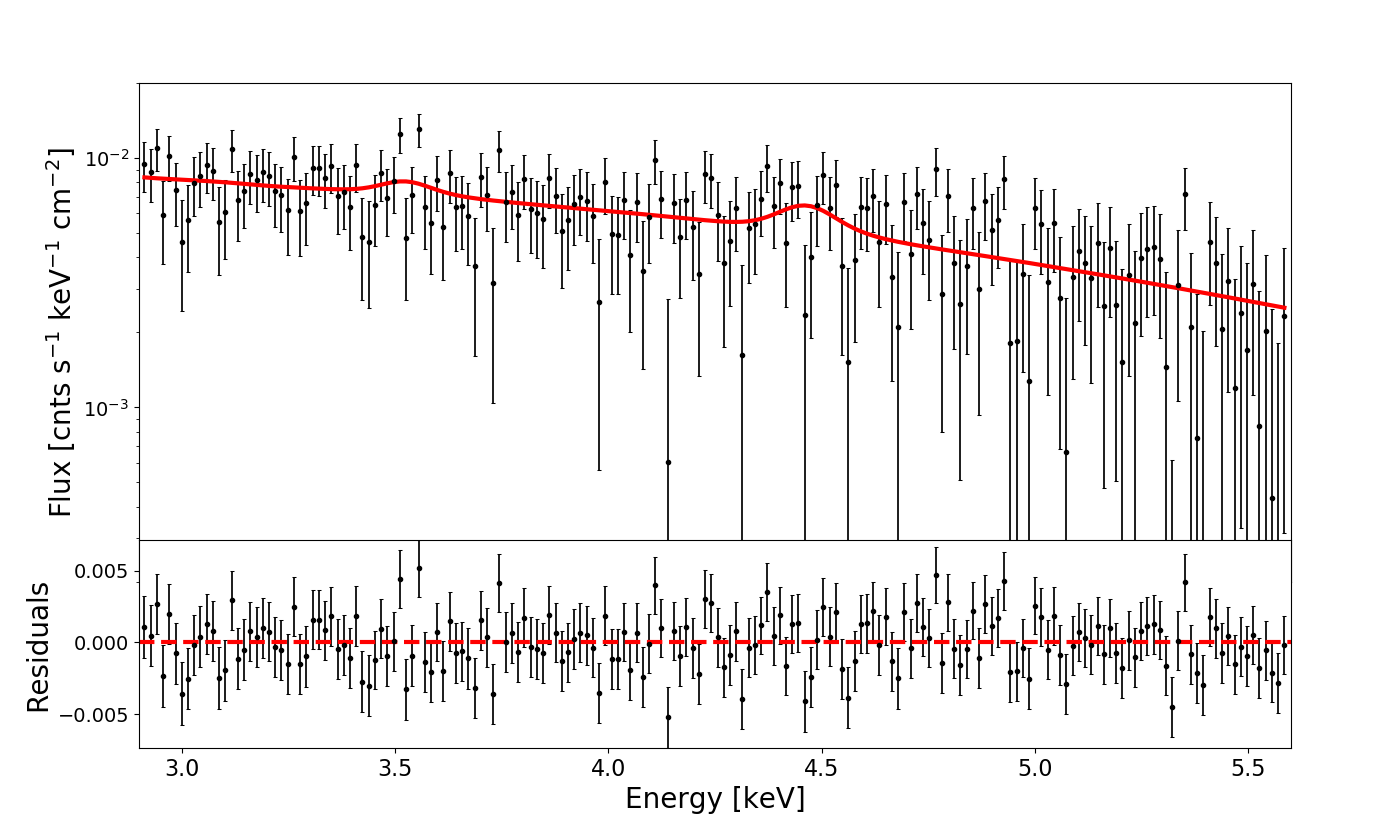}
\includegraphics[width=9.cm]{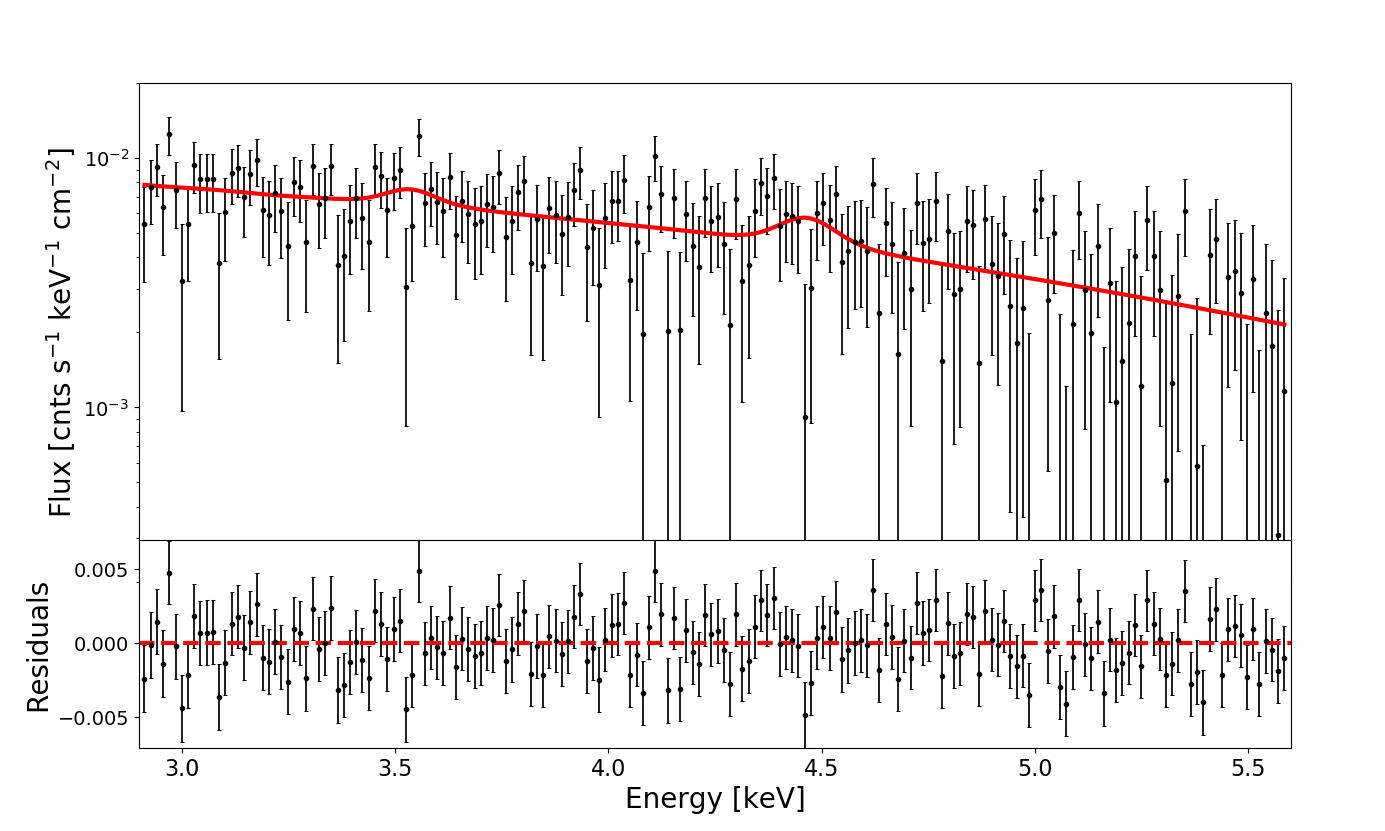}
\includegraphics[width=9.cm]{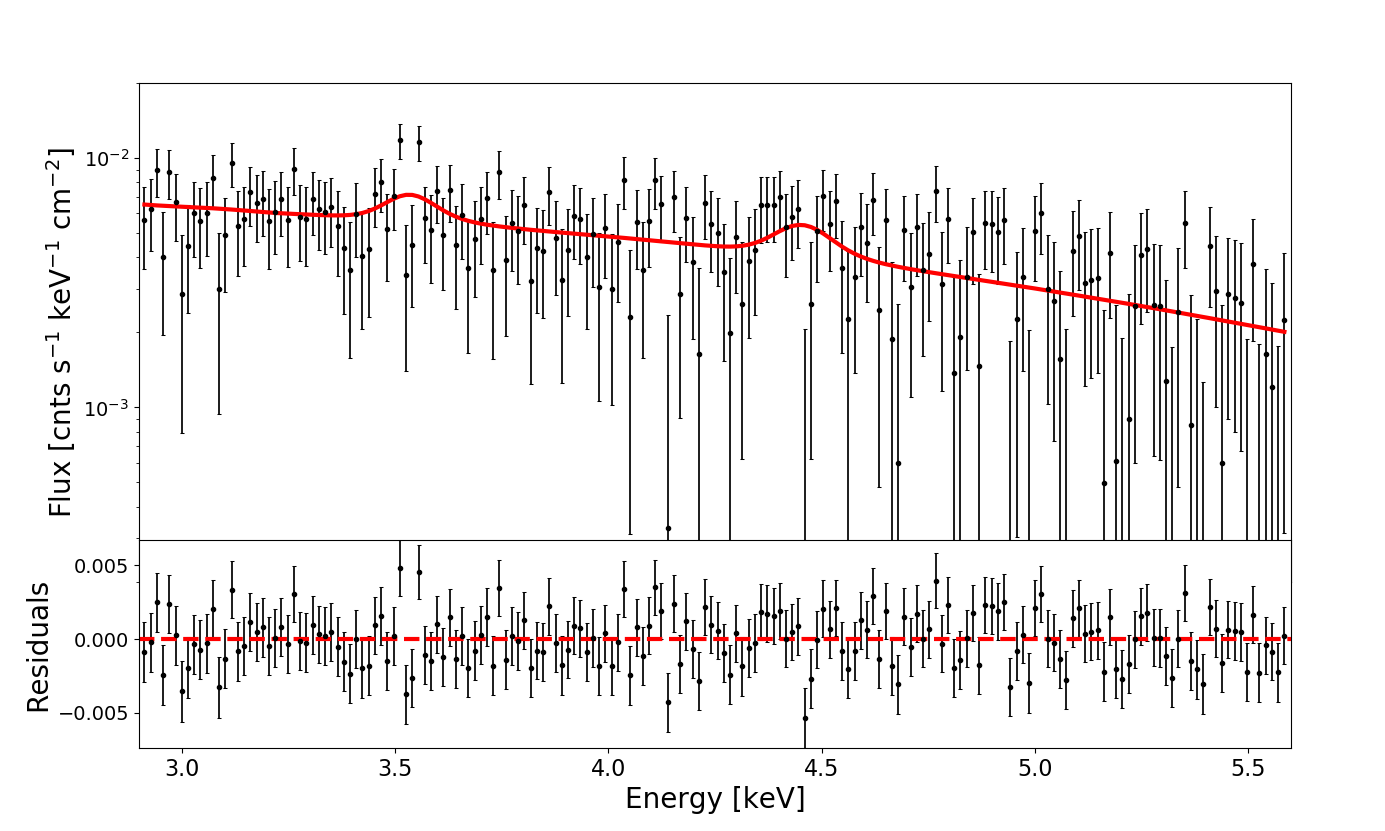}
\includegraphics[width=9.cm]{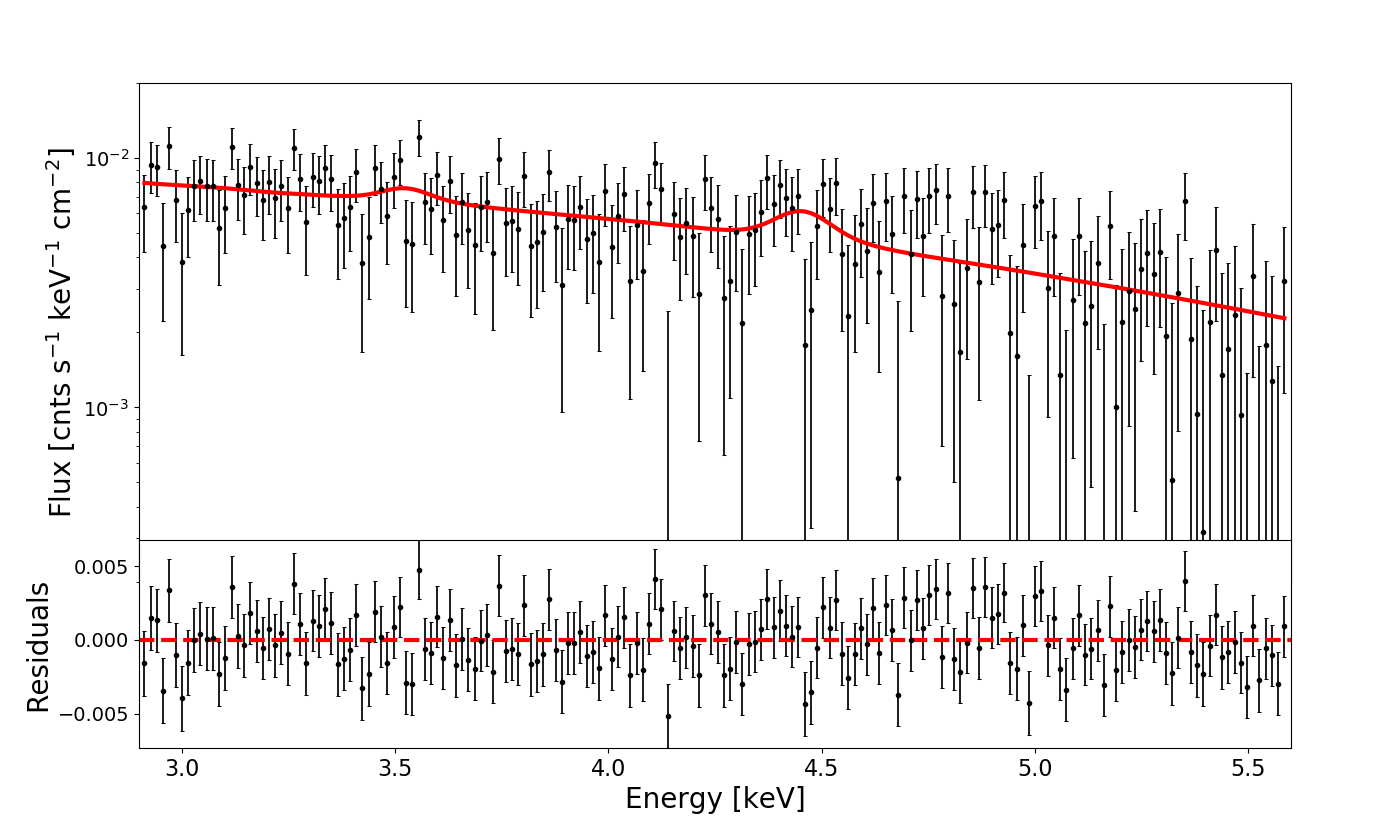}
\caption{Background-subtracted spectra of each angular distance bin, plotted with the $\sim$3.5 keV feature modeled. \textbf{Top Left:} Bin 1,  \textbf{Top Right:} Bin 2, \textbf{Bottom Left:} Bin 3, \textbf{Bottom Right:} Bin 4.} 
\label{fig:bins_sub_line}
\end{figure*}

 
\begin{figure*}[ht]
\includegraphics[width=9.cm]{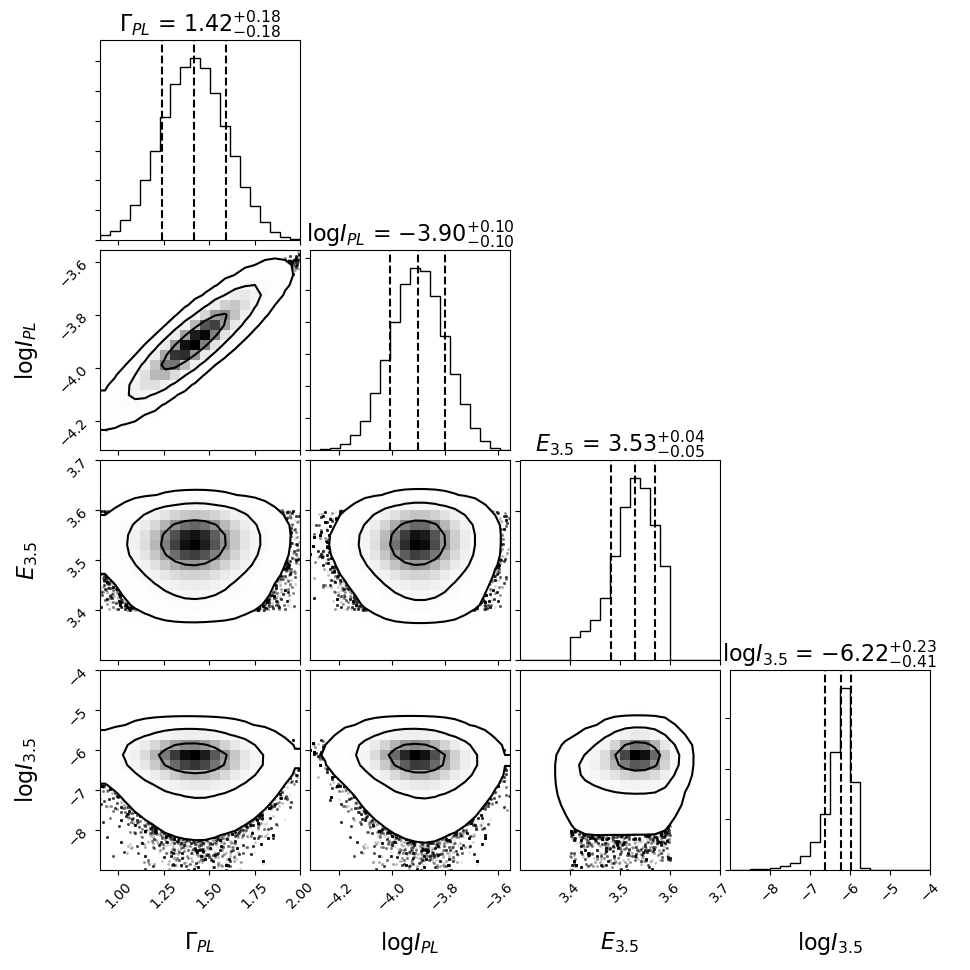}
\caption{MCMC contour plot for the background-subtracted model of the spectrum containing the total data set.}
\label{fig:all_chain}
\end{figure*}

\begin{figure*}[ht]
\includegraphics[width=9.cm]{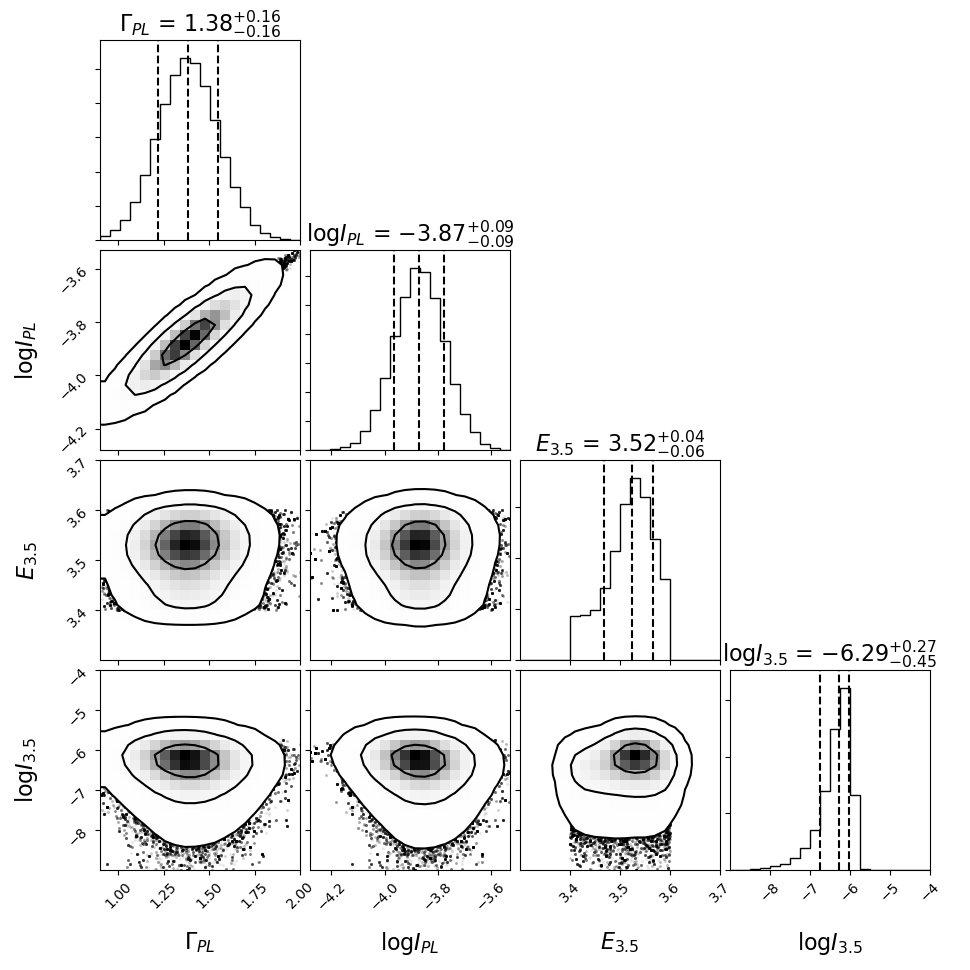}
\includegraphics[width=9.cm]{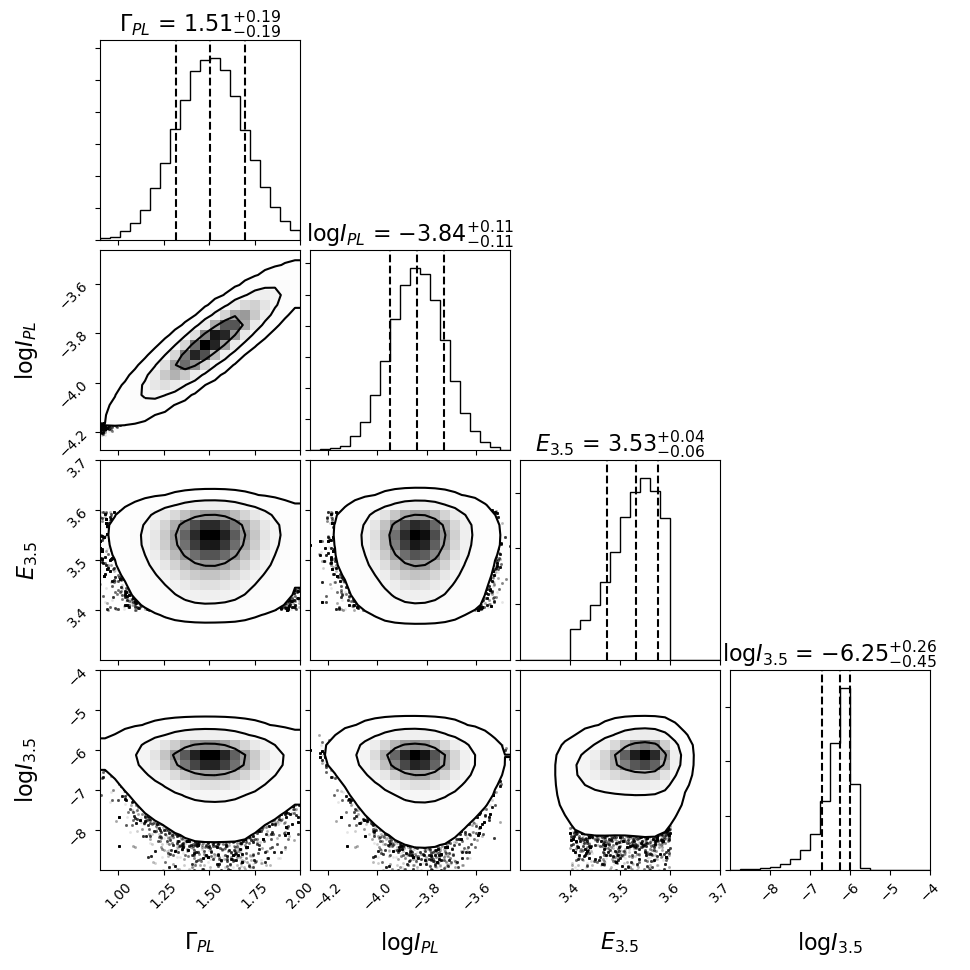}
\includegraphics[width=9.cm]{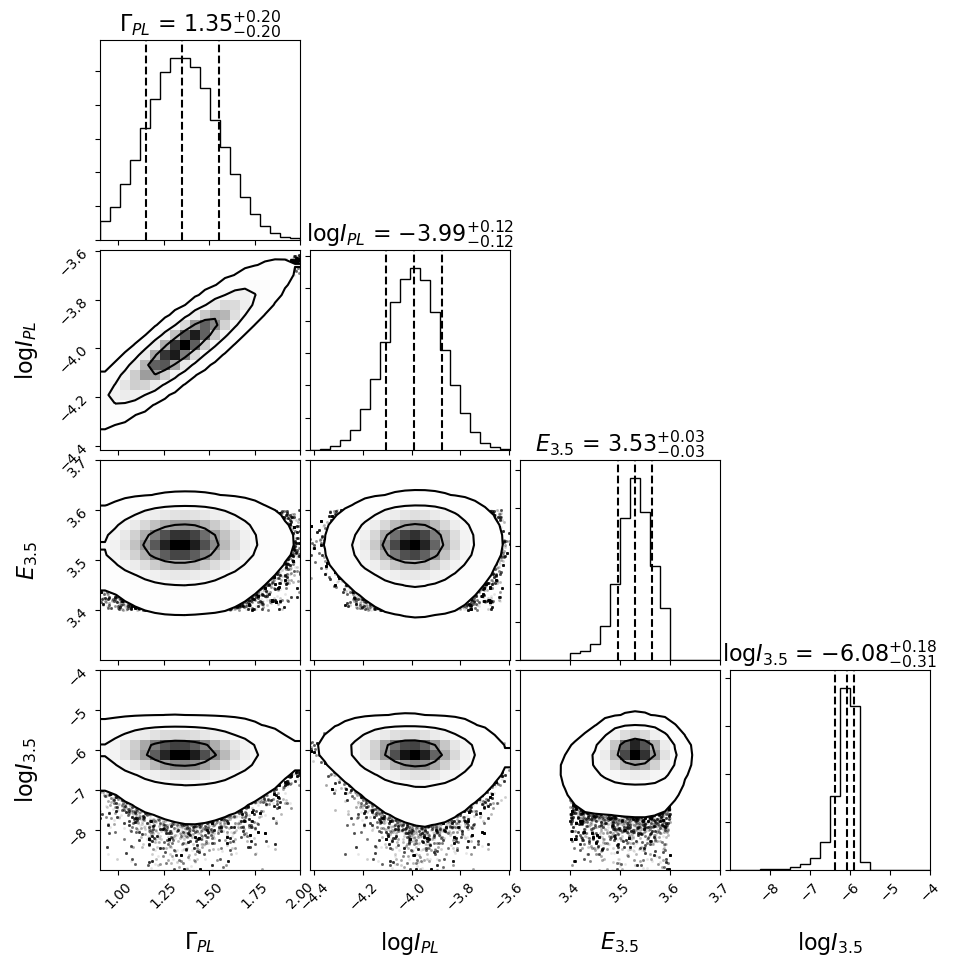}
\includegraphics[width=9.cm]{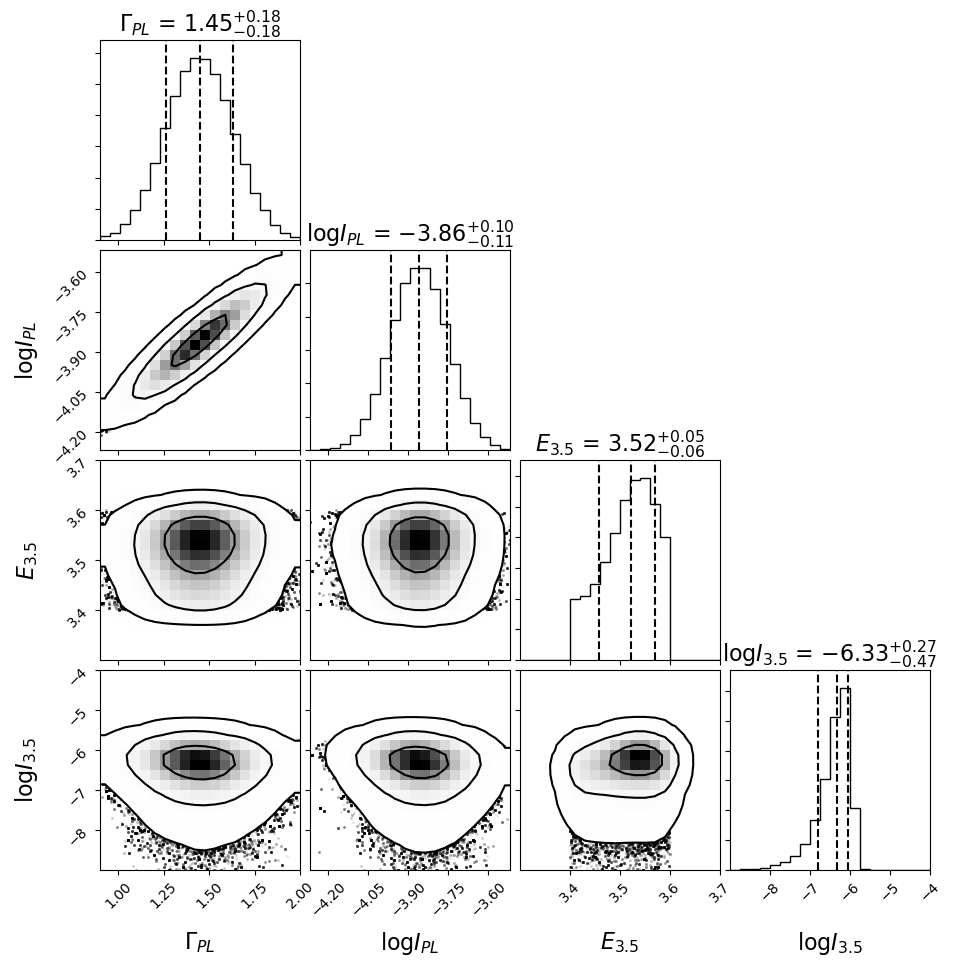}
\caption{MCMC contour plot for the background-subtracted models. \textbf{Top Left:} Bin 1,  \textbf{Top Right:} Bin 2, \textbf{Bottom Left:} Bin 3, \textbf{Bottom Right:} Bin 4.}
\label{fig:bins_chain}
\end{figure*}

\begin{figure*}[ht] 
\includegraphics[width=18.cm]{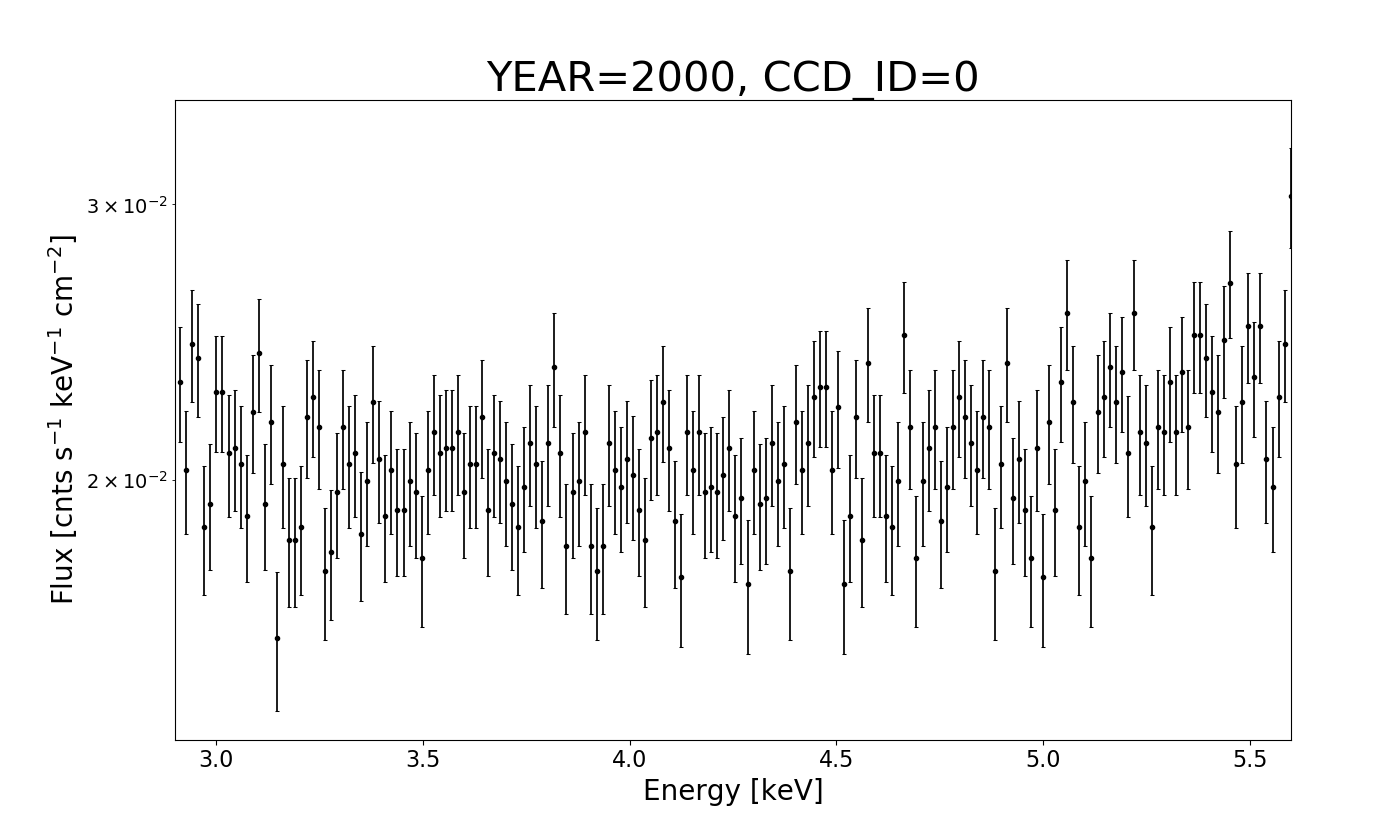}
\includegraphics[width=9.cm]{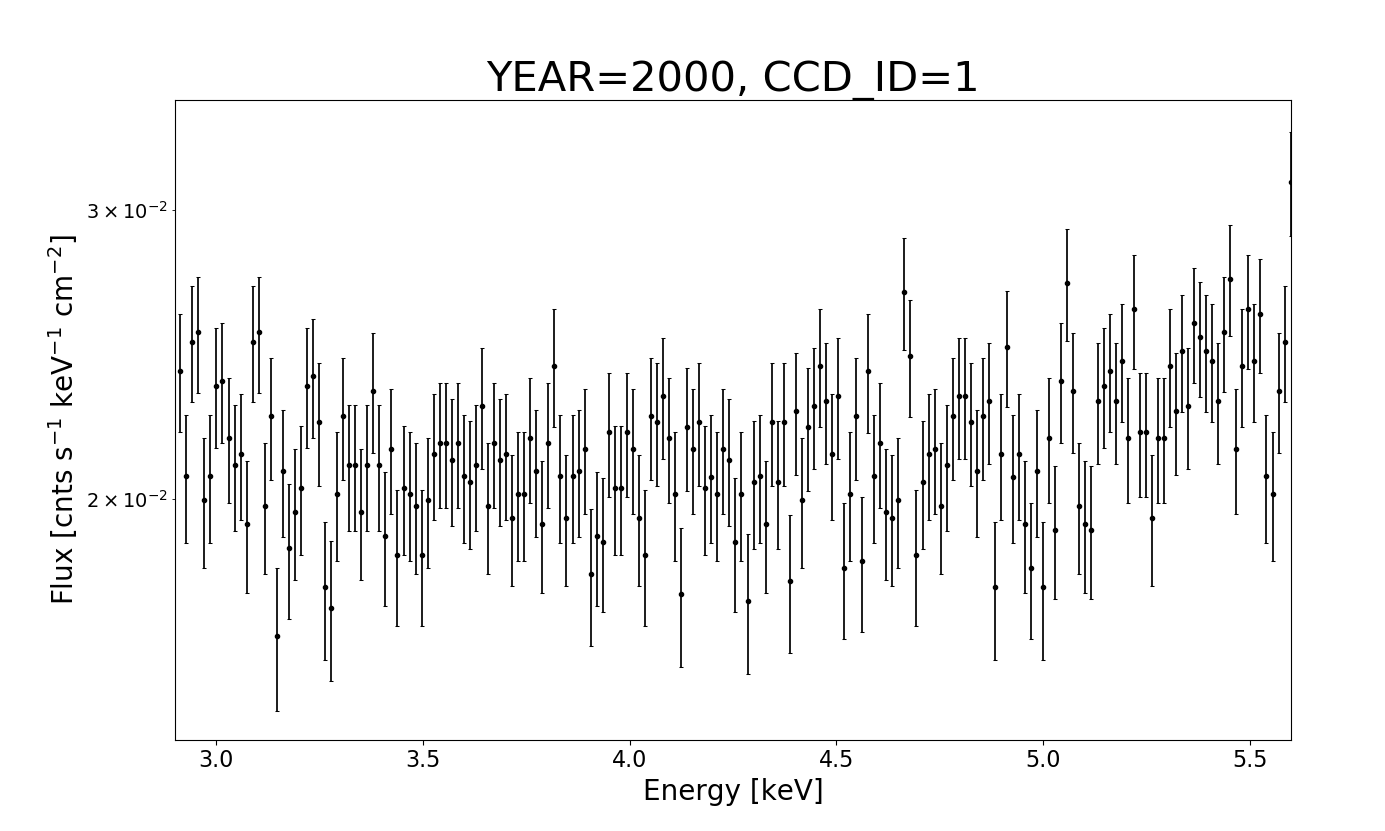}
\includegraphics[width=9.cm]{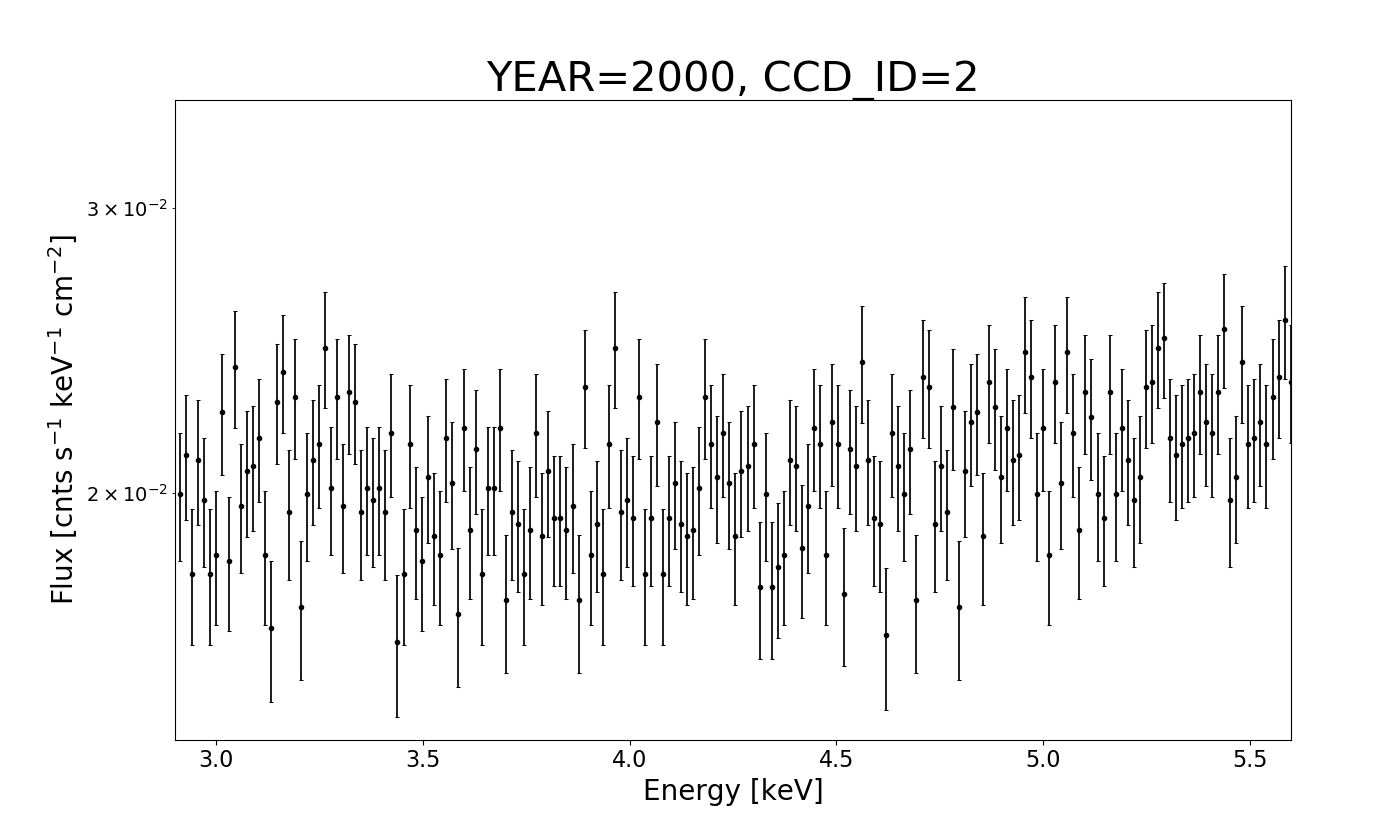}
\includegraphics[width=9.cm]{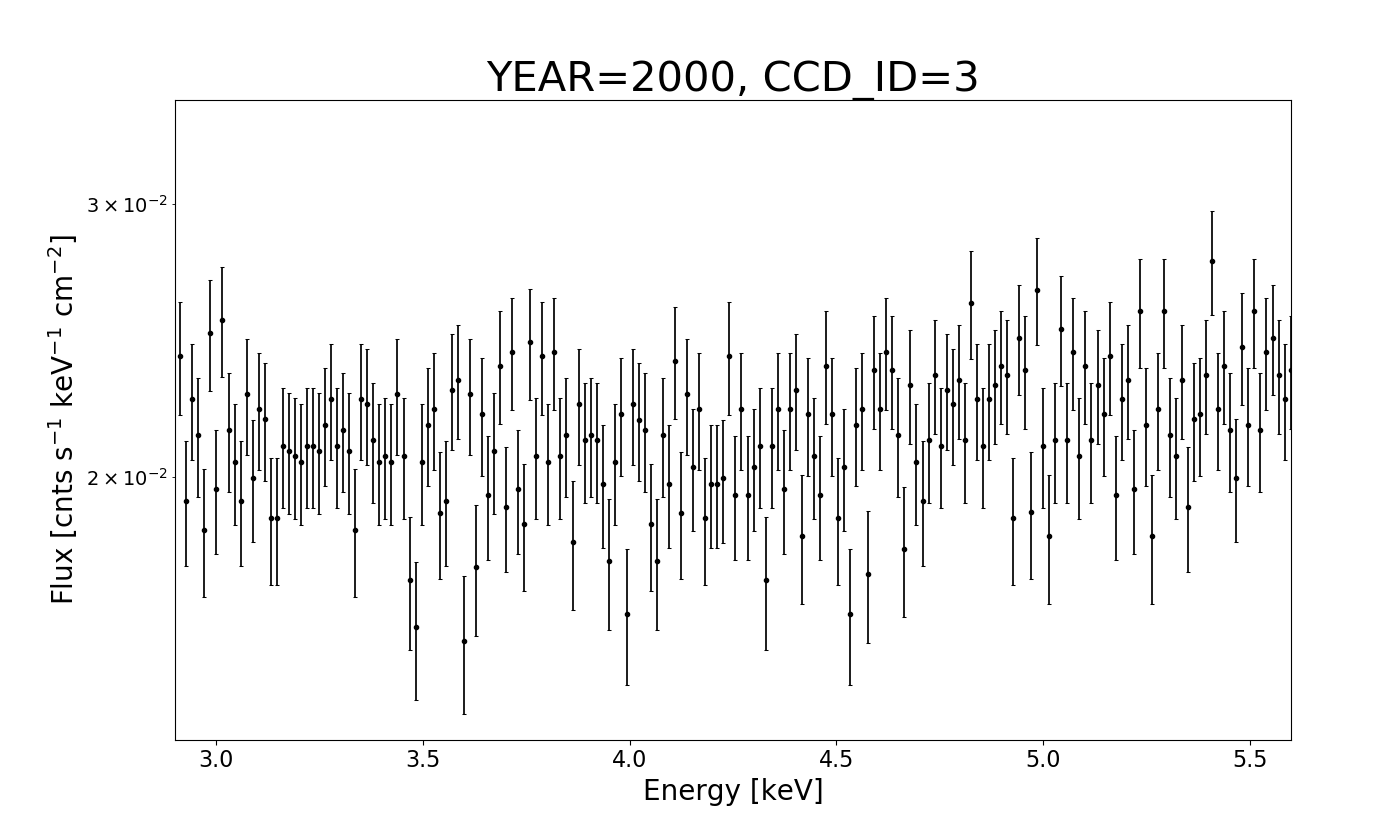}
\includegraphics[width=9.cm]{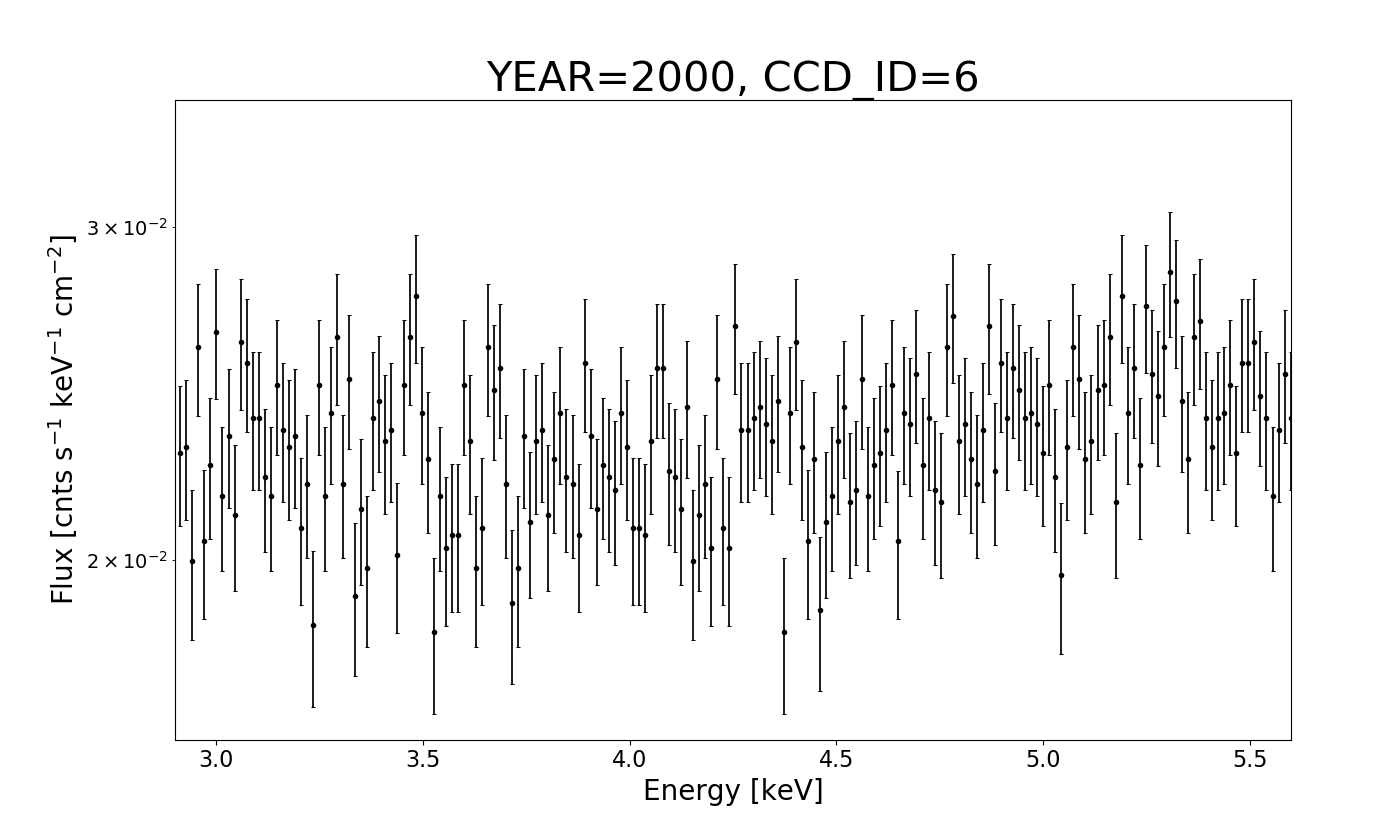}
\caption{Stowed spectrum for each CCD from the stowed observation in 2000.}
\label{fig:stow2000}
\end{figure*}

\begin{figure*}[ht] 
\includegraphics[width=18.cm]{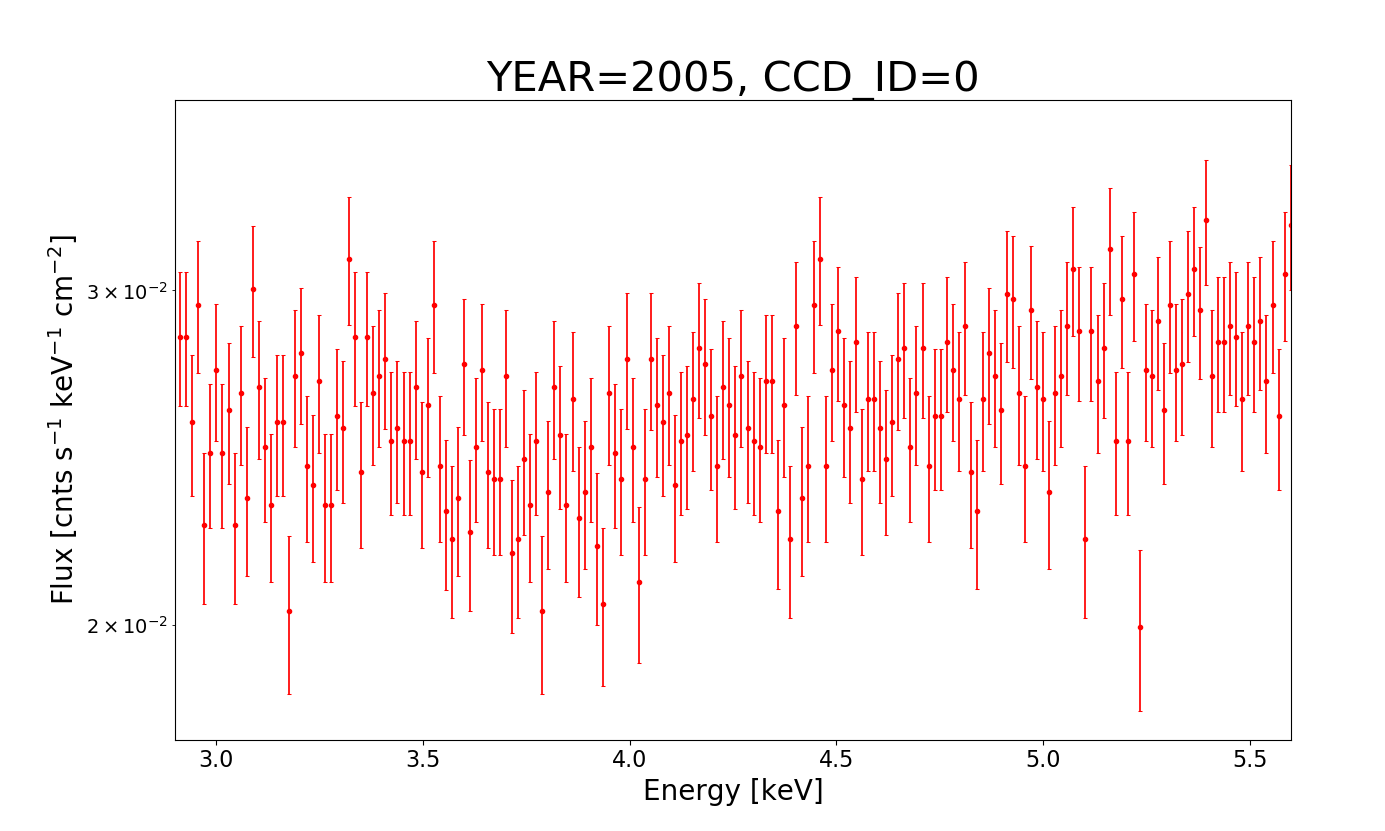}
\includegraphics[width=9.cm]{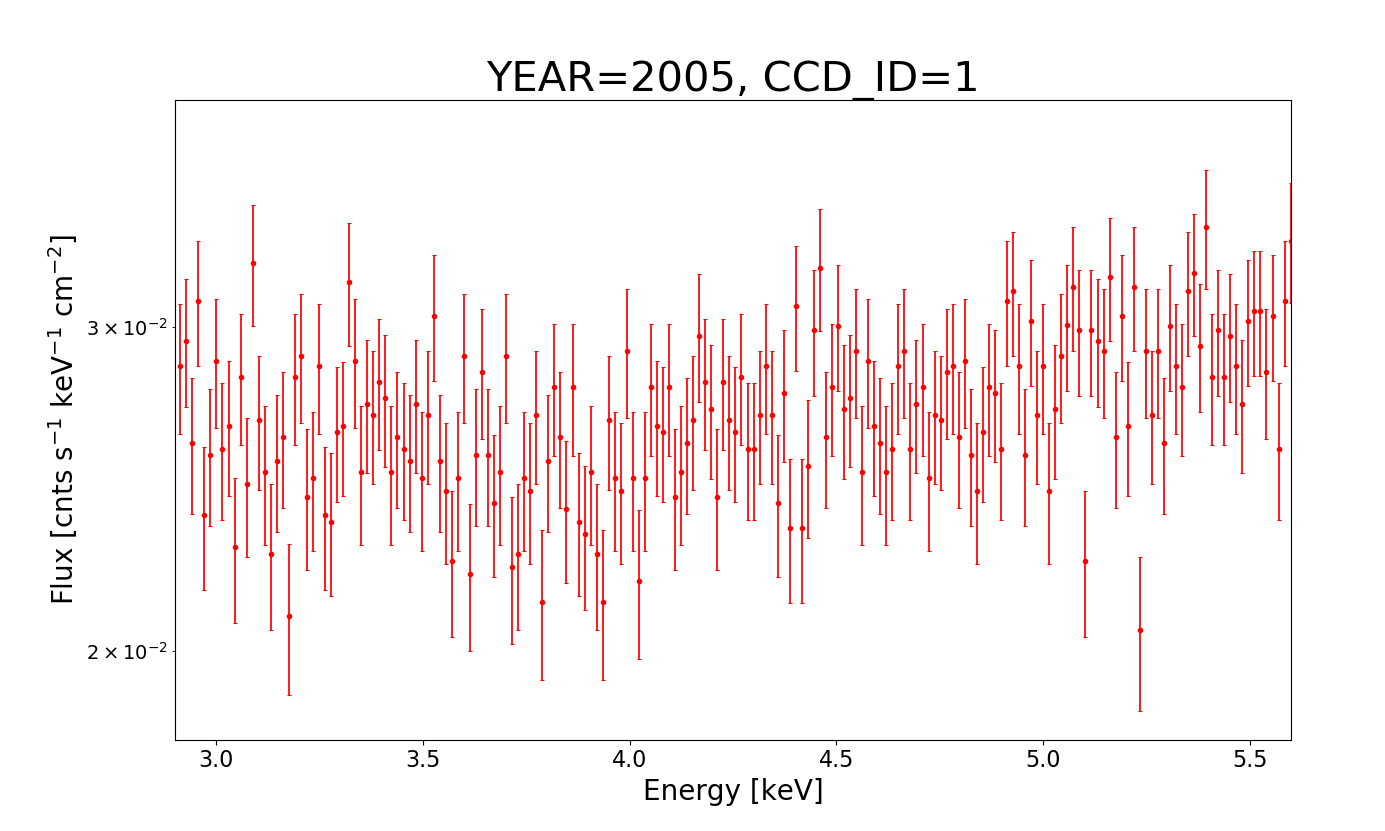}
\includegraphics[width=9.cm]{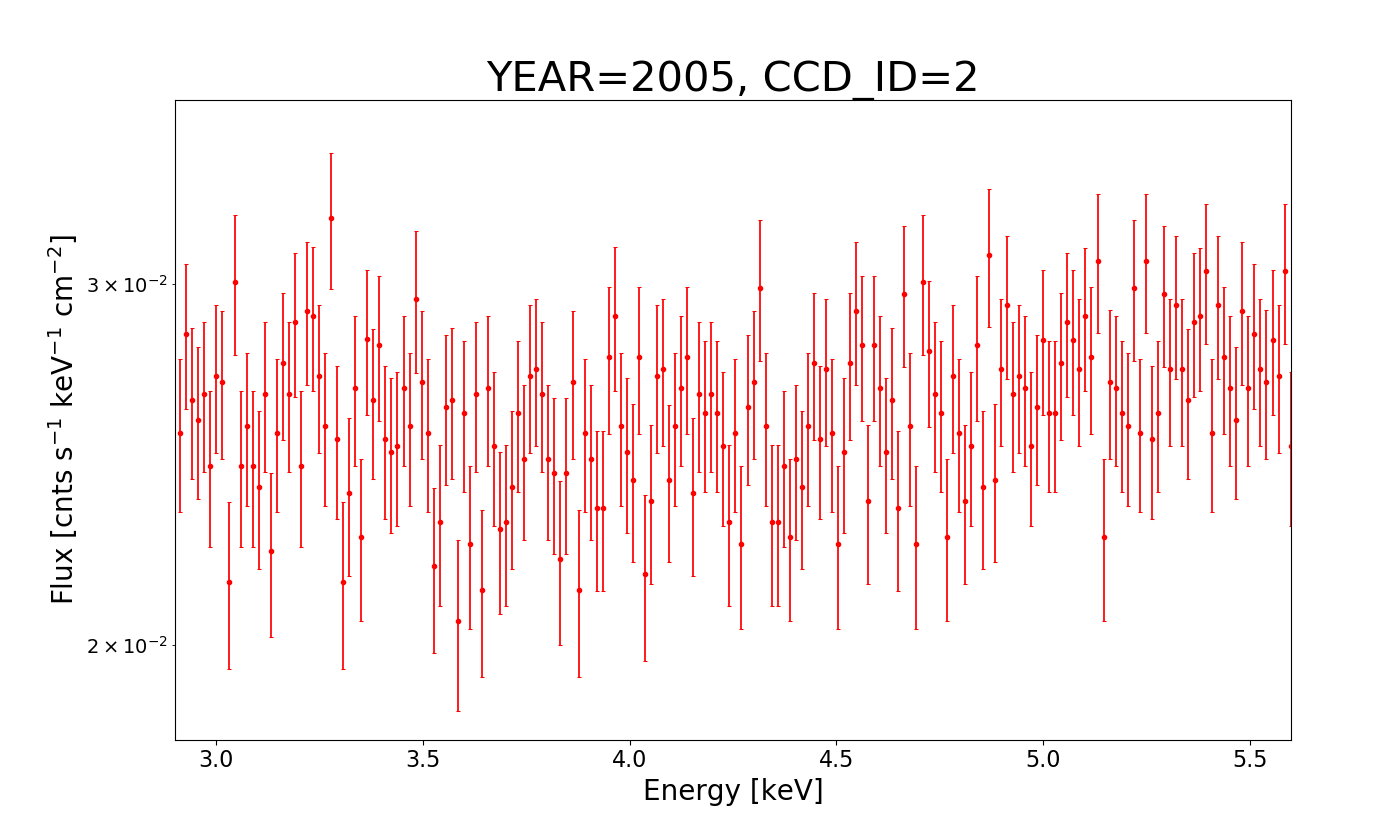}
\includegraphics[width=9.cm]{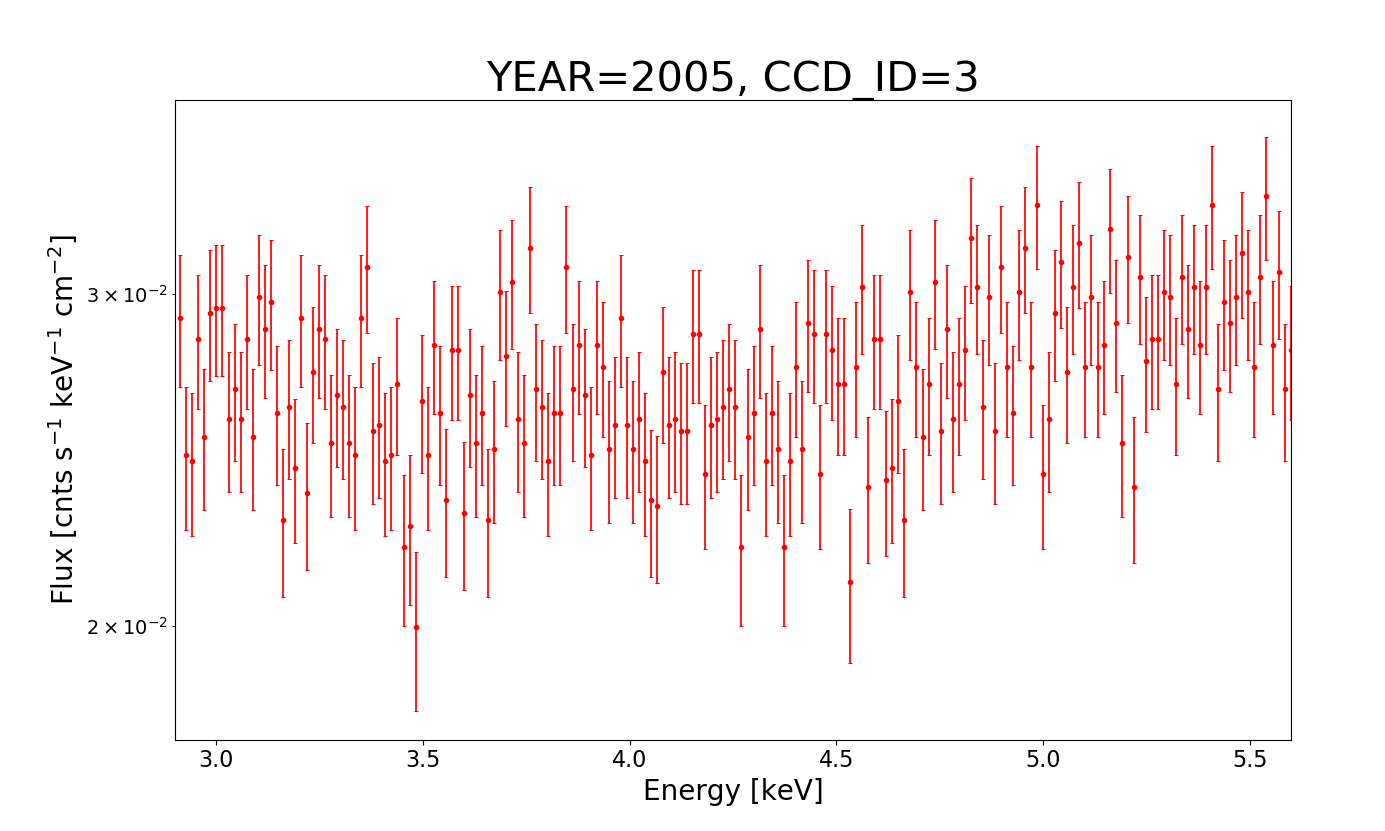}
\includegraphics[width=9.cm]{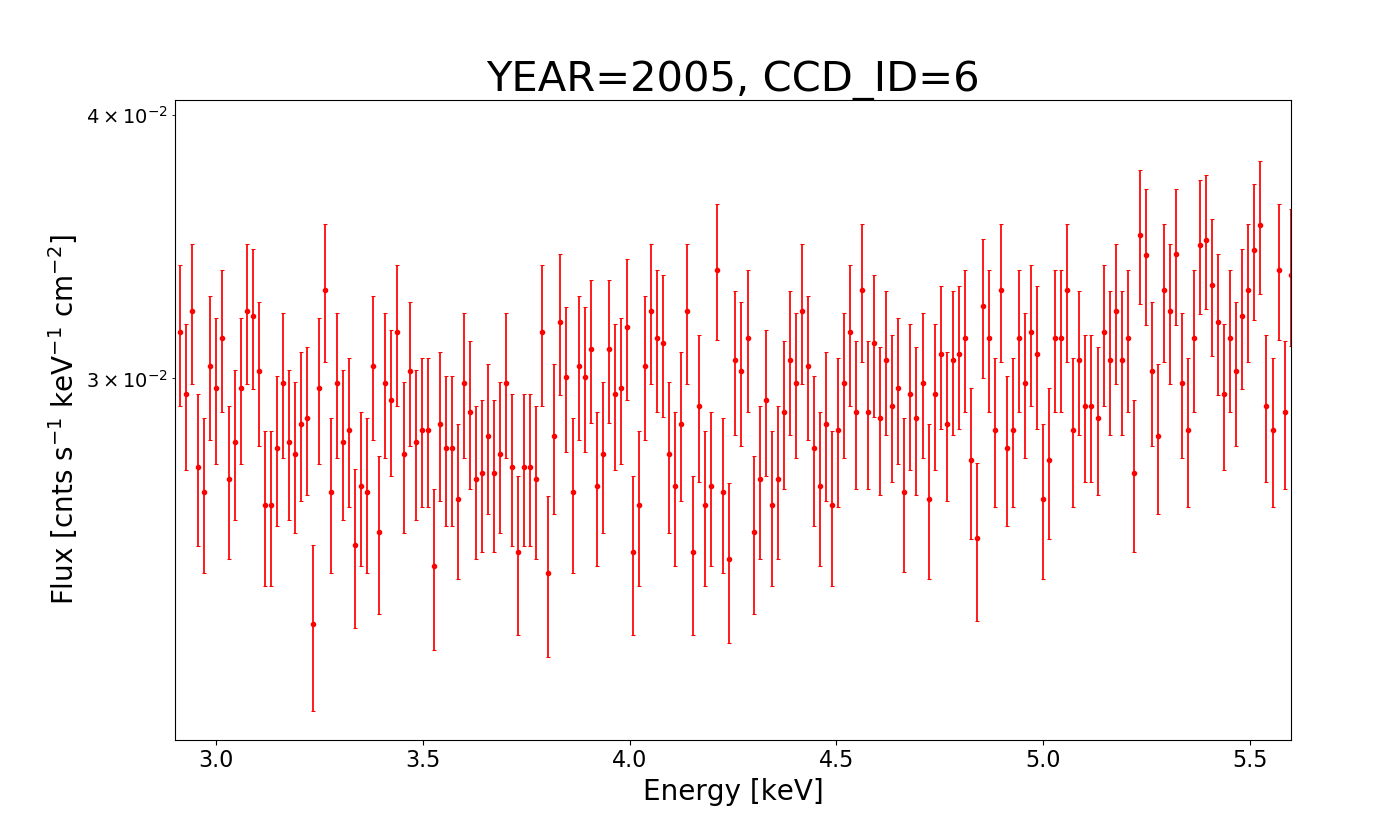}
\caption{Stowed spectrum for each CCD from the stowed observation in 2005.}
\label{fig:stow2005}
\end{figure*}

\begin{figure*}[ht] 
\includegraphics[width=18.cm]{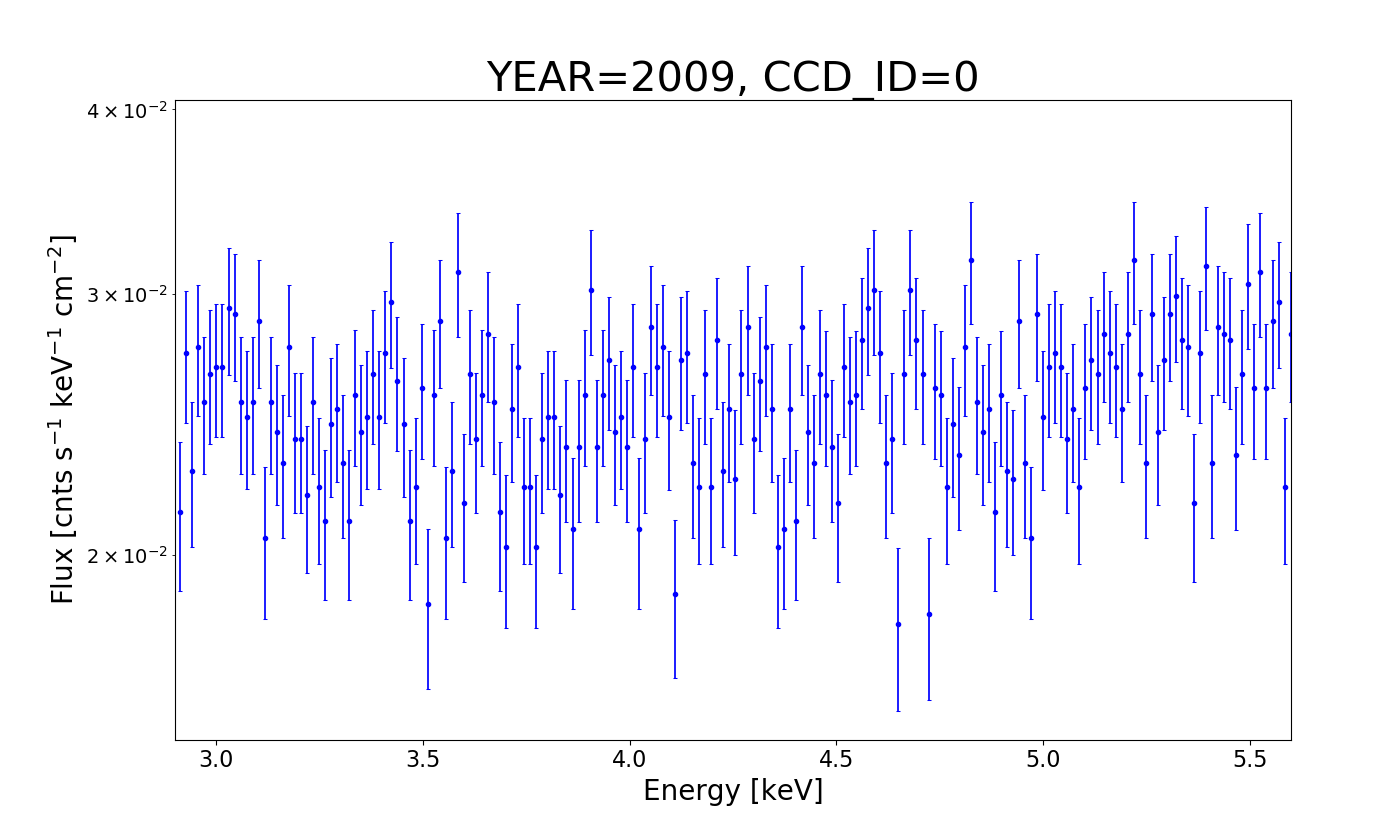}
\includegraphics[width=9.cm]{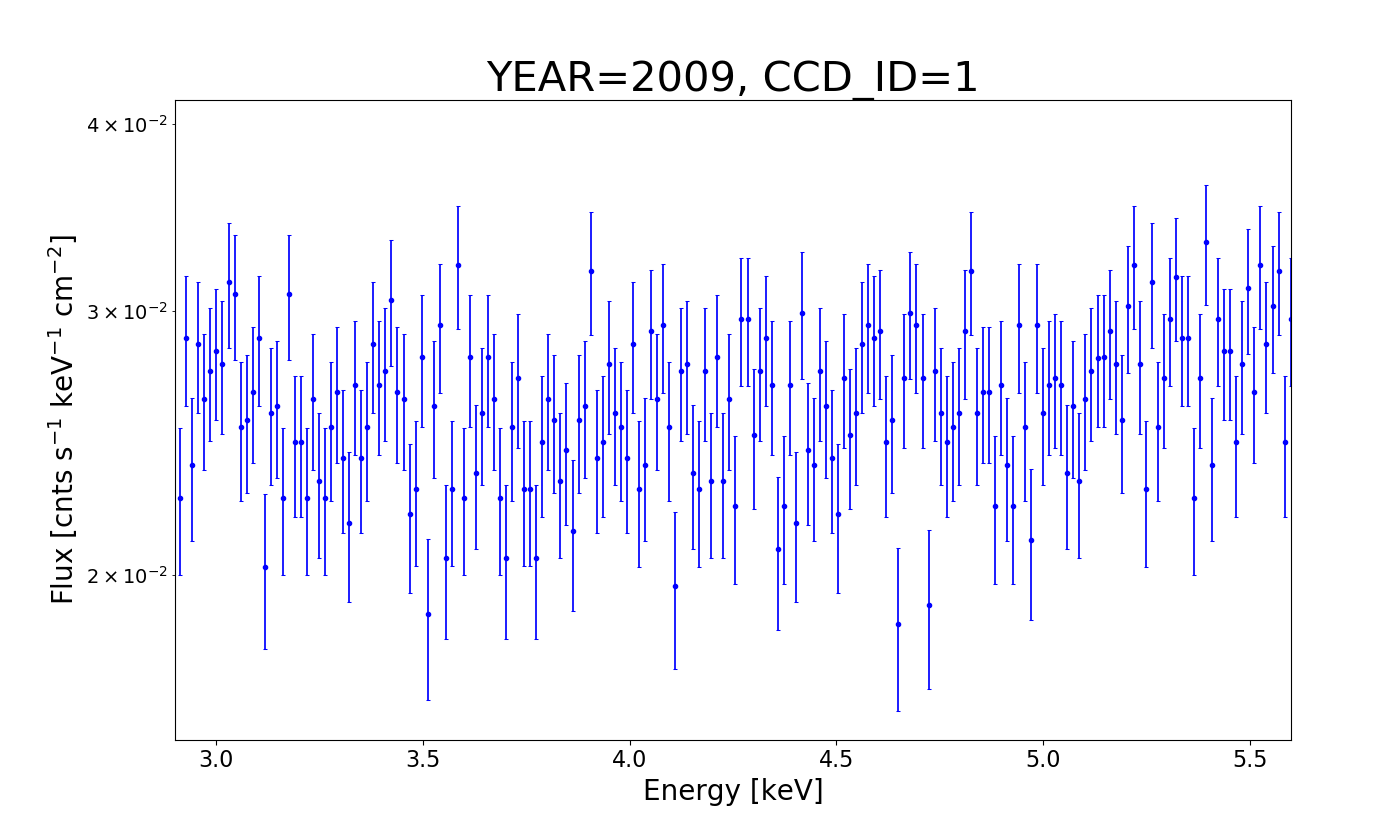}
\includegraphics[width=9.cm]{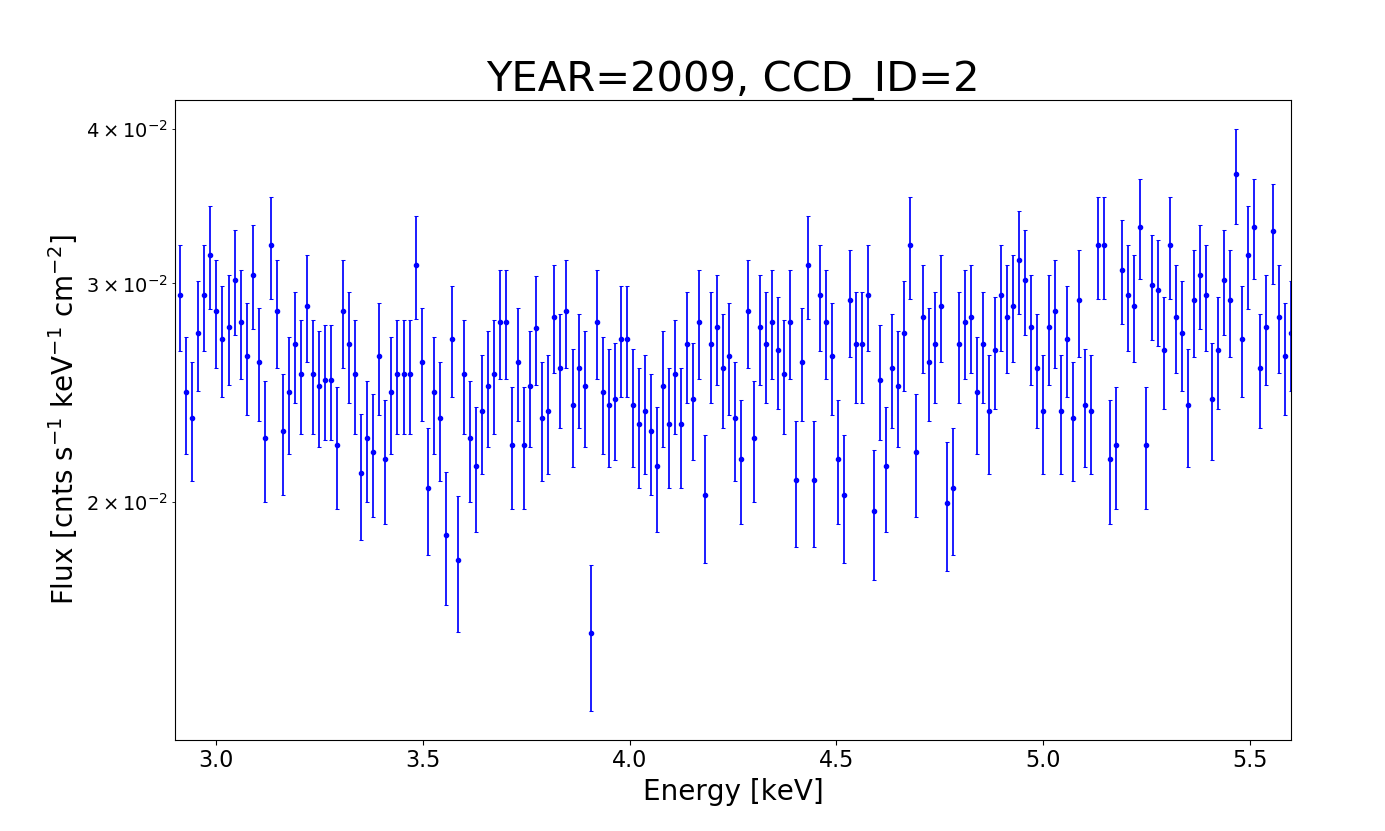}
\includegraphics[width=9.cm]{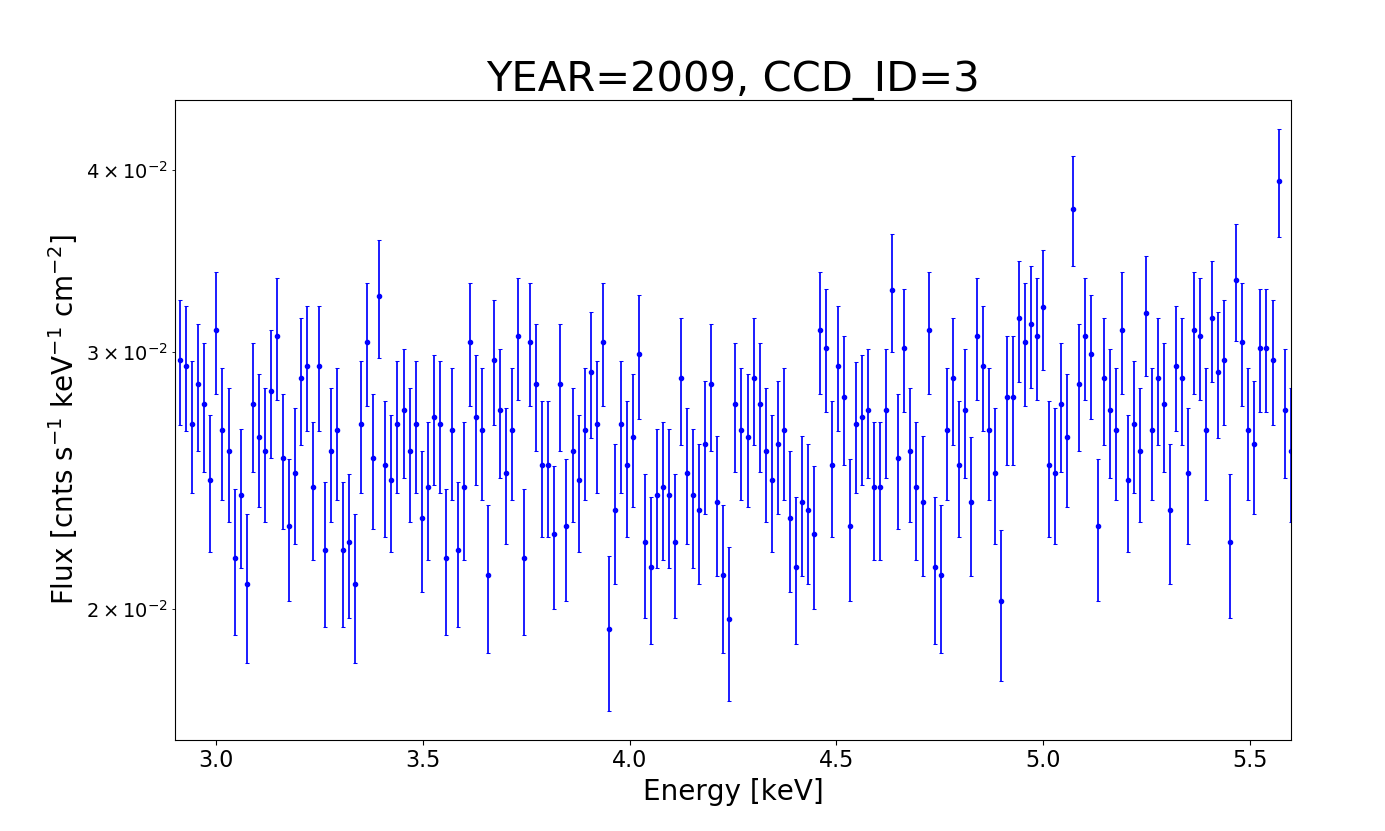}
\includegraphics[width=9.cm]{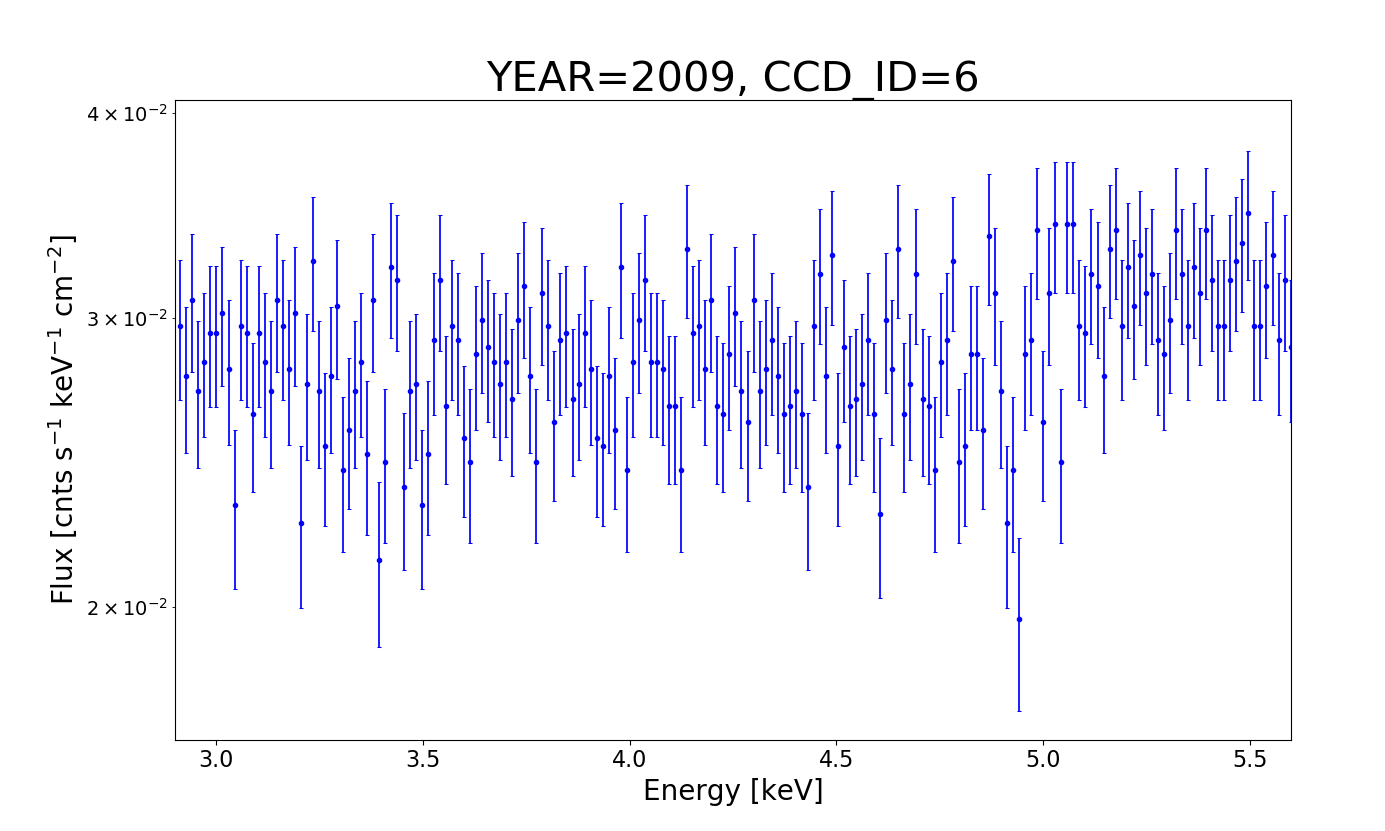}
\caption{Stowed spectrum for each CCD from the stowed observation in 2009.}
\label{fig:stow2009}
\end{figure*}

\begin{figure*}[ht] 
\includegraphics[width=9.cm]{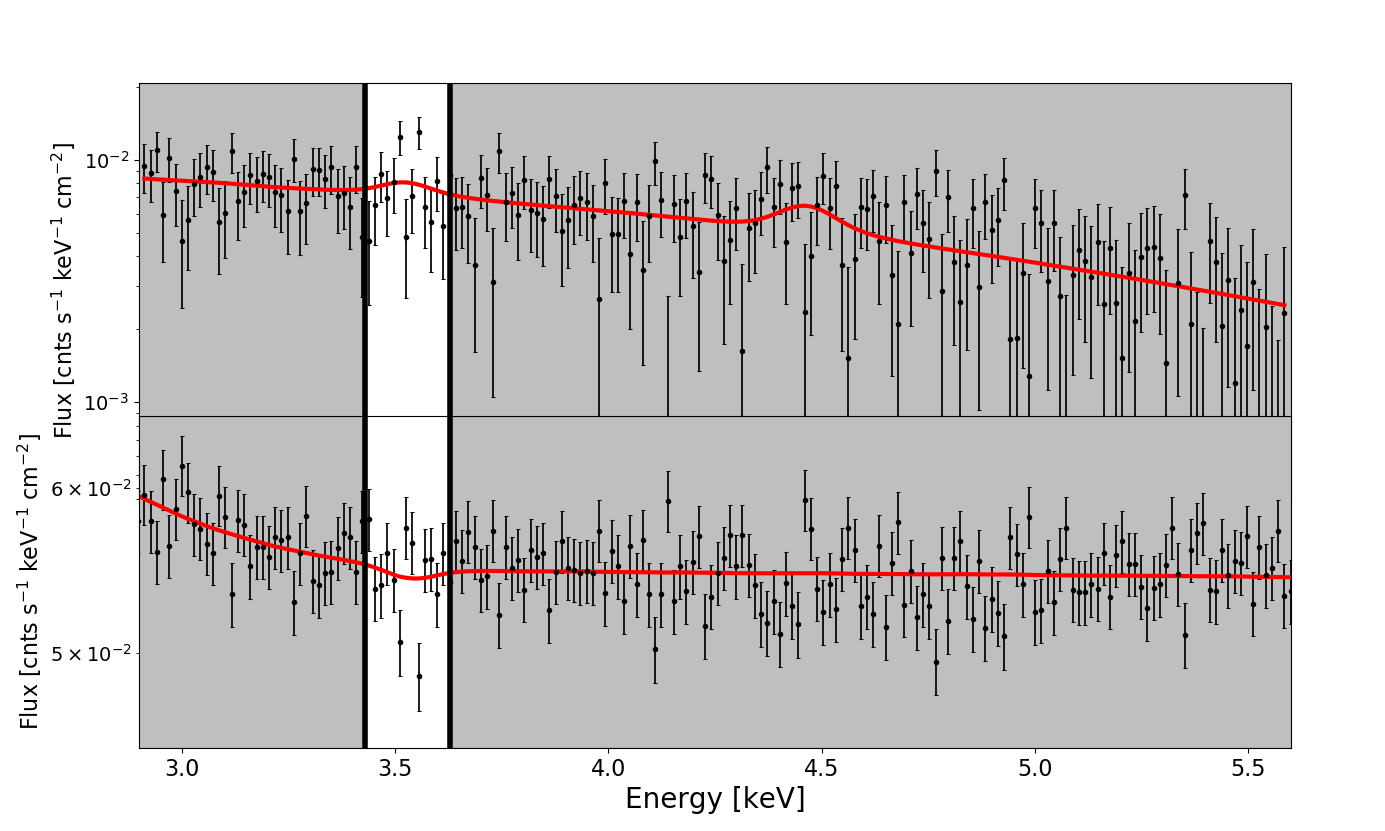}
\includegraphics[width=9.cm]{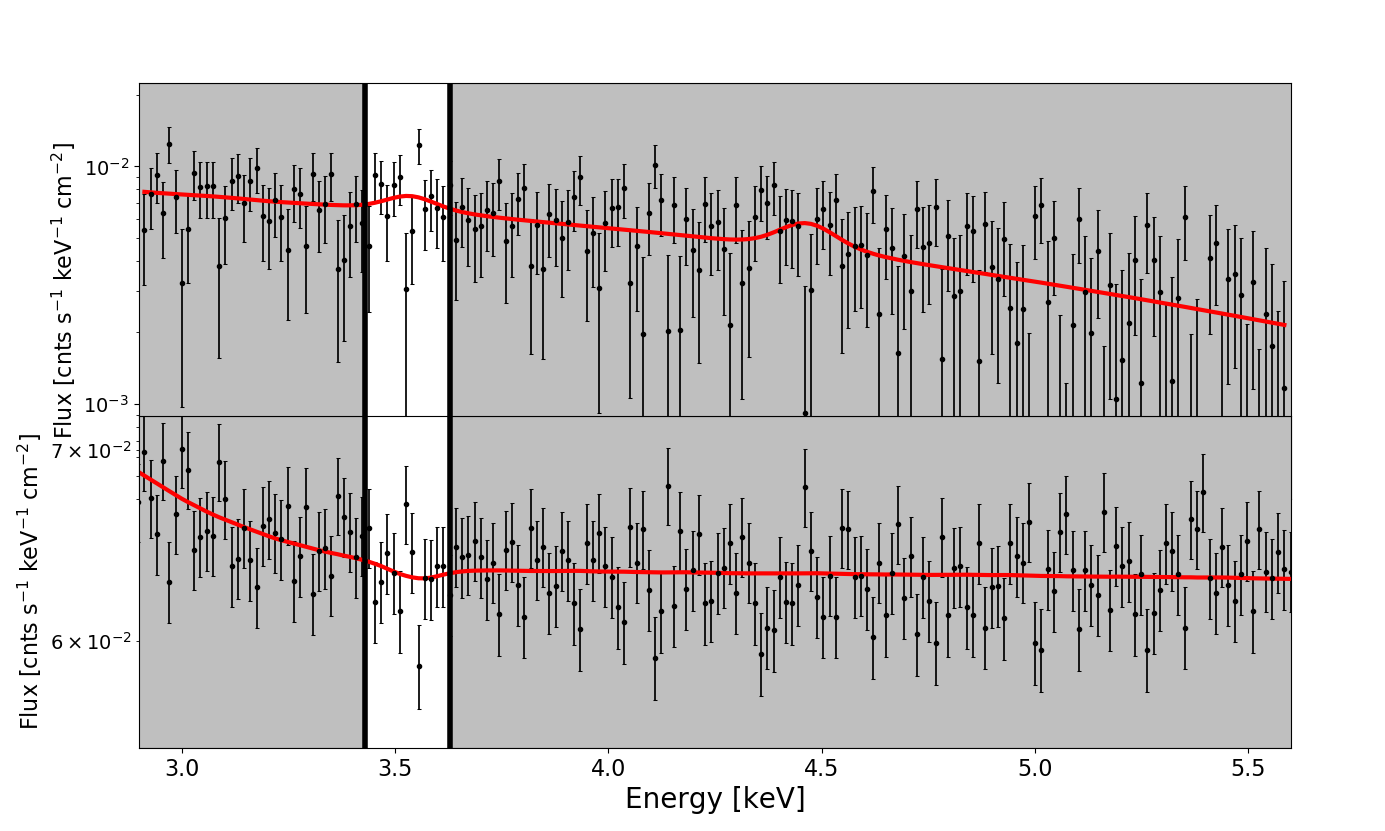}
\includegraphics[width=9.cm]{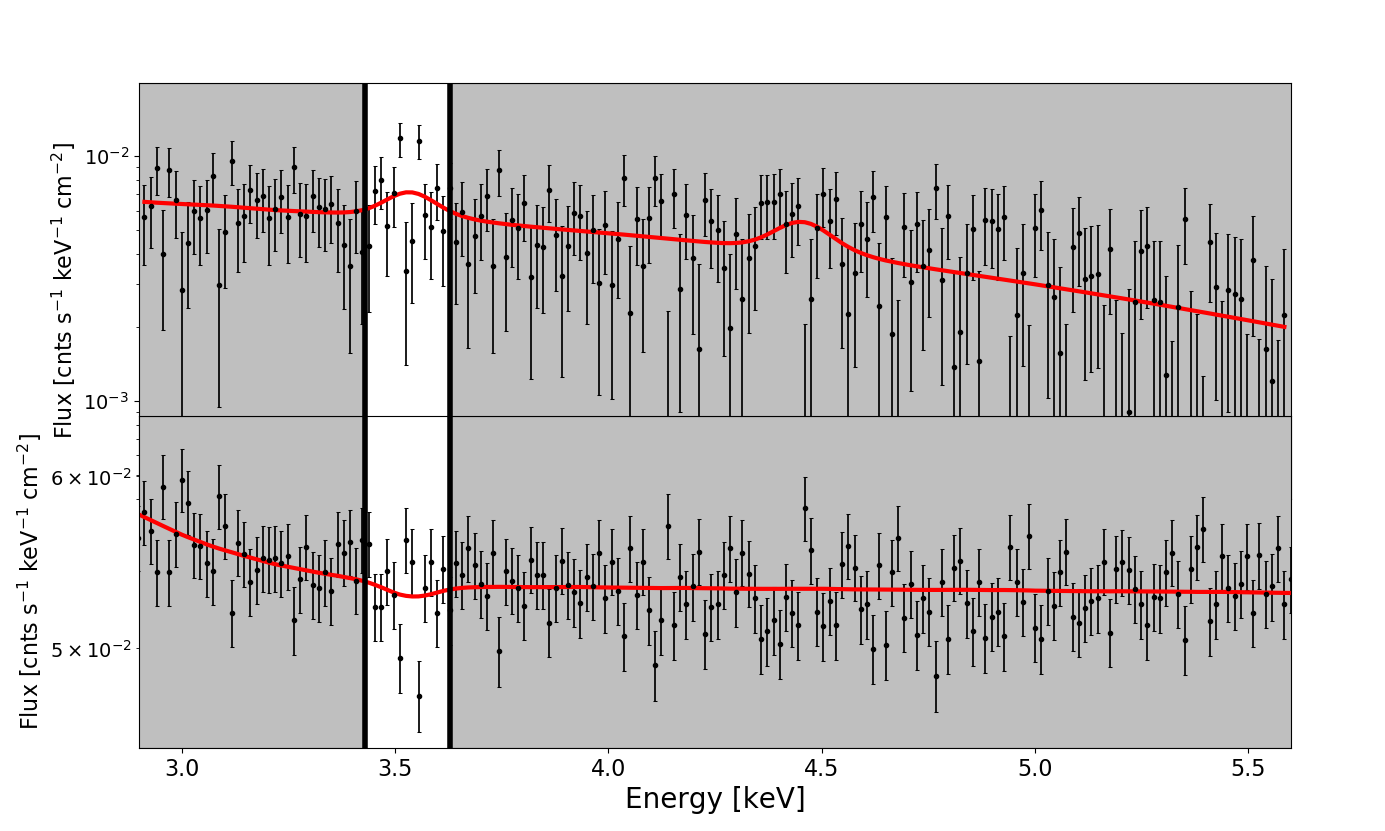}
\includegraphics[width=9.cm]{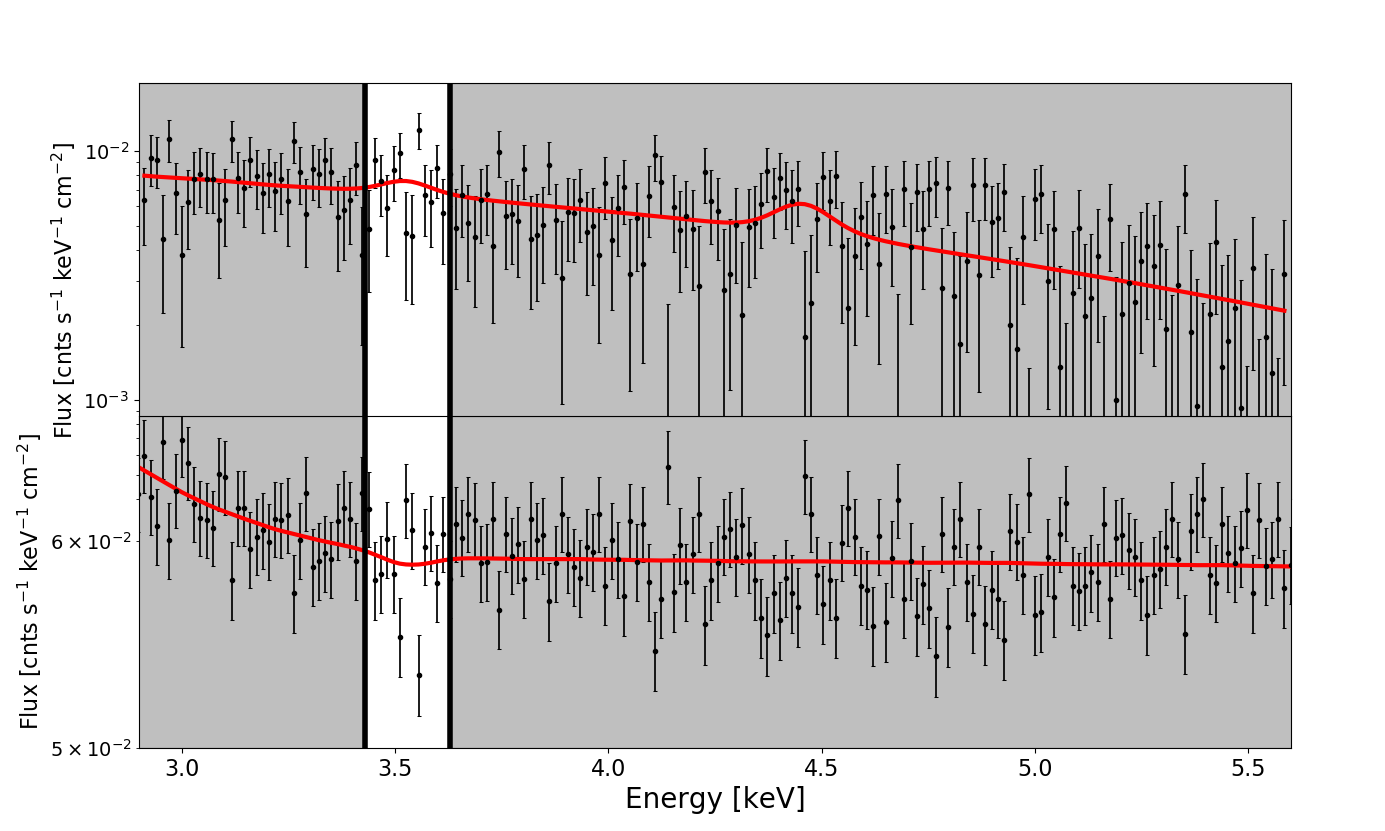}
\caption{Background-subtracted spectrum of each angular distance bin modeled with a 3.5 keV line, compared to the corresponding background spectrum containing the 3.5 keV dip. These further showcase the correspondence between the anomalous, dipped data points in each background and each background-subtracted spectrum's apparent emission line data points. \textbf{Top Left:} Bin 1, \textbf{Top Right:} Bin 2, \textbf{Bottom Left:} Bin 3, \textbf{Bottom Right:} Bin 4.}
\label{fig:dip_bins}
\end{figure*}

\begin{figure*}[ht] 
\includegraphics[width=9.cm]{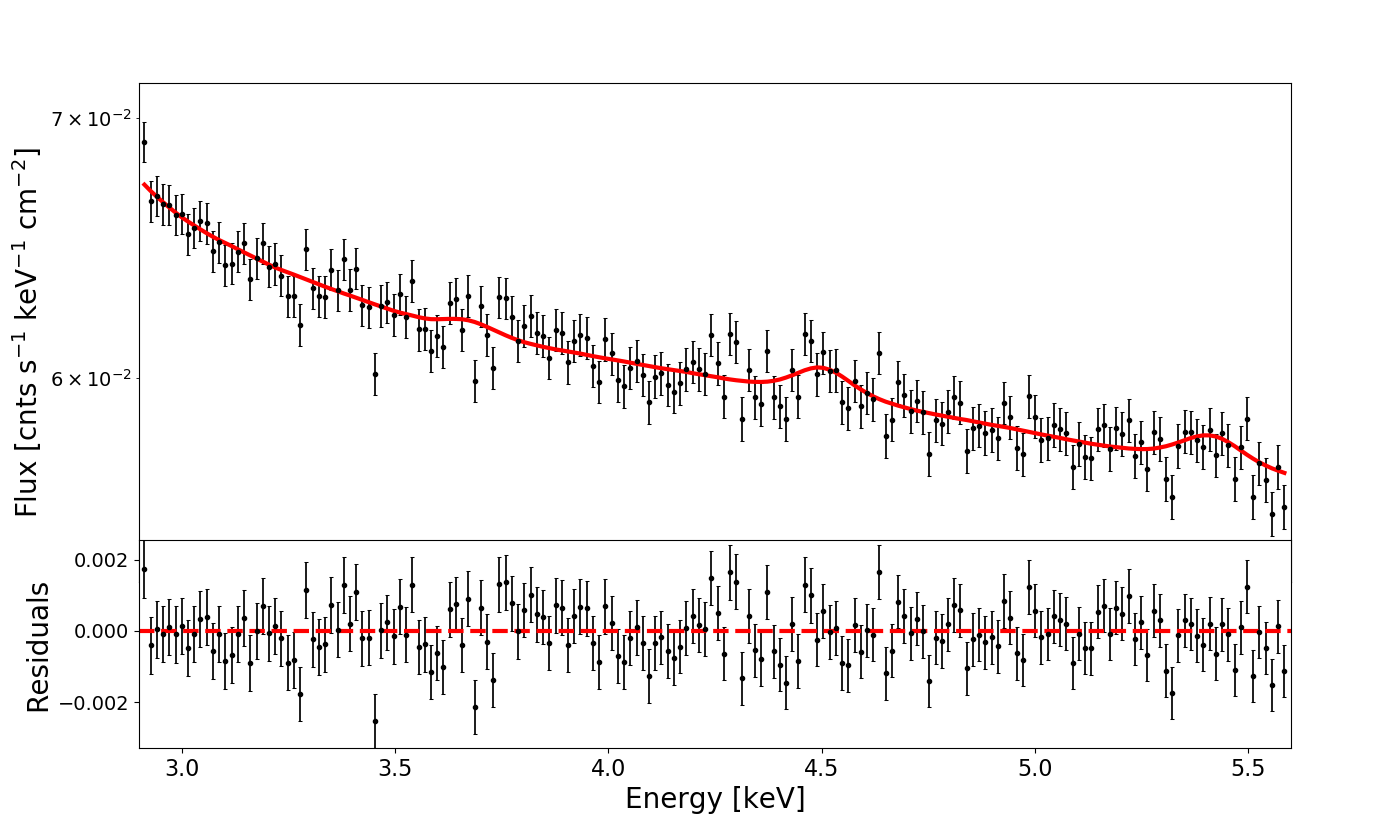}
\includegraphics[width=9.cm]{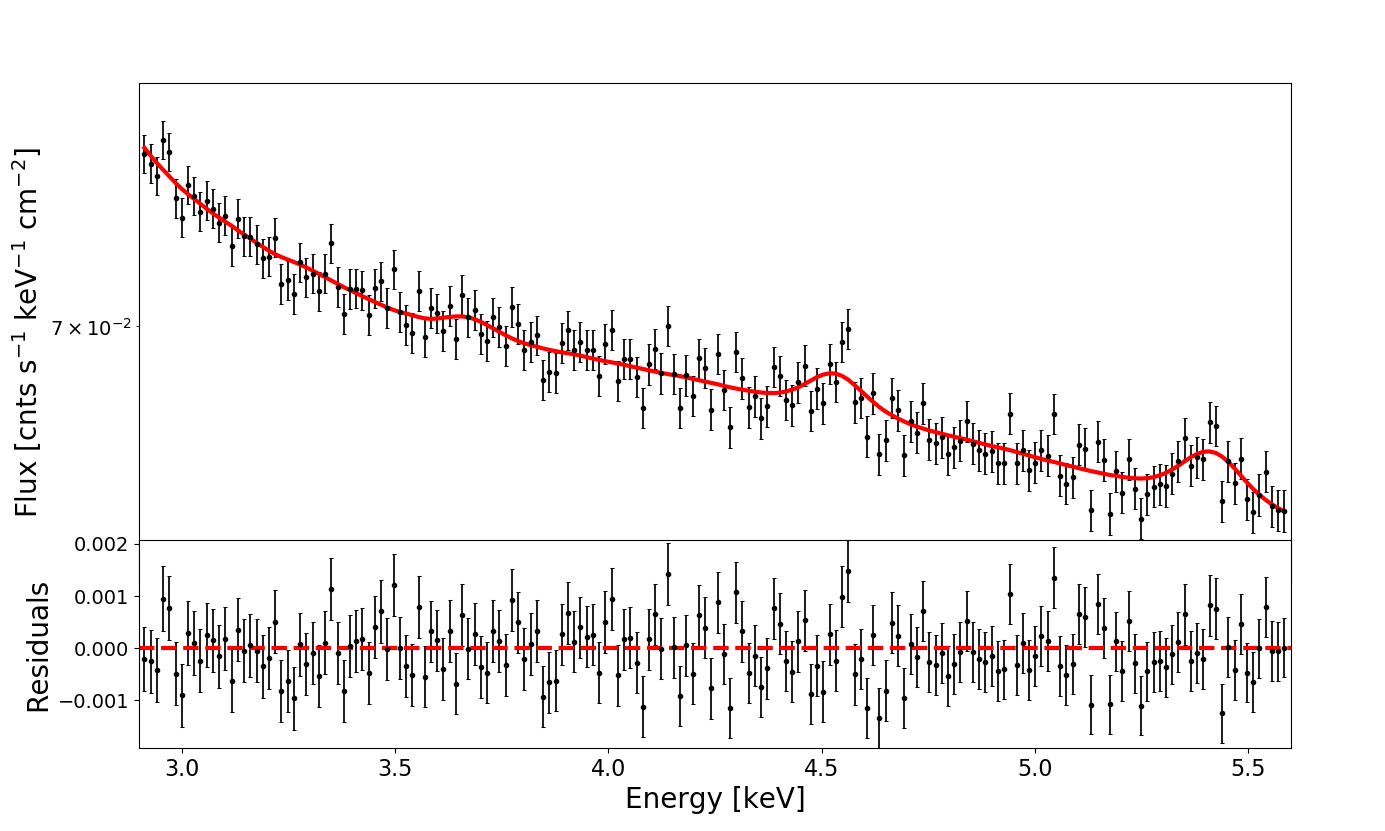}
\includegraphics[width=9.cm]{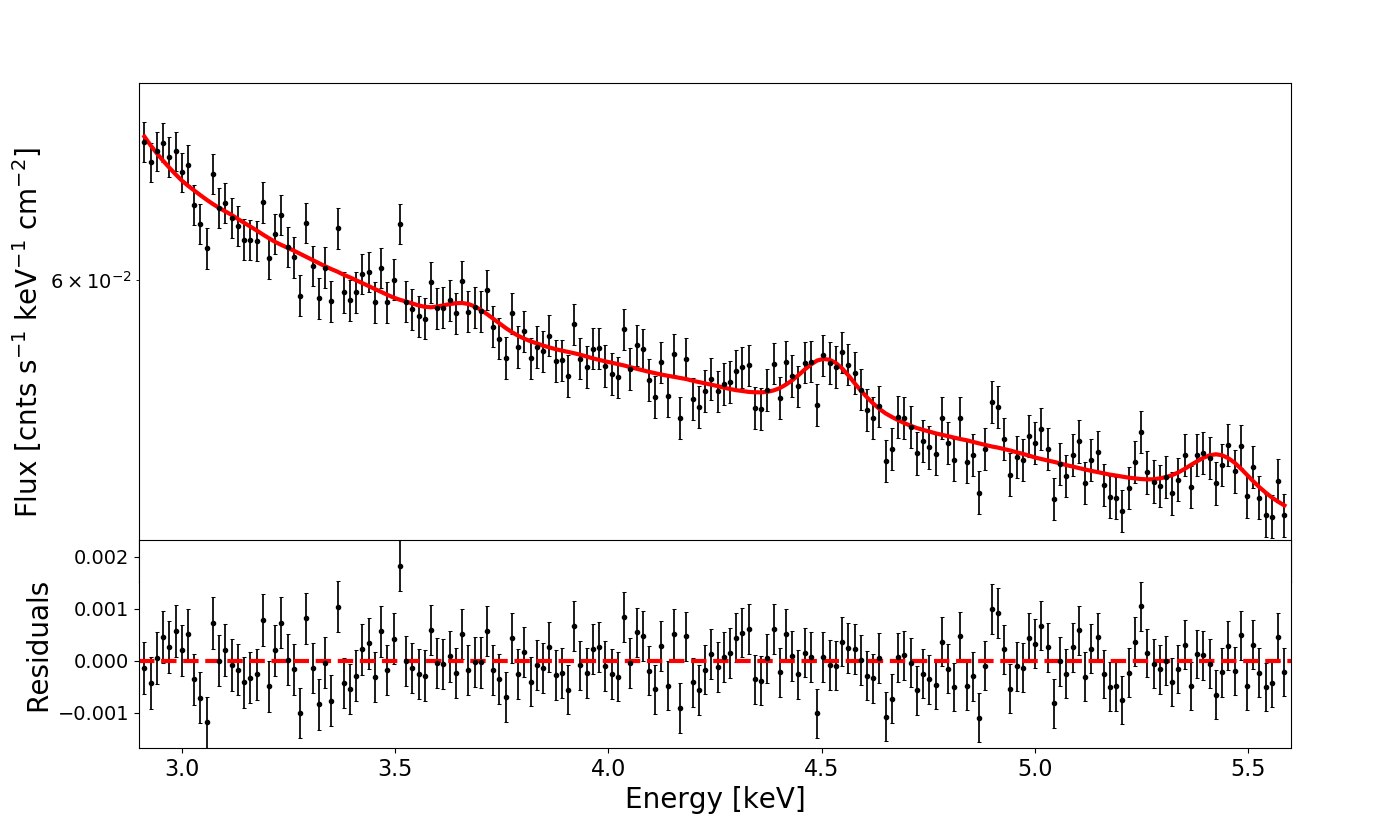}
\includegraphics[width=9.cm]{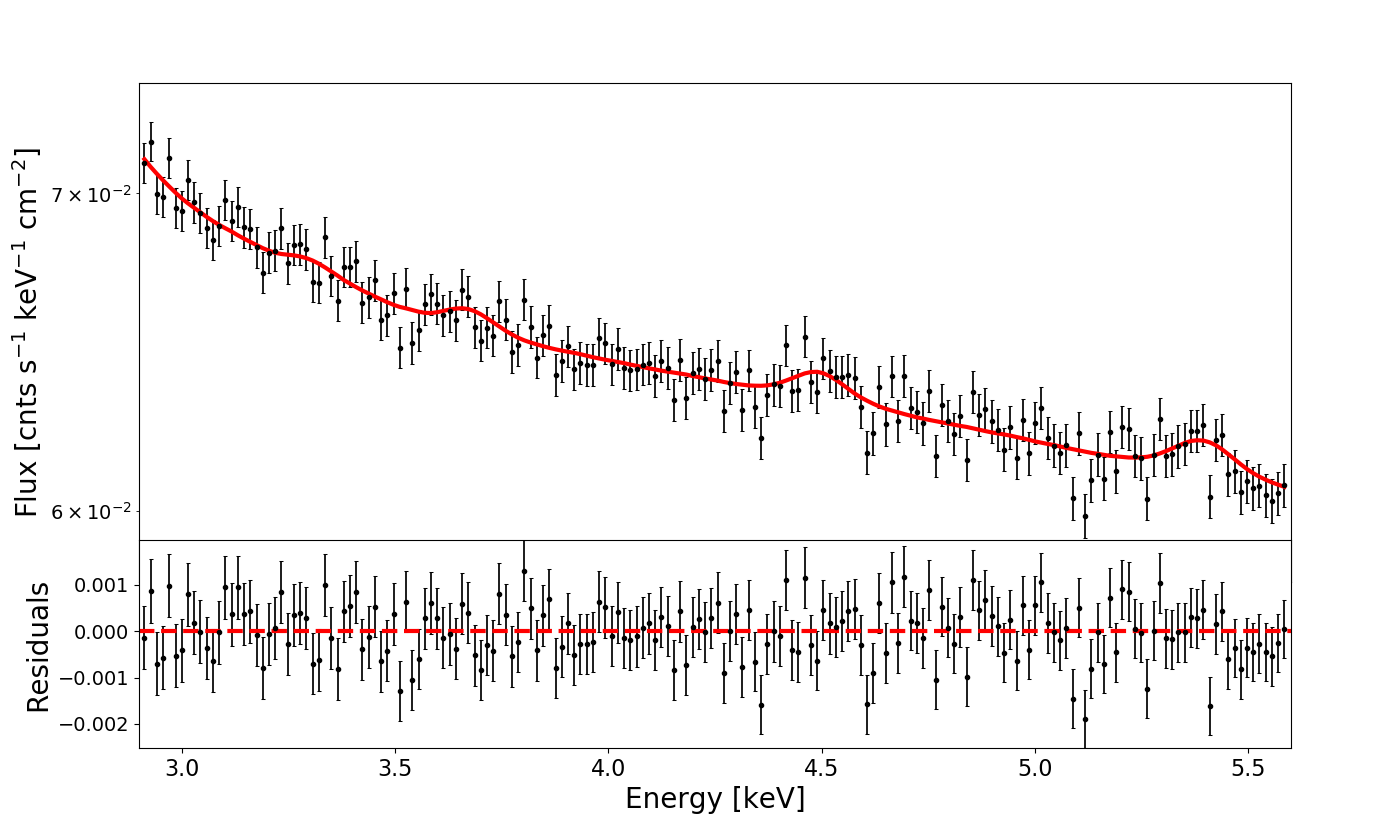}
\caption{Background-modeled spectra. \textbf{Top Left:} Bin 1, \textbf{Top Right:} Bin 2, \textbf{Bottom Left:} Bin 3, \textbf{Bottom Right:} Bin 4.}
\label{fig:bins_mod_line}
\end{figure*}

\begin{figure*}[ht] 
\includegraphics[width=9.cm]{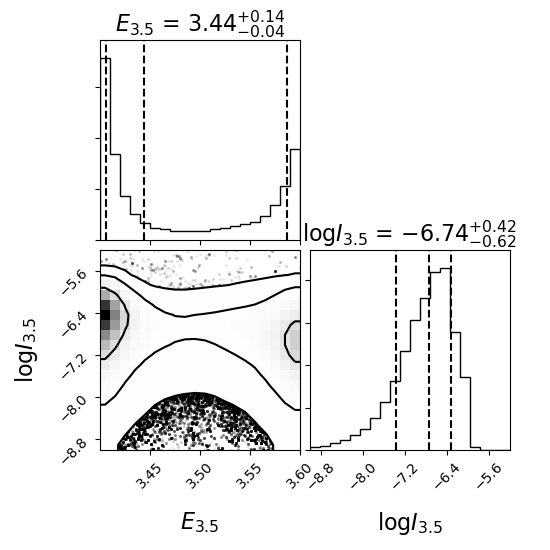}
\includegraphics[width=9.cm]{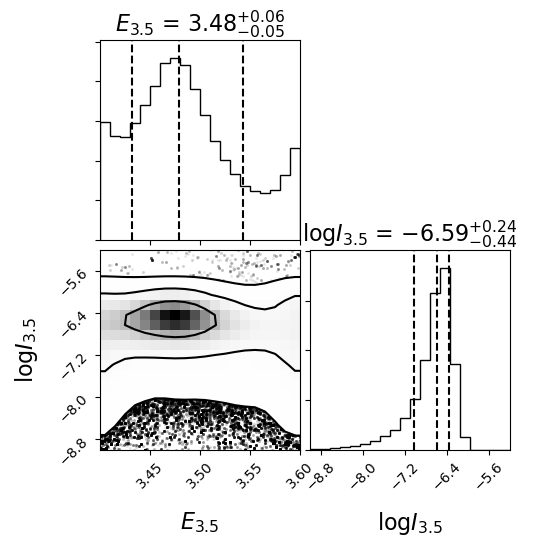}
\includegraphics[width=9.cm]{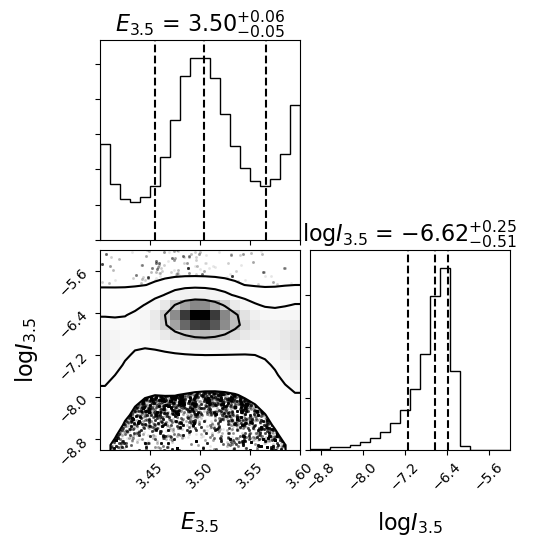}
\includegraphics[width=9.cm]{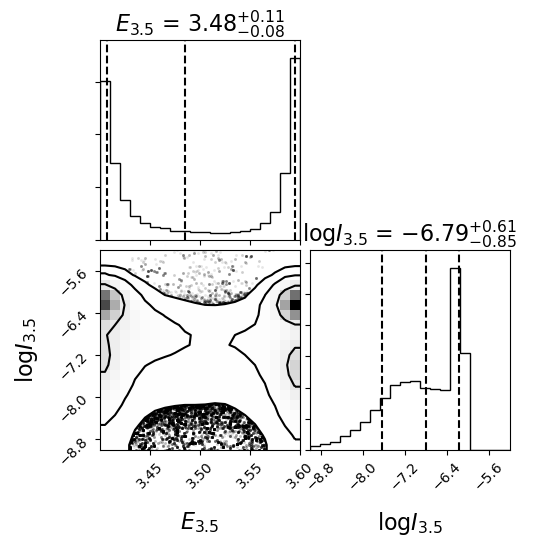}
\caption{MCMC contour plots for the background-modeled spectra with 3.5 keV line energy free to vary between 3.4--3.6 keV. \textbf{Top Left:} Bin 1, \textbf{Top Right:} Bin 2, \textbf{Bottom Left:} Bin 3, \textbf{Bottom Right:} Bin 4.}
\label{fig:bins_chain_bgmodel}
\end{figure*}


\begin{figure*}[t!]
\centering
\includegraphics[width=8.cm]{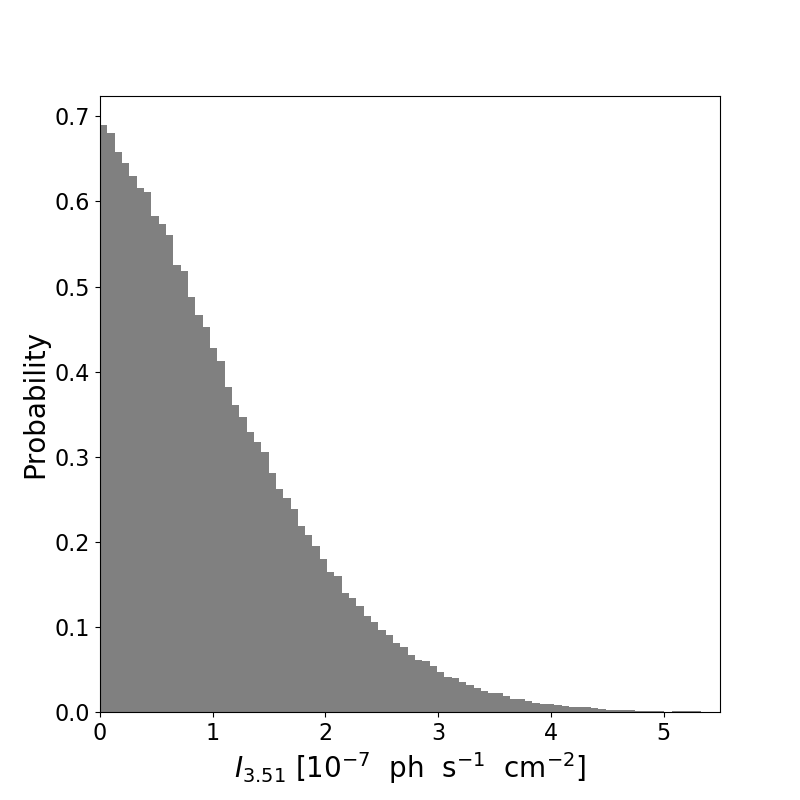}
\includegraphics[width=8.cm]{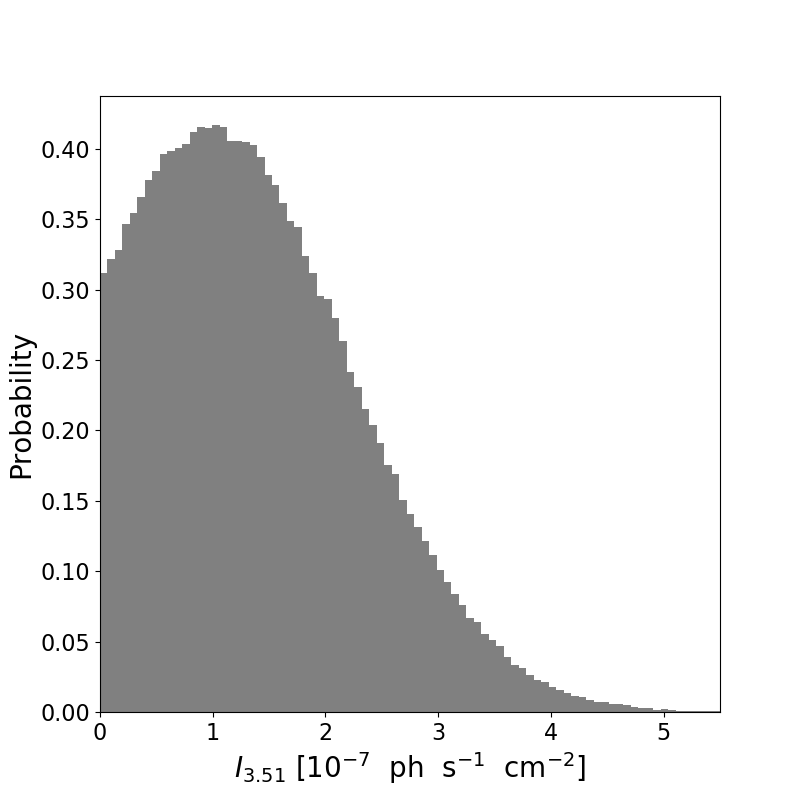}
\includegraphics[width=8.cm]{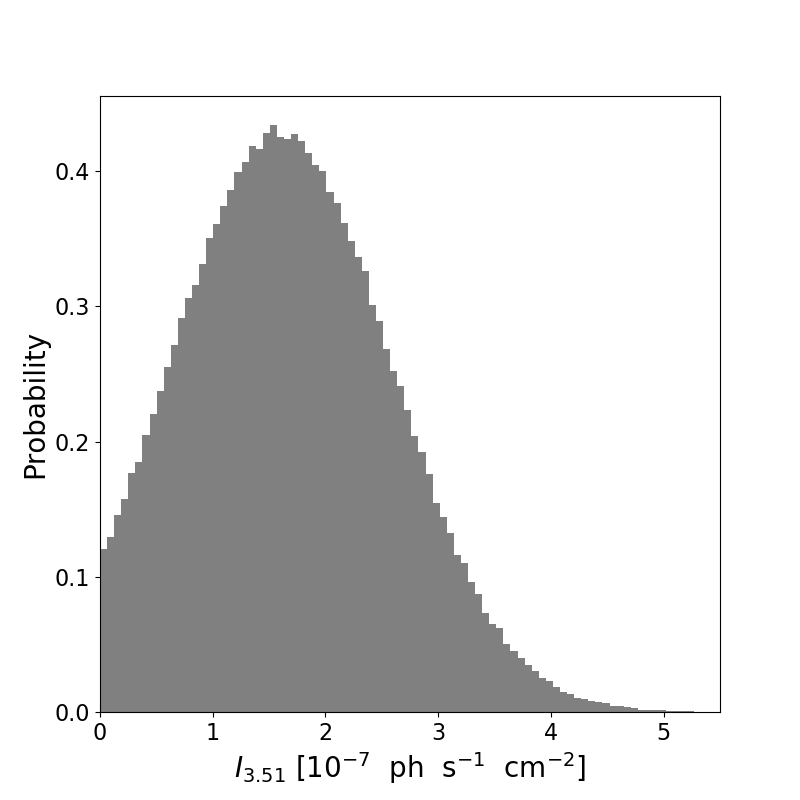}
\includegraphics[width=8.cm]{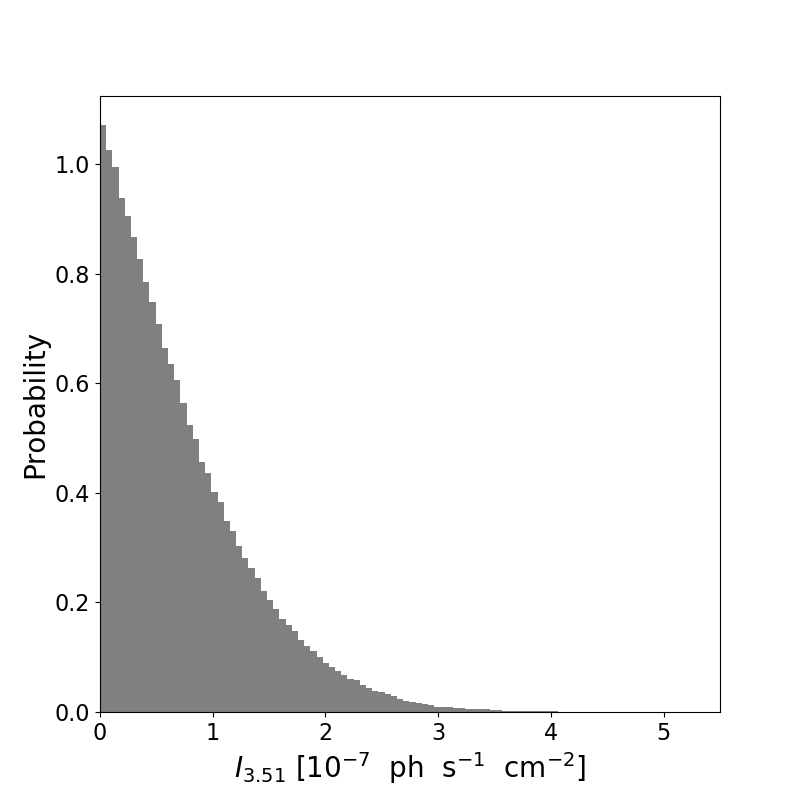}
\caption{Probability density distributions for 3.51 keV line flux. \textbf{Top Left:} Bin 1, \textbf{Top Right:} Bin 2, \textbf{Bottom Left:} Bin 3, \textbf{Bottom Right:} Bin 4.}
\label{fig:bins_lineflux_prob}
\end{figure*}


\begin{figure*}[ht] 
\includegraphics[width=9.cm]{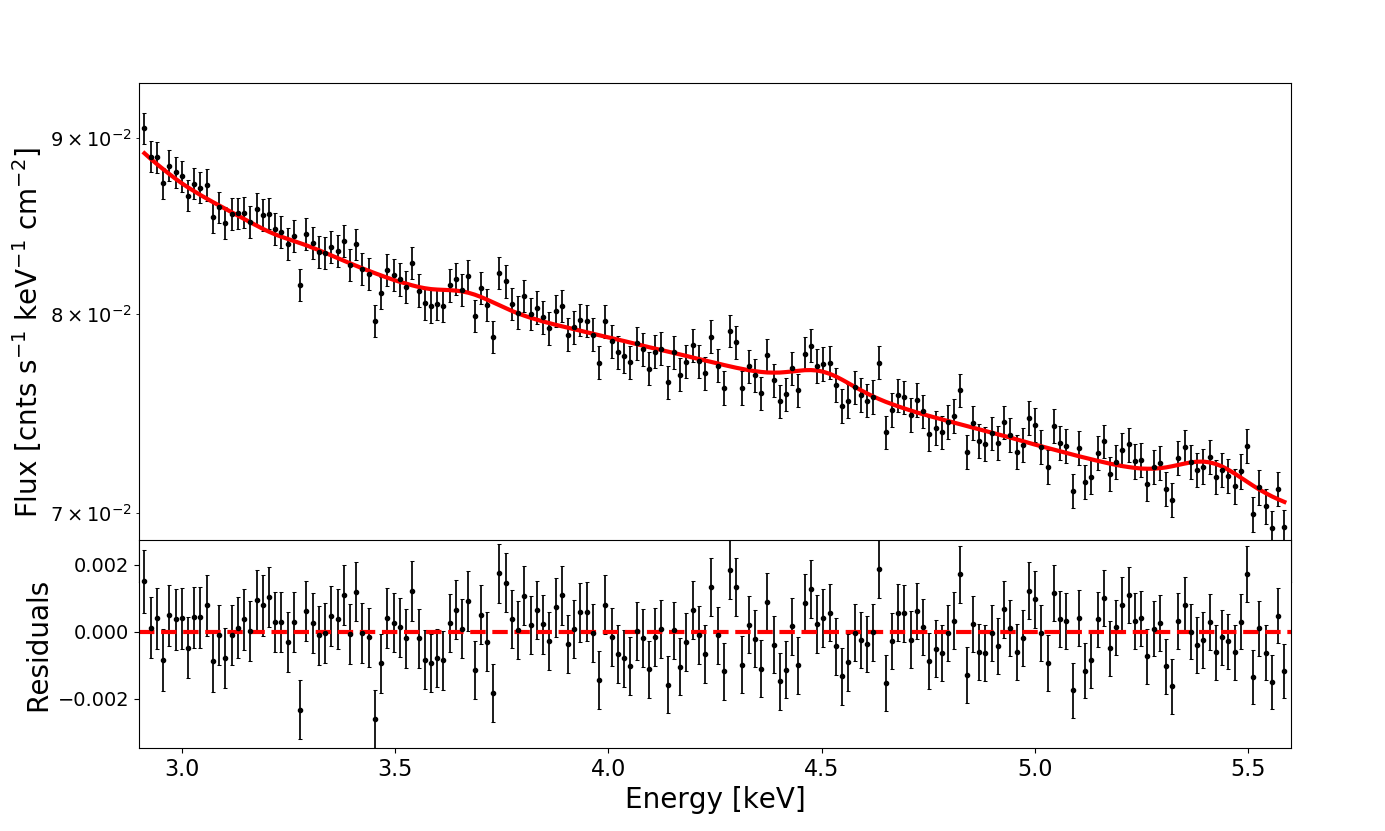}
\includegraphics[width=9.cm]{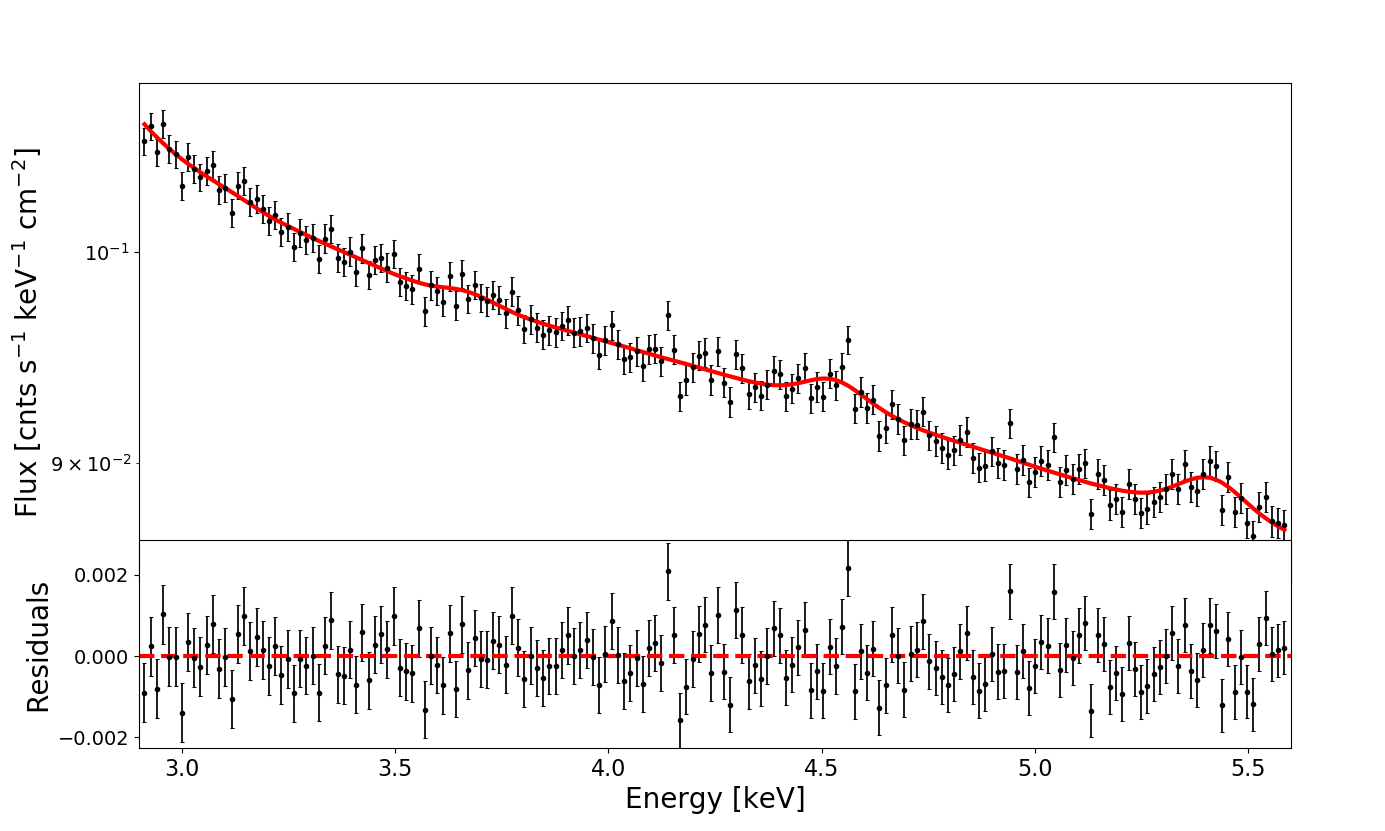}
\includegraphics[width=9.cm]{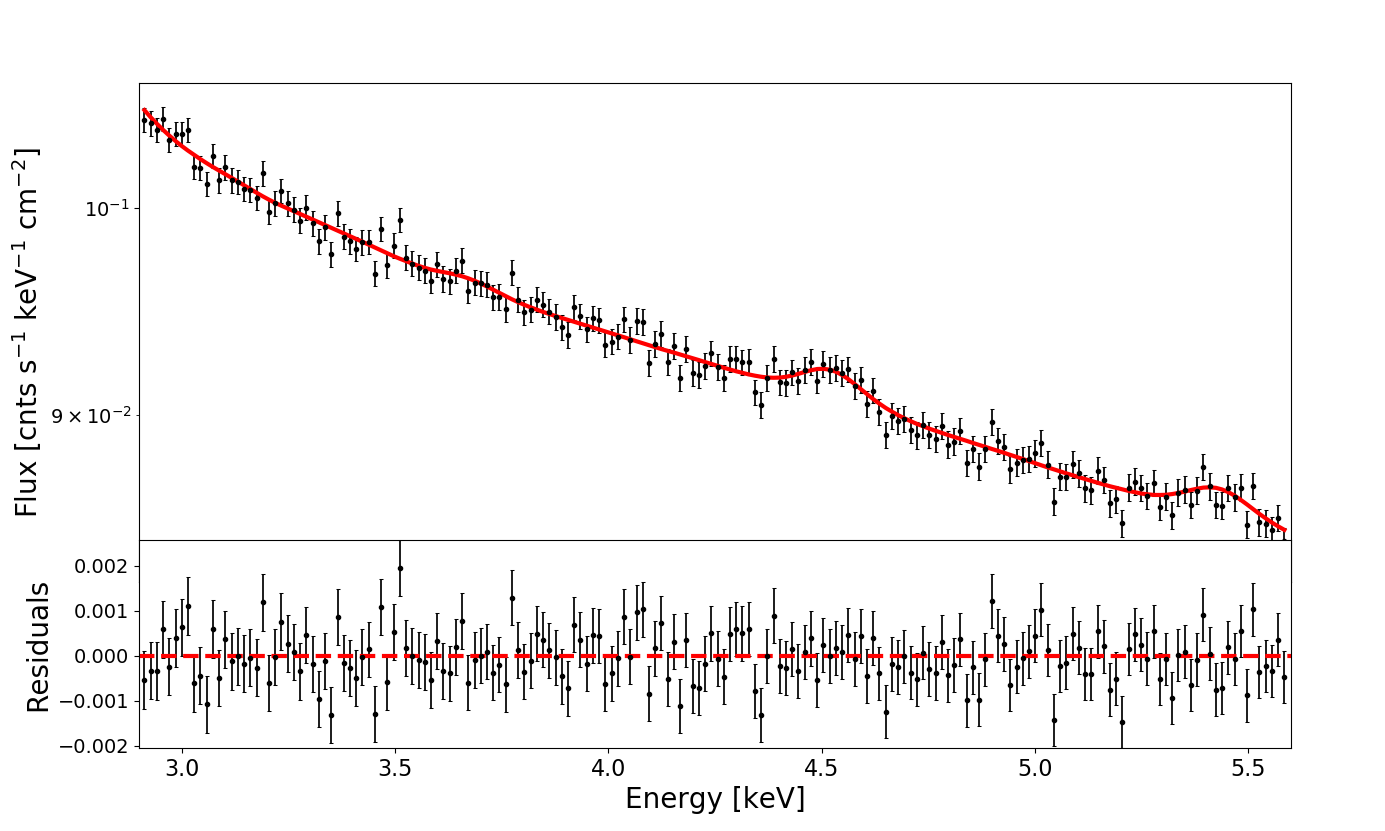}
\includegraphics[width=9.cm]{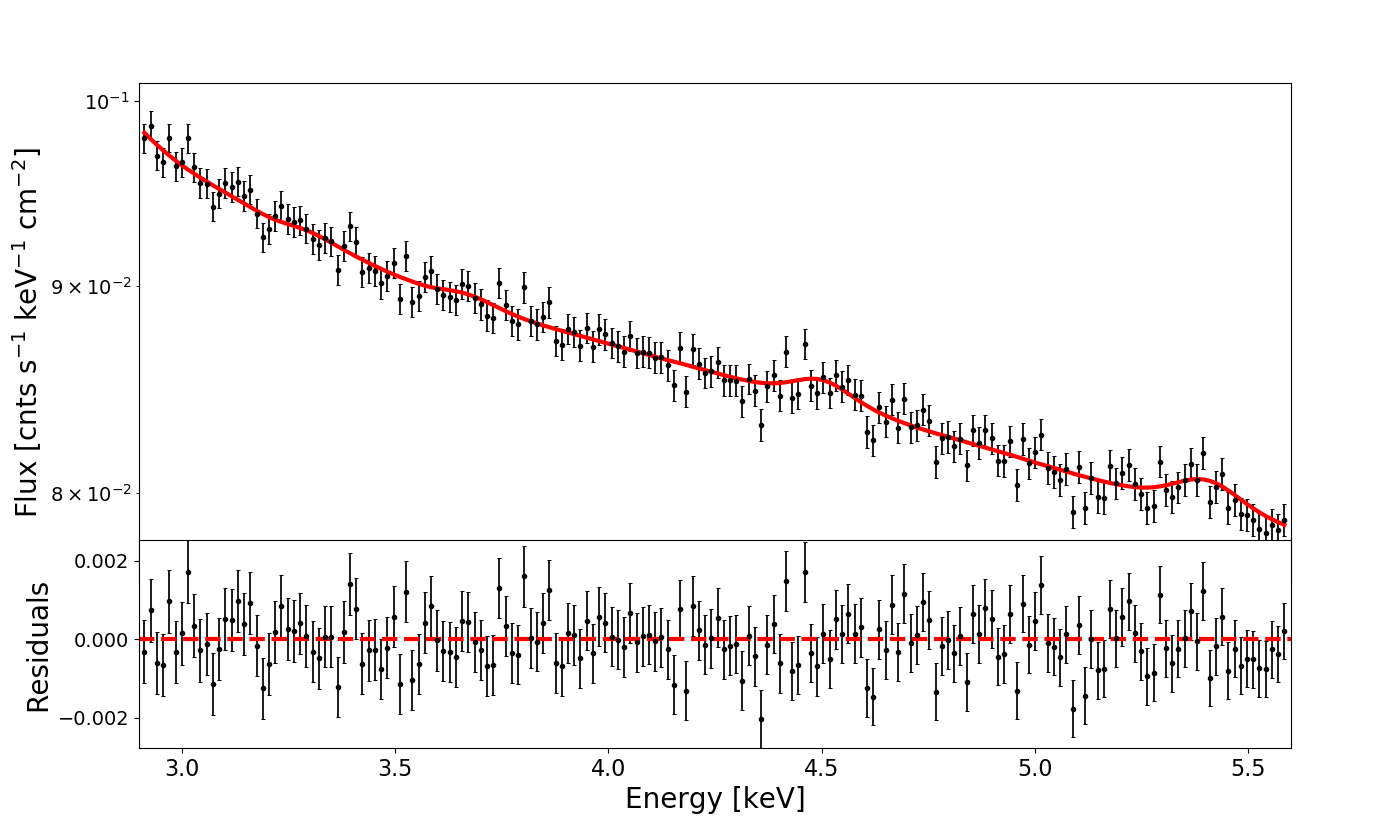}
\caption{\textbf{Without removing point-sources}: Background-modeled spectra. \textbf{Top Left:} Bin 1, \textbf{Top Right:} Bin 2, \textbf{Bottom Left:} Bin 3, \textbf{Bottom Right:} Bin 4.}
\label{fig:bins_cxb}
\end{figure*}

\begin{figure*}[ht] 
\includegraphics[width=9.cm]{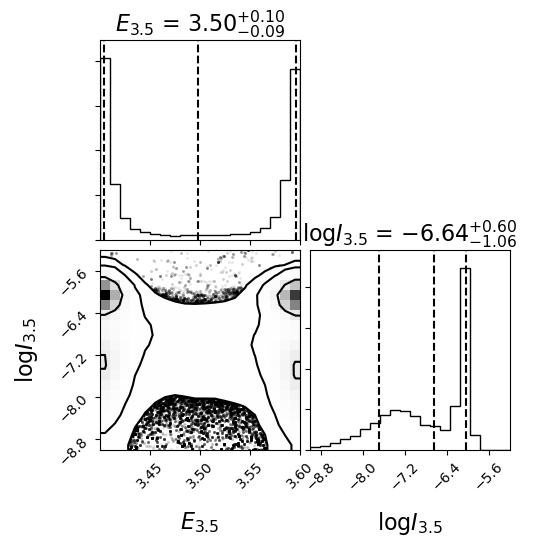}
\includegraphics[width=9.cm]{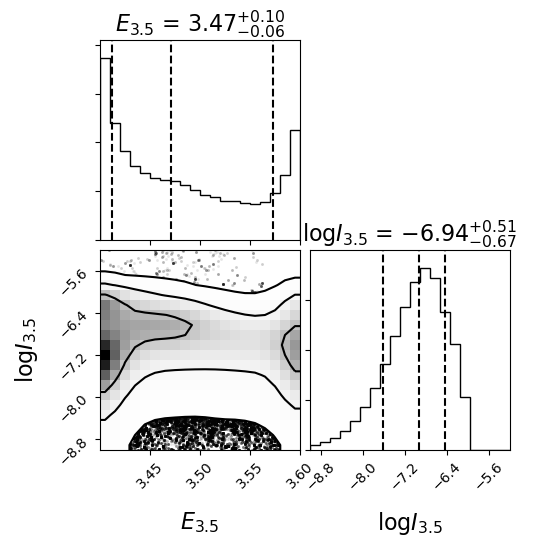}
\includegraphics[width=9.cm]{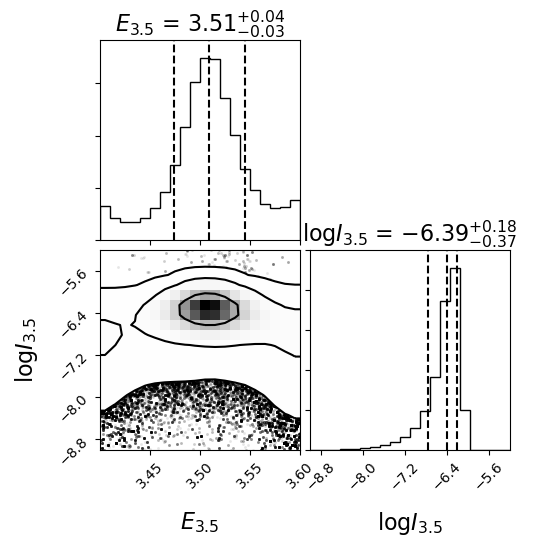}
\includegraphics[width=9.cm]{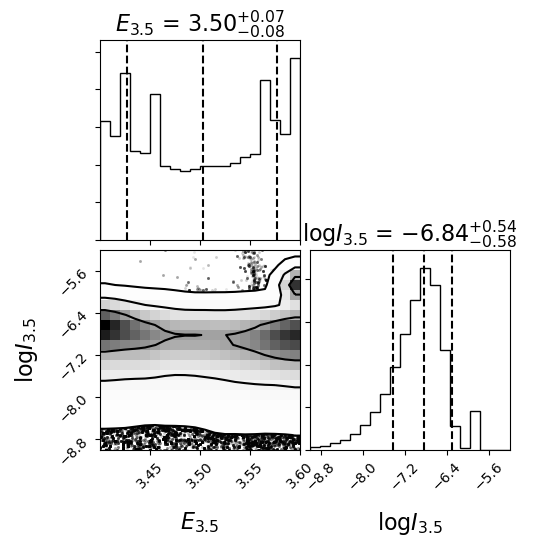}
\caption{\textbf{Without removing point-sources}: MCMC contour plots for the background-modeled spectra with 3.5 keV line energy free to vary between 3.4--3.6 keV. \textbf{Top Left:} Bin 1, \textbf{Top Right:} Bin 2, \textbf{Bottom Left:} Bin 3, \textbf{Bottom Right:} Bin 4.}
\label{fig:bins_chain_cxb}
\end{figure*}


\begin{figure*}[t!]
\centering
\includegraphics[width=8.cm]{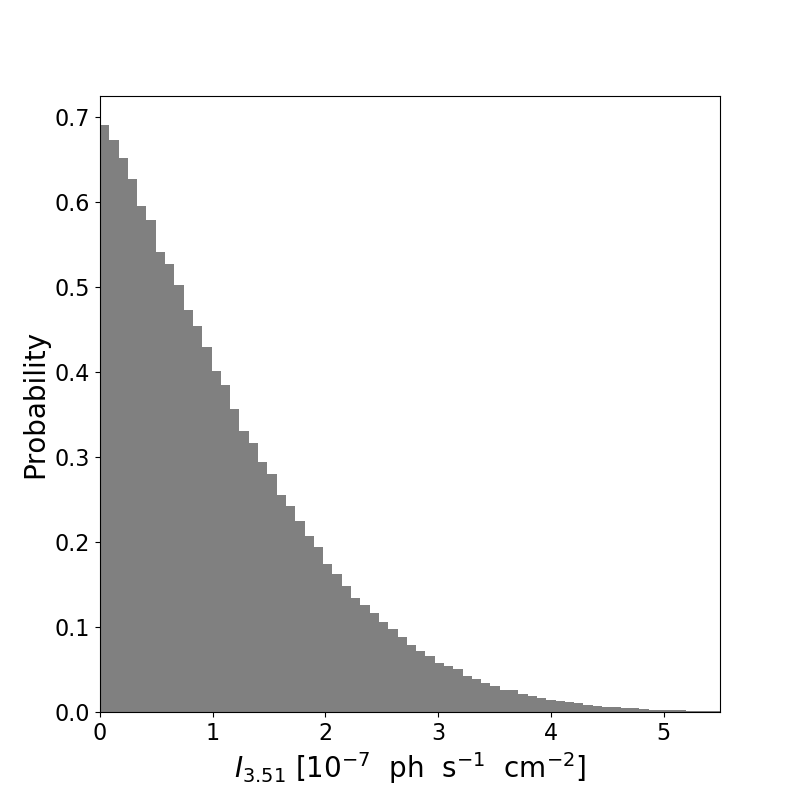}
\includegraphics[width=8.cm]{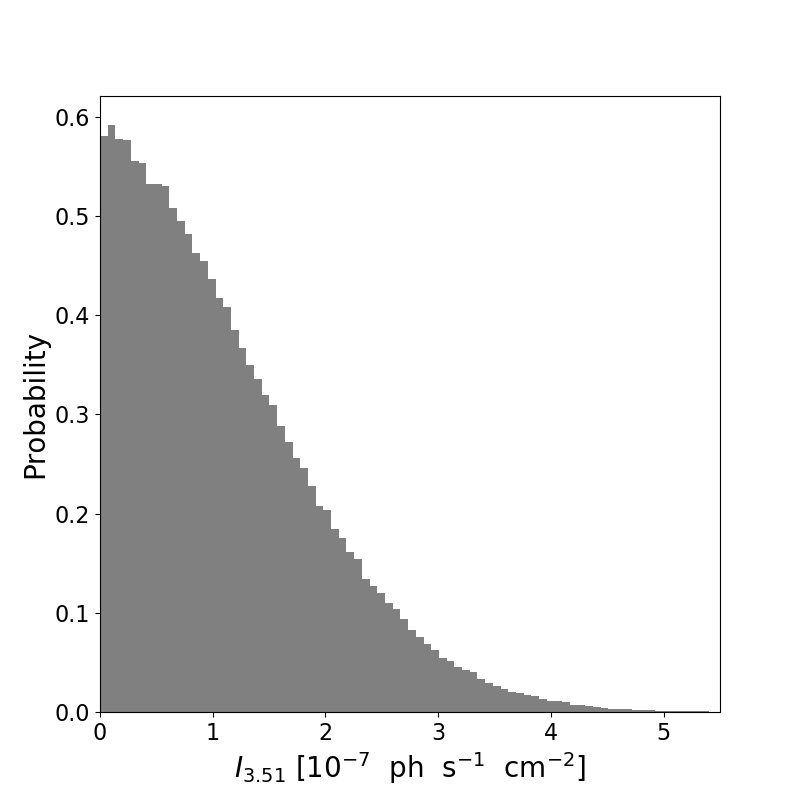}
\includegraphics[width=8.cm]{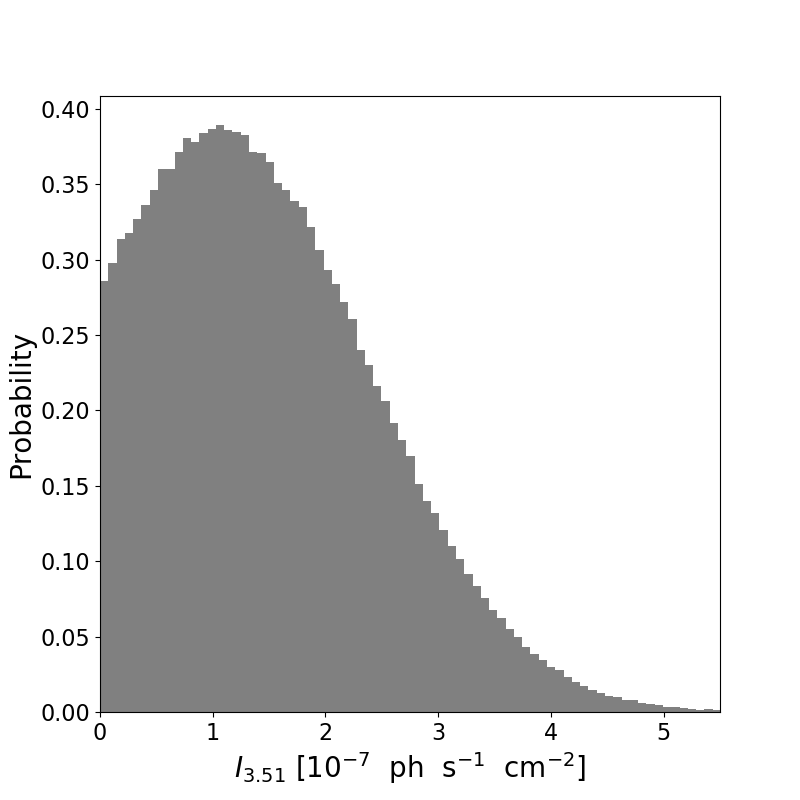}
\includegraphics[width=8.cm]{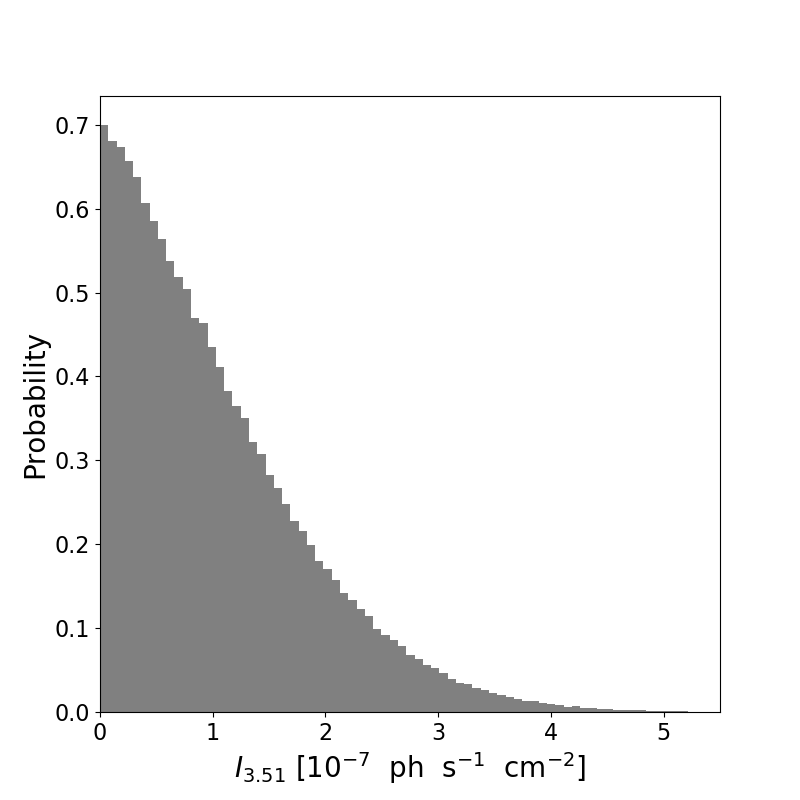}
\caption{\textbf{Without removing point-sources}: Probability density distributions for 3.51 keV line flux. \textbf{Top Left:} Bin 1, \textbf{Top Right:} Bin 2, \textbf{Bottom Left:} Bin 3, \textbf{Bottom Right:} Bin 4.}
\label{fig:cxb_bins_lineflux_prob}
\end{figure*}


\end{document}